\documentclass[11pt]{article}
\pdfoutput=1

\usepackage{jcappub}
\usepackage{lineno}
\usepackage{color, xcolor, appendix,latexsym,amsmath,amssymb,graphicx,booktabs,epsfig,hyperref,url,nicefrac,diagbox,multirow,relsize}
\usepackage{tablefootnote}

\usepackage{dsfont}
\newcommand{\realnum}{\mathds{R}}

\numberwithin{equation}{section}

\definecolor{MyBlue}{rgb}{0.15,0.15,0.70}
\hypersetup{
colorlinks=true,
citecolor=MyBlue,
linkcolor=MyBlue,
urlcolor=MyBlue
}
\definecolor{lightgray}{gray}{0.9}

\graphicspath{figures/} 

\title{Coupling elastic media to gravitational waves: an effective field theory approach}
\author[1,2]{Enis Belgacem,}
\author[1,2]{Michele Maggiore,}
\author[1]{Thomas Moreau}

\affiliation[1]{D\'epartement de Physique Th\'eorique,
Universit\'e de Gen\`eve, 24 quai Ansermet, CH-1211 Gen\`eve 4, Switzerland}
\affiliation[2]{Gravitational Wave Science Center (GWSC), Universit\'e de Gen\`eve, CH-1211 Geneva, Switzerland}

\emailAdd{enis.belgacem@unige.ch}
\emailAdd{michele.maggiore@unige.ch}
\emailAdd{thomas.moreau@etu.unige.ch}

\abstract{The interaction of a gravitational wave (GW) with an elastic body is usually described in terms of a GW ``force" driving the oscillations of the body's normal modes. However, this description is only possible for GW frequencies for which the response of the elastic body is dominated by a few normal modes.  At higher frequencies  the  normal modes  blend into a quasi-continuum  and a field-theoretical description, as pioneered by Dyson already in 1969, becomes necessary. However, since the metric perturbation $\hmn$  is an intrinsically relativistic object, a consistent coupling to GWs can only be obtained within a relativistic (and, in fact generally covariant) theory of elasticity. We develop such a formalism using the methods of modern   effective field theories, and we use it to 
provide a   derivation of the interaction of elastic bodies with GWs valid also in the high-frequency regime, providing a first-principle derivation  of Dyson's result (and partially correcting it).
We also stress that  the field-theoretical results are obtained
working in the TT frame, while the description in terms of a force driving the normal modes is only valid in the proper detector frame. We show how to transform the results between the two frames. Beside an intrinsic conceptual interest, these results are relevant to the computation of  the sensitivity of the recently proposed Lunar Gravitational Wave Antenna.}

\begin{document}


\newcommand{\mmin}{m_{\rm min}}
\newcommand{\mmax}{m_{\rm max}}

\newcommand{\dgw}{d_L^{\,\rm gw}}
\newcommand{\dem}{d_L^{\,\rm em}}
\newcommand{\dcom}{d_{\rm com}}
\newcommand{\dmax}{d_{\rm max}}
\newcommand{\hatO}{\hat{\Omega}}
\newcommand{\ngw}{n_{\rm GW}}
\newcommand{\ngal}{n_{\rm gal}}

\newcommand{\red}{\textcolor{red}} 

\newcommand{\blue}{\textcolor{blue}}
\newcommand{\green}{\textcolor{green}}
\newcommand{\cyan}{\textcolor{cyan}}
\newcommand{\magenta}{\textcolor{magenta}}
\newcommand{\yellow}{\textcolor{yellow}}

\newcommand{\hc}{{\cal H}}

\newcommand{\nn}{\nonumber}

\newcommand{\Lrr}{\Lambda_{\rm\scriptscriptstyle RR}}
\newcommand{\scDE}{{\textsc{DE}}}
\newcommand{\scR}{{\textsc{R}}}
\newcommand{\scM}{{\textsc{M}}}

\newcommand{\hla}{\hat{\lambda}}
\newcommand{\iBox}{\Box^{-1}}
\newcommand{\Stu}{St\"uckelberg }
\newcommand{\phib}{\bar{\phi}}

\newcommand{\Fmn}{F_{\mu\nu}}
\newcommand{\FMN}{F^{\mu\nu}}
\newcommand{\Am}{A_{\mu}}
\newcommand{\An}{A_{\nu}}
\newcommand{\Amu}{A_{\mu}}
\newcommand{\Anu}{A_{\nu}}
\newcommand{\AMU}{A^{\mu}}
\newcommand{\AN}{A^{\nu}}
\newcommand{\ANU}{A^{\nu}}

\renewcommand\({\left(}
\renewcommand\){\right)}
\renewcommand\[{\left[}
\renewcommand\]{\right]}
\newcommand\del{{\mbox {\boldmath $\nabla$}}}
\newcommand\n{{\mbox {\boldmath $\nabla$}}}
\newcommand{\ra}{\rightarrow}

\def\lsim{\raise 0.4ex\hbox{$<$}\kern -0.8em\lower 0.62
ex\hbox{$\sim$}}

\def\gsim{\raise 0.4ex\hbox{$>$}\kern -0.7em\lower 0.62
ex\hbox{$\sim$}}

\def\lbar{{\hbox{$\lambda$}\kern -0.7em\raise 0.6ex
\hbox{$-$}}}

\newcommand\eq[1]{eq.~(\ref{#1})}
\newcommand\eqs[2]{eqs.~(\ref{#1}) and (\ref{#2})}
\newcommand\Eq[1]{Equation~(\ref{#1})}
\newcommand\Eqs[2]{Equations~(\ref{#1}) and (\ref{#2})}
\newcommand\Eqss[3]{Equations~(\ref{#1}), (\ref{#2}) and (\ref{#3})}
\newcommand\eqss[3]{eqs.~(\ref{#1}), (\ref{#2}) and (\ref{#3})}
\newcommand\eqsss[4]{eqs.~(\ref{#1}), (\ref{#2}), (\ref{#3})
and (\ref{#4})}
\newcommand\eqssss[5]{eqs.~(\ref{#1}), (\ref{#2}), (\ref{#3}),
(\ref{#4}) and (\ref{#5})}
\newcommand\eqst[2]{eqs.~(\ref{#1})--(\ref{#2})}
\newcommand\Eqst[2]{Eqs.~(\ref{#1})--(\ref{#2})}
\newcommand\pa{\partial}
\newcommand\p{\partial}
\newcommand\pdif[2]{\frac{\pa #1}{\pa #2}}
\newcommand\pfun[2]{\frac{\delta #1}{\delta #2}}

\newcommand\ee{\end{equation}}
\newcommand\be{\begin{equation}}
\def\bea{\begin{array}}
\def\eea{\end{array}}\def\ea{\end{array}}
\newcommand\ees{\end{eqnarray}}
\newcommand\bees{\begin{eqnarray}}
\def\nn{\nonumber}
\newcommand\sub[1]{_{\rm #1}}
\newcommand\su[1]{^{\rm #1}}

\def\v#1{\hbox{\boldmath$#1$}}
\def\vepsilon{\v{\epsilon}}
\def\vPhi{\v{\Phi}}
\def\vomega{\v{\omega}}
\def\vsigma{\v{\sigma}}
\def\vmu{\v{\mu}}
\def\vxi{\v{\xi}}
\def\vpsi{\v{\psi}}
\def\vth{\v{\theta}}
\def\vphi{\v{\phi}}

\newcommand{\om}{\omega}
\newcommand{\Om}{\Omega}

\def\f{\phi}
\def\D{\Delta}
\def\a{\alpha}
\def\b{\beta}
\def\ab{\alpha\beta}

\def\s{\sigma}
\def\g{\gamma}
\def\G{\Gamma}
\def\d{\delta}
\def\Si{\Sigma}
\def\eps{\epsilon}
\def\veps{\varepsilon}
\def\Ups{\Upsilon}
\def\Upsun{{\Upsilon}_{\odot}}

\def\dslash{\hspace{-1mm}\not{\hbox{\kern-2pt $\partial$}}}
\def\Dslash{\not{\hbox{\kern-2pt $D$}}}
\def\pslash{\not{\hbox{\kern-2.1pt $p$}}}
\def\kslash{\not{\hbox{\kern-2.3pt $k$}}}
\def\qslash{\not{\hbox{\kern-2.3pt $q$}}}


\newcommand{\vac}{|0\rangle}
\newcommand{\cav}{\langle 0|}
\newcommand{\hint}{H_{\rm int}}
\newcommand{\va}{{\bf a}}
\newcommand{\vb}{{\bf b}}
\newcommand{\vp}{{\bf p}}
\newcommand{\vq}{{\bf q}}
\newcommand{\vk}{{\bf k}}
\newcommand{\vx}{{\bf x}}
\newcommand{\xp}{{\bf x}_{\perp}}
\newcommand{\vy}{{\bf y}}
\newcommand{\vz}{{\bf z}}
\newcommand{\vu}{{\bf u}}

\def\p1{{\bf p}_1}
\def\p2{{\bf p}_2}
\def\k1{{\bf k}_1}
\def\k2{{\bf k}_2}

\newcommand{\emn}{\eta_{\mu\nu}}
\newcommand{\ers}{\eta_{\rho\sigma}}
\newcommand{\emr}{\eta_{\mu\rho}}
\newcommand{\ens}{\eta_{\nu\sigma}}
\newcommand{\ems}{\eta_{\mu\sigma}}
\newcommand{\enr}{\eta_{\nu\rho}}
\newcommand{\eMN}{\eta^{\mu\nu}}
\newcommand{\eRS}{\eta^{\rho\sigma}}
\newcommand{\eMR}{\eta^{\mu\rho}}
\newcommand{\eNS}{\eta^{\nu\sigma}}
\newcommand{\eMS}{\eta^{\mu\sigma}}
\newcommand{\eNR}{\eta^{\nu\rho}}
\newcommand{\ema}{\eta_{\mu\alpha}}
\newcommand{\emb}{\eta_{\mu\beta}}
\newcommand{\ena}{\eta_{\nu\alpha}}
\newcommand{\enb}{\eta_{\nu\beta}}
\newcommand{\eab}{\eta_{\alpha\beta}}
\newcommand{\eAB}{\eta^{\alpha\beta}}

\newcommand{\gmn}{g_{\mu\nu}}
\newcommand{\grs}{g_{\rho\sigma}}
\newcommand{\gmr}{g_{\mu\rho}}
\newcommand{\gns}{g_{\nu\sigma}}
\newcommand{\gms}{g_{\mu\sigma}}
\newcommand{\gnr}{g_{\nu\rho}}
\newcommand{\gsn}{g_{\sigma\nu}}
\newcommand{\gsm}{g_{\sigma\mu}}
\newcommand{\gMN}{g^{\mu\nu}}
\newcommand{\gRS}{g^{\rho\sigma}}
\newcommand{\gMR}{g^{\mu\rho}}
\newcommand{\gNS}{g^{\nu\sigma}}
\newcommand{\gMS}{g^{\mu\sigma}}
\newcommand{\gNR}{g^{\nu\rho}}
\newcommand{\gLR}{g^{\lambda\rho}}
\newcommand{\gSN}{g^{\sigma\nu}}
\newcommand{\gSM}{g^{\sigma\mu}}
\newcommand{\gAB}{g^{\alpha\beta}}
\newcommand{\gab}{g_{\alpha\beta}}

\newcommand{\gBmn}{\bar{g}_{\mu\nu}}
\newcommand{\gBrs}{\bar{g}_{\rho\sigma}}
\newcommand{\gBMN}{\bar{g}^{\mu\nu}}
\newcommand{\gBRS}{\bar{g}^{\rho\sigma}}
\newcommand{\gBMS}{\bar{g}^{\mu\sigma}}
\newcommand{\gBAB}{\bar{g}^{\alpha\beta}}
\newcommand{\gBma}{\bar{g}_{\mu\alpha}}
\newcommand{\gBnb}{\bar{g}_{\nu\beta}}
\newcommand{\gBab}{\bar{g}_{\a\b}}
\newcommand{\gbmn}{\bar{g}_{\mu\nu}}
\newcommand{\gbrs}{\bar{g}_{\rho\sigma}}
\newcommand{\gbMN}{\bar{g}^{\mu\nu}}
\newcommand{\gbRS}{\bar{g}^{\rho\sigma}}
\newcommand{\gbMS}{\bar{g}^{\mu\sigma}}
\newcommand{\gbAB}{\bar{g}^{\alpha\beta}}
\newcommand{\gbma}{\bar{g}_{\mu\alpha}}
\newcommand{\gbnb}{\bar{g}_{\nu\beta}}
\newcommand{\gbab}{\bar{g}_{\a\b}}

\newcommand{\hmn}{h_{\mu\nu}}
\newcommand{\hrs}{h_{\rho\sigma}}
\newcommand{\hmr}{h_{\mu\rho}}
\newcommand{\hns}{h_{\nu\sigma}}
\newcommand{\hms}{h_{\mu\sigma}}
\newcommand{\hnr}{h_{\nu\rho}}
\newcommand{\hrn}{h_{\rho\nu}}
\newcommand{\hra}{h_{\rho\alpha}}
\newcommand{\hsb}{h_{\sigma\beta}}
\newcommand{\hma}{h_{\mu\alpha}}
\newcommand{\hna}{h_{\nu\alpha}}
\newcommand{\hmb}{h_{\mu\beta}}
\newcommand{\has}{h_{\alpha\sigma}}
\newcommand{\hab}{h_{\alpha\beta}}
\newcommand{\hnb}{h_{\nu\beta}}
\newcommand{\hcr}{h_{\times}}

\newcommand{\hMN}{h^{\mu\nu}}
\newcommand{\hRS}{h^{\rho\sigma}}
\newcommand{\hMR}{h^{\mu\rho}}
\newcommand{\hRM}{h^{\rho\mu}}
\newcommand{\hRN}{h^{\rho\nu}}
\newcommand{\hNS}{h^{\nu\sigma}}
\newcommand{\hMS}{h^{\mu\sigma}}
\newcommand{\hNR}{h^{\nu\rho}}
\newcommand{\hAB}{h^{\alpha\beta}}
\newcommand{\hij}{h_{ij}}
\newcommand{\hIJ}{h^{ij}}
\newcommand{\hkl}{h_{kl}}
\newcommand{\hTTij}{h_{ij}^{\rm TT}}
\newcommand{\HTTij}{H_{ij}^{\rm TT}}
\newcommand{\dhTTij}{\dot{h}_{ij}^{\rm TT}}
\newcommand{\hTTab}{h_{ab}^{\rm TT}}

\newcommand{\thmn}{\tilde{h}_{\mu\nu}}
\newcommand{\thrs}{\tilde{h}_{\rho\sigma}}
\newcommand{\thmr}{\tilde{h}_{\mu\rho}}
\newcommand{\thns}{\tilde{h}_{\nu\sigma}}
\newcommand{\thms}{\tilde{h}_{\mu\sigma}}
\newcommand{\thnr}{\tilde{h}_{\nu\rho}}
\newcommand{\thrn}{\tilde{h}_{\rho\nu}}
\newcommand{\thab}{\tilde{h}_{\alpha\beta}}
\newcommand{\thMN}{\tilde{h}^{\mu\nu}}
\newcommand{\thRS}{\tilde{h}^{\rho\sigma}}
\newcommand{\thMR}{\tilde{h}^{\mu\rho}}
\newcommand{\thRM}{\tilde{h}^{\rho\mu}}
\newcommand{\thRN}{\tilde{h}^{\rho\nu}}
\newcommand{\thNS}{\tilde{h}^{\nu\sigma}}
\newcommand{\thMS}{\tilde{h}^{\mu\sigma}}
\newcommand{\thNR}{\tilde{h}^{\nu\rho}}
\newcommand{\thAB}{\tilde{h}^{\alpha\beta}}

\newcommand{\vvarphi}{\hat{\varphi}}
\newcommand{\hhmn}{\hat{h}_{\mu\nu}}
\newcommand{\hhrs}{\hat{h}_{\rho\sigma}}
\newcommand{\hhmr}{\hat{h}_{\mu\rho}}
\newcommand{\hhns}{\hat{h}_{\nu\sigma}}
\newcommand{\hhms}{\hat{h}_{\mu\sigma}}
\newcommand{\hhnr}{\hat{h}_{\nu\rho}}
\newcommand{\hhra}{\hat{h}_{\rho\alpha}}

\newcommand{\hhMN}{\hat{h}^{\mu\nu}}
\newcommand{\hhRS}{\hat{h}^{\rho\sigma}}
\newcommand{\hhMR}{\hat{h}^{\mu\rho}}
\newcommand{\hhNS}{\hat{h}^{\nu\sigma}}
\newcommand{\hhMS}{\hat{h}^{\mu\sigma}}
\newcommand{\hhNR}{\hat{h}^{\nu\rho}}
\newcommand{\hhAB}{\hat{h}^{\alpha\beta}}

\newcommand{\sh}{\mathsf{h}}
\newcommand{\shmn}{\mathsf{h}_{\mu\nu}}
\newcommand{\shrs}{\mathsf{h}_{\rho\sigma}}
\newcommand{\shmr}{\mathsf{h}_{\mu\rho}}
\newcommand{\shns}{\mathsf{h}_{\nu\sigma}}
\newcommand{\shms}{\mathsf{h}_{\mu\sigma}}
\newcommand{\shnr}{\mathsf{h}_{\nu\rho}}
\newcommand{\shra}{\mathsf{h}_{\rho\alpha}}
\newcommand{\shsb}{\mathsf{h}_{\sigma\beta}}
\newcommand{\shma}{\mathsf{h}_{\mu\alpha}}
\newcommand{\shna}{\mathsf{h}_{\nu\alpha}}
\newcommand{\shmb}{\mathsf{h}_{\mu\beta}}
\newcommand{\shas}{\mathsf{h}_{\alpha\sigma}}
\newcommand{\shab}{\mathsf{h}_{\alpha\beta}}
\newcommand{\shnb}{\mathsf{h}_{\nu\beta}}
\newcommand{\shcr}{\mathsf{h}_{\times}}
\newcommand{\shMN}{\mathsf{h}^{\mu\nu}}
\newcommand{\shRS}{\mathsf{h}^{\rho\sigma}}
\newcommand{\shMR}{\mathsf{h}^{\mu\rho}}
\newcommand{\shNS}{\mathsf{h}^{\nu\sigma}}
\newcommand{\shMS}{\mathsf{h}^{\mu\sigma}}
\newcommand{\shNR}{\mathsf{h}^{\nu\rho}}
\newcommand{\shAB}{\mathsf{h}^{\alpha\beta}}
\newcommand{\shij}{\mathsf{h}_{ij}}
\newcommand{\shIJ}{\mathsf{h}^{ij}}
\newcommand{\shkl}{\mathsf{h}_{kl}}
\newcommand{\shTTij}{\mathsf{h}_{ij}^{\rm TT}}
\newcommand{\shTTab}{\mathsf{h}_{ab}^{\rm TT}}

\newcommand{\bhmn}{\bar{h}_{\mu\nu}}
\newcommand{\bhrs}{\bar{h}_{\rho\sigma}}
\newcommand{\bhmr}{\bar{h}_{\mu\rho}}
\newcommand{\bhns}{\bar{h}_{\nu\sigma}}
\newcommand{\bhms}{\bar{h}_{\mu\sigma}}
\newcommand{\bhnr}{\bar{h}_{\nu\rho}}
\newcommand{\bhRS}{\bar{h}^{\rho\sigma}}
\newcommand{\bhMN}{\bar{h}^{\mu\nu}}
\newcommand{\bhNR}{\bar{h}^{\nu\rho}}
\newcommand{\bhMR}{\bar{h}^{\mu\rho}}
\newcommand{\bhAB}{\bar{h}^{\alpha\beta}}

\newcommand{\hax}{h^{\rm ax}}
\newcommand{\haxmn}{h^{\rm ax}_{\mu\nu}}
\newcommand{\hpol}{h^{\rm pol}}
\newcommand{\hpolmn}{h^{\rm pol}_{\mu\nu}}

\newcommand{\dgzz}{{^{(2)}g_{00}}}
\newcommand{\qgzz}{{^{(4)}g_{00}}}
\newcommand{\tgzi}{{^{(3)}g_{0i}}}
\newcommand{\dgij}{{^{(2)}g_{ij}}}
\newcommand{\zTzz}{{^{(0)}T^{00}}}
\newcommand{\dTzz}{{^{(2)}T^{00}}}
\newcommand{\dTii}{{^{(2)}T^{ii}}}
\newcommand{\uTzi}{{^{(1)}T^{0i}}}

\newcommand{\xm}{x^{\mu}}
\newcommand{\xn}{x^{\nu}}
\newcommand{\xr}{x^{\rho}}
\newcommand{\xs}{x^{\sigma}}
\newcommand{\xa}{x^{\a}}
\newcommand{\xb}{x^{\b}}

\newcommand{\hatk}{\hat{\bf k}}
\newcommand{\hatn}{\hat{\bf n}}
\newcommand{\hatx}{\hat{\bf x}}
\newcommand{\haty}{\hat{\bf y}}
\newcommand{\hatz}{\hat{\bf z}}
\newcommand{\hatr}{\hat{\bf r}}
\newcommand{\hatu}{\hat{\bf u}}
\newcommand{\hatv}{\hat{\bf v}}
\newcommand{\xim}{\xi_{\mu}}
\newcommand{\xin}{\xi_{\nu}}
\newcommand{\xia}{\xi_{\a}}
\newcommand{\xib}{\xi_{\b}}
\newcommand{\xiM}{\xi^{\mu}}
\newcommand{\xiN}{\xi^{\nu}}

\newcommand{\tA}{\tilde{\bf A} ({\bf k})}

\newcommand{\pam}{\pa_{\mu}}
\newcommand{\pal}{\pa_{\mu}}
\newcommand{\pan}{\pa_{\nu}}
\newcommand{\parho}{\pa_{\rho}}
\newcommand{\pas}{\pa_{\sigma}}
\newcommand{\paM}{\pa^{\mu}}
\newcommand{\paN}{\pa^{\nu}}
\newcommand{\paR}{\pa^{\rho}}
\newcommand{\paS}{\pa^{\sigma}}
\newcommand{\paa}{\pa_{\alpha}}
\newcommand{\pab}{\pa_{\beta}}
\newcommand{\pat}{\pa_{\theta}}
\newcommand{\paf}{\pa_{\phi}}

\newcommand{\Dam}{D_{\mu}}
\newcommand{\Dan}{D_{\nu}}
\newcommand{\Dar}{D_{\rho}}
\newcommand{\Das}{D_{\sigma}}
\newcommand{\DaM}{D^{\mu}}
\newcommand{\DaN}{D^{\nu}}
\newcommand{\DaR}{D^{\rho}}
\newcommand{\DaS}{D^{\sigma}}
\newcommand{\Daa}{D_{\alpha}}
\newcommand{\Dab}{D_{\beta}}

\newcommand{\DBm}{\bar{D}_{\mu}}
\newcommand{\DBn}{\bar{D}_{\nu}}
\newcommand{\DBr}{\bar{D}_{\rho}}
\newcommand{\DBs}{\bar{D}_{\sigma}}
\newcommand{\DBt}{\bar{D}_{\tau}}
\newcommand{\DBa}{\bar{D}_{\alpha}}
\newcommand{\DBb}{\bar{D}_{\beta}}
\newcommand{\DBM}{\bar{D}^{\mu}}
\newcommand{\DBN}{\bar{D}^{\nu}}
\newcommand{\DBR}{\bar{D}^{\rho}}
\newcommand{\DBS}{\bar{D}^{\sigma}}
\newcommand{\DBA}{\bar{D}^{\alpha}}

\newcommand{\GMnr}{{\Gamma}^{\mu}_{\nu\rho}}
\newcommand{\Glmn}{{\Gamma}^{\lambda}_{\mu\nu}}
\newcommand{\barGMnr}{{\bar{\Gamma}}^{\mu}_{\nu\rho}}
\newcommand{\GMns}{{\Gamma}^{\mu}_{\nu\sigma}}
\newcommand{\GInr}{{\Gamma}^{i}_{\nu\rho}}
\newcommand{\Rmn}{R_{\mu\nu}}
\newcommand{\Gmn}{G_{\mu\nu}}
\newcommand{\RMN}{R^{\mu\nu}}
\newcommand{\GMN}{G^{\mu\nu}}
\newcommand{\Rmnrs}{R_{\mu\nu\rho\sigma}}
\newcommand{\RMnrs}{{R^{\mu}}_{\nu\rho\sigma}}
\newcommand{\Tmn}{T_{\mu\nu}}
\newcommand{\Smn}{S_{\mu\nu}}
\newcommand{\Tab}{T_{\a\b}}
\newcommand{\TMN}{T^{\mu\nu}}
\newcommand{\TAB}{T^{\a\b}}
\newcommand{\TBmn}{\bar{T}_{\mu\nu}}
\newcommand{\TBMN}{\bar{T}^{\mu\nu}}
\newcommand{\TRS}{T^{\rho\sigma}}
\newcommand{\tmn}{t_{\mu\nu}}
\newcommand{\tMN}{t^{\mu\nu}}
\newcommand{\RUmn}{R_{\mu\nu}^{(1)}}
\newcommand{\RDmn}{R_{\mu\nu}^{(2)}}
\newcommand{\RTmn}{R_{\mu\nu}^{(3)}}
\newcommand{\RBmn}{\bar{R}_{\mu\nu}}
\newcommand{\RBmr}{\bar{R}_{\mu\rho}}
\newcommand{\RBnr}{\bar{R}_{\nu\rho}}

\newcommand{\dddM}{\kern 0.2em \raise 1.9ex\hbox{$...$}\kern -1.0em \hbox{$M$}}
\newcommand{\dddQ}{\kern 0.2em \raise 1.9ex\hbox{$...$}\kern -1.0em \hbox{$Q$}}
\newcommand{\dddI}{\kern 0.2em \raise 1.9ex\hbox{$...$}\kern -1.0em\hbox{$I$}}
\newcommand{\dddJ}{\kern 0.2em \raise 1.9ex\hbox{$...$}\kern-1.0em
\hbox{$J$}}
\newcommand{\dddcalJ}{\kern 0.2em \raise 1.9ex\hbox{$...$}\kern-1.0em
\hbox{${\cal J}$}}

\newcommand{\dddO}{\kern 0.2em \raise 1.9ex\hbox{$...$}\kern -1.0em
\hbox{${\cal O}$}}
\def\dddz{\raise 1.5ex\hbox{$...$}\kern -0.8em \hbox{$z$}}
\def\dddd{\raise 1.8ex\hbox{$...$}\kern -0.8em \hbox{$d$}}
\def\dddbd{\raise 1.8ex\hbox{$...$}\kern -0.8em \hbox{${\bf d}$}}
\def\ddbd{\raise 1.8ex\hbox{$..$}\kern -0.8em \hbox{${\bf d}$}}
\def\dddx{\raise 1.6ex\hbox{$...$}\kern -0.8em \hbox{$x$}}

\newcommand{\hti}{\tilde{h}}
\newcommand{\hf}{\tilde{h}_{ab}(f)}
\newcommand{\Hti}{\tilde{H}}
\newcommand{\fmin}{f_{\rm min}}
\newcommand{\fmax}{f_{\rm max}}
\newcommand{\frot}{f_{\rm rot}}
\newcommand{\fpol}{f_{\rm pole}}
\newcommand{\omax}{\o_{\rm max}}
\newcommand{\orot}{\o_{\rm rot}}
\newcommand{\op}{\o_{\rm p}}
\newcommand{\tmax}{t_{\rm max}}
\newcommand{\tobs}{t_{\rm obs}}
\newcommand{\fobs}{f_{\rm obs}}
\newcommand{\temis}{t_{\rm emis}}
\newcommand{\DE}{\D E_{\rm rad}}
\newcommand{\DEm}{\D E_{\rm min}}
\newcommand{\msun}{M_{\odot}}
\newcommand{\rsun}{R_{\odot}}
\newcommand{\ogw}{\omega_{\rm gw}}
\newcommand{\fgw}{f_{\rm gw}}
\newcommand{\oL}{\omega_{\rm L}}
\newcommand{\kL}{k_{\rm L}}
\newcommand{\lL}{\l_{\rm L}}
\newcommand{\mns}{M_{\rm NS}}
\newcommand{\rns}{R_{\rm NS}}
\newcommand{\tret}{t_{\rm ret}}
\newcommand{\Sch}{Schwarzschild }
\newcommand{\rtid}{r_{\rm tidal}}

\newcommand{\ot}{\o_{\rm t}}
\newcommand{\mt}{m_{\rm t}}
\newcommand{\gt}{\g_{\rm t}}
\newcommand{\xit}{\tilde{\xi}}
\newcommand{\xtr}{\xi_{\rm t}}
\newcommand{\xtj}{\xi_{{\rm t},j}}
\newcommand{\dxtj}{\dot{\xi}_{{\rm t},j}}
\newcommand{\ddxtj}{\ddot{\xi}_{{\rm t},j}}
\newcommand{\teff}{T_{\rm eff}}
\newcommand{\samp}{S_{\xi_{\rm t}}^{\rm ampl}}

\newcommand{\mpl}{M_{\rm Pl}}
\newcommand{\mplr}{m_{\rm Pl}}
\newcommand{\mgut}{M_{\rm GUT}}
\newcommand{\lpl}{l_{\rm Pl}}
\newcommand{\tpl}{t_{\rm Pl}}
\newcommand{\ls}{\lambda_{\rm s}}
\newcommand{\Ogw}{\Omega_{\rm gw}}
\newcommand{\hogw}{h_0^2\Omega_{\rm gw}}
\newcommand{\hn}{h_n(f)}

\newcommand{\sinc}{{\rm sinc}\, }
\newcommand{\Ein}{E_{\rm in}}
\newcommand{\Eout}{E_{\rm out}}
\newcommand{\Et}{E_{\rm t}}
\newcommand{\Er}{E_{\rm refl}}
\newcommand{\lm}{\l_{\rm mod}}

\newcommand{\mrI}{\mathrm{I}}
\newcommand{\mrJ}{\mathrm{J}}
\newcommand{\mrW}{\mathrm{W}}
\newcommand{\mrX}{\mathrm{X}}
\newcommand{\mrY}{\mathrm{Y}}
\newcommand{\mrZ}{\mathrm{Z}}
\newcommand{\mrM}{\mathrm{M}}
\newcommand{\mrS}{\mathrm{S}}

\newcommand{\rmI}{\mathrm{I}}
\newcommand{\rmJ}{\mathrm{J}}
\newcommand{\rmW}{\mathrm{W}}
\newcommand{\rmX}{\mathrm{X}}
\newcommand{\rmY}{\mathrm{Y}}
\newcommand{\rmZ}{\mathrm{Z}}
\newcommand{\rmM}{\mathrm{M}}
\newcommand{\rmS}{\mathrm{S}}
\newcommand{\rmU}{\mathrm{U}}
\newcommand{\rmV}{\mathrm{V}}


\newcommand{\et}{{{\bf e}^t}}
\newcommand{\etm}{{\bf e}^t_{\mu}}
\newcommand{\etn}{{\bf e}^t_{\nu}}

\newcommand{\er}{{{\bf e}^r}}
\newcommand{\erm}{{\bf e}^r_{\mu}}
\newcommand{\ern}{{\bf e}^r_{\nu}}

\newcommand{\hz}{H^{(0)}}
\newcommand{\hu}{H^{(1)}}
\newcommand{\hd}{H^{(2)}}
\newcommand{\thz}{\tilde{H}^{(0)}}
\newcommand{\thu}{\tilde{H}^{(1)}}
\newcommand{\thd}{\tilde{H}^{(2)}}
\newcommand{\tK}{\tilde{K}}
\newcommand{\tZ}{\tilde{Z}}
\newcommand{\tQ}{\tilde{Q}}

\newcommand{\inT}{\int_{-\infty}^{\infty}}
\newcommand{\intz}{\int_{0}^{\infty}}
\newcommand{\Dl}{\int{\cal D}\lambda}

\newcommand{\fnl}{f_{\rm NL}}

\newcommand{\ode}{\Omega_{\rm DE}}
\newcommand{\oma}{\Omega_{M}}
\newcommand{\ora}{\Omega_{R}}
\newcommand{\ovac}{\Omega_{\rm vac}}
\newcommand{\ola}{\Omega_{\Lambda}}
\newcommand{\oxi}{\Omega_{\xi}}
\newcommand{\oga}{\Omega_{\gamma}}

\newcommand{\lc}{\Lambda_c}
\newcommand{\rde}{\rho_{\rm DE}}
\newcommand{\wde}{w_{\rm DE}}
\newcommand{\rvac}{\rho_{\rm vac}}
\newcommand{\rlam}{\rho_{\Lambda}}

\maketitle
\flushbottom

\section{Introduction}

The response of elastic bodies to gravitational waves (GWs) was at the basis of the first concepts for GW detectors. 
Resonant-mass detectors, pioneered by Weber in the 1960s, were further developed by various groups around the world and remained in operation for several decades (see e.g. chapter 8 of \cite{Maggiore:2007ulw} for review and references). Weber was  also  the first to point out that the elastic vibrations of the Earth or the Moon could be used to detect GWs (see \cite{LGWA:2020mma} for a summary of related historical developments and  references). 
Recently, building on new technological advances and on the resurgence of interest for Lunar missions, the idea of deploying an array of seismometers on the Moon to monitor the vibrations induced by GWs has been revived, and has led to the proposal of the 
Lunar Gravitational Wave Antenna (LGWA) \cite{LGWA:2020mma,Cozzumbo:2023gzs,LGWA:2024inprep}. This  project could possibly be operative by the next decade, and  could bridge the frequency gap between the LISA space interferometer \cite{LISA:2017pwj}
and the next generation of ground-based detectors such as the Einstein Telescope \cite{Punturo:2010zz,Maggiore:2019uih,Branchesi:2023mws} and Cosmic Explorer~\cite{Evans:2021gyd}, providing complementary multi-band observations.

When studying the response of an elastic body to GWs, there are two quite distinct regimes. The first and simpler regime takes place when the response of the body is dominated by just a single normal mode, or at most a few of them.
The lowest-order modes in general  have  damping times long compared to the inverse of their frequency, and therefore their response to an external force is a narrow resonance; then, only modes sufficiently close to the characteristic frequencies of the external perturbation, such as an incoming GW, are excited, and the dynamics of the system is fully described in terms of these few modes. This is essentially the reason for the usefulness of a normal mode description: as we can see for instance from the standard  example of a  one-dimensional chain of harmonic oscillators with nearest-neighbor coupling, mathematically the normal modes  are just a change of basis with respect to  the displacements of the individual elementary constituents. However, while an external perturbation excites  all individual displacements, so that we would have to deal with an infinity of dynamical variables, once one passes to the basis of normal modes only a few of them are excited, or even just one,  and the dynamics of the system is fully encoded in  a few collective degrees of freedom, such as the amplitude $\xi_0(t)$ associated to the fundamental normal mode.
 
As we increase the frequency $f$ of the external perturbation, however, the usefulness of the normal-mode description gradually disappears. As we move toward frequencies $f$ much higher than the frequency $f_0$ of the fundamental mode, the external perturbation (in our case, the GW) interacts with modes whose response  is no longer a narrow resonance, but a broader and broader peak that eventually blends with the response of all other modes;  the response of the system is now determined  by a quasi-continuum of normal modes and therefore  depends on a large number of microscopic parameters, the frequencies and damping times of all the large-$n$ modes (where $n$ labels the tower of modes, in general for fixed values of other indices, as for instance the spherical harmonics indices $(l,m)$ in spherical symmetry).  However, in general these are not really known, neither observationally nor theoretically;  in the large-$n$ limit
the frequencies  could still be computed in highly idealized cases such as a uniform one-dimensional bar, but not really in realistic cases, since  they depend on all small-scale details,  and even less is known, in general, for the damping times of the large-$n$ modes, that depend on the details of the dissipation mechanisms inside the body. So, we end up with a description which involves a large number (actually, an infinity) of unknown parameters, the frequencies and damping times  of the large-$n$ modes. 

In this regime a more convenient approach is  based on field theory that, through the logic of long-distance effective field theory (EFT), provides a different and efficient organizing principle. One can introduce a vector field $\vu(t,\vx)$ that describes the displacement, at time $t$, of an infinitesimal volume element that, in the absence of perturbations, would be located at a given position $\vx$, and study its dynamics.  The advantage of switching to a field theory description is that we can construct the long-distance dynamics (i.e. the dynamics at distances much larger than the scale of the elementary constituents of the elastic body) using the methods of  effective field theory. The dynamics is then significantly constrained by the symmetries of the theory, including Lorentz invariance. Naively, one might think that Lorentz invariance is not relevant here, because  the elastic vibrations  that we wish to consider, such as  the elastic vibrations of the Moon surface, are obviously non-relativistic. However,   a  Lorentz-invariant, and in fact even a generally-covariant  theory of elasticity is necessary to write consistently a coupling to GWs, since the metric perturbation $\hmn$ is  intrinsically a general-relativistic object. The consistent way of coupling an elastic medium to GWs, from this field-theoretical perspective, is then to construct a  diffeomorphism-invariant relativistic theory of elasticity involving  a generic metric $\gmn$, and then take the linearized limit for the metric, $\gmn=\emn+\hmn$; only after, if appropriate, we can eventually take also the non-relativistic limit for the velocities of the volume elements of the material. 

A second independent reason why the standard computation of the interaction of an elastic body with GWs, based on normal modes driven by a GW ``force", eventually breaks down is that this Newtonian picture only holds in the proper detector frame. As we will recall in Section~\ref{sect:PDFvsTT}, this frame can only be used when the reduced wavelength  $\lbar\equiv \lambda/(2\pi)$ of the GW satisfies $\lbar\gg L$, where $L$ is the typical linear size of the body. This is indeed the case for the lowest normal modes of typical elastic materials. 
For instance, consider a one-dimensional bar of length $L$ and speed of sound $v_s$. The frequency of its fundamental mode is $f_0=v_s/(2L)$; the  GW interacts significantly with this mode only if its frequency $f$   is of order $f_0$. Since the   (reduced) wavelength of a GW with frequency $f$ is $\lbar =c/(2\pi f)$,
the condition $f\simeq f_0$ implies
\be\label{Lsulbar}
\frac{L}{\lbar}\simeq \frac{\pi v_s}{c}\, .
\ee 
In any ``normal'' elastic material  the speed of sound is many orders of magnitude smaller than the speed of light $c$ (this is no longer the case, however, in neutron stars \cite{Altiparmak:2022bke}), and therefore $\lbar\gg L$. As we will briefly review in Section~\ref{sect:PDFvsTT}, this is the regime where we can use the proper detector frame, where the Newtonian intuition applies, and the effect of a GW can be described as a Newtonian force in flat space-time. The corresponding computation is then a standard textbook exercise, that we will review in Section~\ref{sect:normal modes}. However, as $\lbar$ becomes comparable to $L$, the whole approach based on the proper detector frame breaks down, since it receives general-relativistic corrections of order $(L/\lbar)^2$. For GWs with frequency $f\sim f_0$ this is of order $v_s^2/c^2$, as we see from \eq{Lsulbar}, but as $f$ grows and $\lbar$ becomes as small as $L$, the whole approach breaks down.  In this regime, we must use full general-relativistic concepts, rather than the Newtonian picture of a force acting in flat space-time.
In contrast, as we will recall in Section~\ref{sect:PDFvsTT}, the TT gauge can be used without any restriction on the GW frequency, but in this case the effect of the GW cannot be described as a Newtonian force.

The standard computation based on the response of the normal modes, and using the proper detector frame, therefore breaks down as we reach a  GW frequency such that one of the two conditions, normal mode blending or $\lbar\sim L$, sets in, whichever comes first.\footnote{In typical materials, such as the Aluminium of which some resonant-bar detectors were made, or rocks on Earth and the Moon, $v_s$ can be of order $2-5$~km/s, resulting in $\pi v_s/c\simeq (3-6)\times 10^{-5}$. Therefore the proper detector frame can no longer be used for  normal modes with $f_n \, \gsim\, 10^4 f_0$; in general, we expect that the blending of the response of the normal modes will have already taken place at much smaller frequencies, so the latter effect should set in first.}
At  larger frequencies, the  best strategy is then  to use a field-theoretical formulation of elasticity theory (rather than using the normal modes) and couple it to GWs in the TT gauge (rather than using the proper detector frame). This was indeed the spirit of a 1969 paper by Dyson~\cite{Dyson:1969zgf}, which has  become a standard reference on this subject. However, analyzing this paper, we have found some problem with it; Dyson considered a non-relativistic field theory of elasticity, and  couples  the metric perturbation $\hmn$ to  an energy--momentum tensor $\TMN$ that he writes down, without actually providing any explicit derivation of it; however, we will see, first of all, that the  energy--momentum tensor that he writes down does not  really follow from any application of Noether's theorem  to his non-relativistic Lagrangian; and,  second, that his expression is not completely correct. To correctly identify the energy--momentum tensor $\TMN$ that couples to the metric perturbation   $\hmn$,  a first principle approach is to develop a relativistic (and, in fact, even general-relativistic) formulation of elasticity theory.
A relativistic formulation of elasticity theory was  developed in ref.~\cite{Carter:1977qf}. This  work uses classic notions of elasticity theory and generalizes them to arbitrary curved space, using the language of differential geometry 
(see \cite{Hudelist:2022ixo} for a clear presentation, and \cite{Magli:1992,Beig:2002pk,Brown:2020pav} for a Lagrangian approach to  relativistic elasticity in this geometric language) and,
for isotropic and  homogeneous elastic media, recovers Dyson's result for the coupling to GWs in the TT gauge.\footnote{Actually, as we will discuss in App.~\ref{sect:Dyson}, the energy--momentum tensor proposed by Dyson for the theory of elasticity is not really correct even in the homogeneous case, although this eventually does not influence the coupling to GWs in the TT gauge (but this would affect the results for more general gravitational fields). In the inhomogeneous case (which is not considered in  \cite{Carter:1977qf,Hudelist:2022ixo}) we will see that there are further complications. This is  important since, as found by Dyson and as we will review below, in the TT gauge the whole coupling to GWs comes either  from the boundary conditions or from bulk inhomogeneities.}
 Here we will rather take a different path and elaborate on the  EFT formalism developed in \cite{Dubovsky:2005xd,Endlich:2010hf}  in the context of relativistic classical and quantum fluids, and further discussed for elastic  solids in \cite{Endlich:2012pz,Nicolis:2013lma,Alberte:2018doe,Baggioli:2019elg}. This will allow us to make  contact with  the modern  language of effective field theories, that can be physically illuminating,  and to provide what seems to us a streamlined and more transparent derivation of various results. 

Our original interest in the problem was spurred by the fact that the regime where the GW frequency is much larger than the frequencies of the lowest normal modes is very relevant for LGWA, since it is precisely for those frequencies that  the current estimate for the sensitivity curve of LGWA goes below the sensitivity curve of LISA (see Fig.~3 of ref.~\cite{LGWA:2020mma}). However, we also found that the EFT  formulation of the problem is quite interesting and elegant by itself. The paper is organized as follows. In Section~\ref{sect:EFT_NR} we discuss non-relativistic elasticity from the Lagrangian point of view, while the EFT formulation of relativistic elasticity is discussed in 
Section~\ref{sect:EFT_Rel}. In Section~\ref{sect:coupling_to_GWs} we discuss how to couple elasticity to GWs. We begin, in Section~\ref{sect:PDFvsTT}, with  a quick reminder of the definition and properties of the proper detector frame and of the TT frame, while in Section~\ref{sect:normal modes} we provide a reminder of the  computation based on the normal modes of the elastic body, which is valid only in the proper detector frame. The coupling of elasticity to GWs is discussed, in the EFT approach, in Section~\ref{sect:Coupling}. In Section~\ref{sect:Comparison} we show how, in the low-frequency limit, taking into account properly the transformation from the TT frame to the proper detector frame, the general field-theoretical results can be matched to those obtained in the proper detector frame. Section~\ref{sect:Conclusions} contains our conclusions. Some technical material is discussed in the appendices and, in particular,  in App.~\ref{sect:Dyson}  we compare our results with Dyson's paper. In the field theoretical discussion, in Section~\ref{sect:EFT_Rel},
we use units $c=1$, otherwise we write $c$ explicitly. We use
the signature $\emn=(-,+,+,+)$.

\section{Field-theory description of non-relativistic elasticity}\label{sect:EFT_NR}

As a warm-up, we begin by formulating the non-relativistic   theory of elasticity in the language of  field theory and effective field theory  (although the full constraining power  of effective field theory will only emerge in the relativistic formulation discussed in the next section).  We assume that, in the absence of external perturbations, the body has   a static equilibrium configuration, that we take as a reference configuration.
The basic field-theoretical variable is the vector field $\vu(t,\vx)$, with components $u_i(t,\vx)$, that represents the displacement,  at time $t$,  of an infinitesimal volume element which, in the reference configuration, is located at the position $\vx$. Denoting by $\vy$ the perturbed position, we then have
\be\label{dydx}
\vy (t,\vx)= \vx +\vu(t,\vx)\, .
\ee
We  restrict to  linear elasticity, so to terms linear in $\vu$ in the equations of motions (and, therefore, terms quadratic in $\vu$ in the Lagrangian), and neglect for the moment dissipative effects. We also restrict to
elastic bodies with no  plasticity (i.e., whose dynamics is local in time, rather than retaining memory of the past history).

\subsection{Lagrangian and symmetries}\label{sect:Lagandsym}

Consider  an elastic body idealized, at first, as spatially uniform and filling all space.
The structure of its Lagrangian is fixed  as follows. The kinetic term is easily found. If  $\rho$ is the mass density of the material, an infinitesimal  volume element $d^3x$  located at $(t,\vx)$  has a 
kinetic energy  $(1/2) dm\,  \dot{u}_i\dot{u}_i$, where
$dm=\rho d^3x$ (and the sum over repeated indices is understood). Therefore, the kinetic term in the Lagrangian is
\be\label{Lkin}
L_{\rm kin}=\frac{\rho}{2} \int d^3x\,  \dot{u}_i\dot{u}_i\, .
\ee
As usual, there is more freedom in the choice of the interaction terms. In the logic of low-energy effective field theories, we only restrict to the terms with the lowest number of spatial derivatives, since these determine the long-distance behavior and, again in the spirit of an EFT treatment, in order to restrict the possible structures that can appear in the Lagrangian  we begin by looking at the symmetries that we want to impose on the theory.
Consider  the transformation
\be\label{transu}
\begin{array}{rl}
\vx&\ra\vx'=\vx\, ,\\
\vu(t,\vx)&\ra \vu'(t',\vx')=\vu(t,\vx)+\va\, ,
\end{array}
\ee
where $\va$ is a constant vector, and time transforms as $t\ra t'=t$. In 
a field-theoretical language, \eq{transu} defines a global internal transformation: it is a global transformation because the parameter $\va$ is a constant, independent of the space-time coordinates, and it is an internal transformation since it is a transformation in the space of fields, that does not touch the space-time coordinates. However, under this transformation, the perturbed positions (\ref{dydx}) of all points of the body change rigidly, $\vy\ra \vy+\va$, so this transformation is equivalent to a rigid displacement of the elastic body. Unless the elastic body is subject to an external potential that depends on the position of its center--of--mass, such a rigid displacement will have no effect on its dynamics, and therefore in the absence of external forces \eq{transu} must be a symmetry of the action. This is indeed the case for the kinetic term (\ref{Lkin}), and we must impose that the same will hold for the terms in the Lagrangian that describes the elastic forces. This means that the latter will depend on $\vu$ only through its derivatives and, confining the time derivatives to the kinetic term, the elastic term can then depend only on  combinations made with $\pa_i u_j$, rather than on the field $u_i$ itself.
It is useful to split $\pa_i u_j$ into its symmetric and anti-symmetric parts,
\bees
\pa_i u_j&=&\frac{1}{2}(\pa_i u_j+\pa_j u_i) +\frac{1}{2}(\pa_i u_j-\pa_j u_i)\nn\\
&\equiv& u_{ij}+\tilde{u}_{ij}\, .\label{defuutilde}
\ees
To understand the meaning of this separation we observe, from \eq{dydx}, that given two infinitesimally close points whose unperturbed 
distance is $d\vx$, the distance between the corresponding perturbed points will be $d\vy$, with
\bees
dy_i&=&dx_i+(\pa_j u_i) dx_j \nn\\
&=&dx_i+(u_{ij}-\tilde{u}_{ij}) dx_j\, .
\ees
The term $u_{ij}$ corresponds to an actual local stretching of the elastic body; for instance,  the part of $u_{ij}$ proportional to its trace, $u_{ij}=\kappa\delta_{ij}$, corresponds to an isotropic dilatation, $dy_i=(1+\kappa) dx_i$, and an elastic material always has an energy associated to it. In contrast, $\tilde{u}_{ij}$ corresponds to a local rotation. This can be seen observing 
 that any antisymmetric tensor $\tilde{u}_{ij}$ in three dimensions can be written as $\tilde{u}_{ij}=\eps_{ijk}\tilde{u}_k$, where $\tilde{u}_k$ is a vector. Writing further $\tilde{u}_k=\theta n_k$, where $n_k$ is the unit vector in the direction of $\tilde{u}_k$, and $\theta$ is the modulus of $\tilde{u}_k$, and considering a transformation with $\theta$ infinitesimal, we have
$\tilde{u}_{ij}=\theta \eps_{ijk} n_k$.   
The effect of the anti-symmetric part is therefore
\be\label{dyirodxi}
dy_i=dx_i- \theta(x) \eps_{ijk} dx_j n_k(x) \, ,
\ee
and this is  a local  rotation by an infinitesimal angle $\theta(x)$ around the $\hatn(x)$ axis. Here  we have stressed  that $\theta$ and $\hatn$, just as $\tilde{u}_{ij}$, are functions of the space-time point $x$, so \eq{dyirodxi} is a local, rather than a rigid rotation.
The antisymmetric tensor $\tilde{u}_{ij}(x)$ is called the vorticity tensor (see e.g. Section~2.7 of \cite{Holzapfel}). 

In the simplest approach to elasticity, corresponding to ``hyper-elastic materials", one associates an elastic energy to local stretching, but not to local vorticity (see  eq.~(6.8) of \cite{Holzapfel}).
In that case,  the interaction Lagrangian  depends only on the symmetric combination $u_{ij}$. Then, at quadratic order, the most general non-relativistic Lagrangian for a uniform elastic body  (assumed for the moment to fill all space) can be written as
\be\label{LfiniteV}
L= \int d^3x\, {\cal L}\, ,
\ee
where the Lagrangian density ${\cal L}$ is  
\be\label{LDyson}
{\cal L}= \frac{1}{2} \rho\,\dot{u}_i\dot{u}_i- \frac{1}{2} c_{ijkl}\pa_i u_j\pa_k u_l\, ,
\ee
and, for a uniform body, $\rho$ and  $c_{ijkl}$ are constant in space (and we also take them time independent); $c_{ijkl}$  is a tensor with the symmetry properties
\be\label{csym1}
c_{ijkl}=c_{klij}\, ,
\ee
\be\label{csym2}
c_{ijkl}=c_{jikl}\, ,\qquad c_{ijkl}=c_{ijlk}\, .
\ee
\Eq{csym1} trivially expresses the fact that, in \eq{LDyson}, $(\pa_i u_j)(\pa_k u_l)$ is symmetric under the simultaneous exchanges $i\leftrightarrow k$ and  $j\leftrightarrow l$, so also $c_{ijkl}$ can be taken to be symmetric under this exchange, since anyhow any anti-symmetric part would not contribute. This equation is therefore  always valid. In contrast, \eq{csym2} expresses the fact that we are neglecting any contribution to elastic energy from local vorticity, i.e. we are assuming that the Lagrangian depends on $\pa_i u_j$ only through its symmetric part $u_{ij}$. \Eq{csym2} must therefore  be dropped if we want to include the effect of vorticity.\footnote{In particular, we will see in Section~\ref{sect:homanis} that, when constructing a relativistic effective action for an anisotropic material, a dependence on the vorticity automatically appears.} Assuming  the validity of \eq{csym2}, we can  rewrite the Lagrangian  density as
\be\label{LDysonuij}
{\cal L}= \frac{1}{2} \rho\,\dot{u}_i\dot{u}_i- \frac{1}{2} c_{ijkl}u_{ij}u_{kl}\, .
\ee
Having assumed that the body is homogeneous and fills all space, the Lagrangian (\ref{LfiniteV}, \ref{LDyson}) is also invariant under spatial translations,
under which the vector field $\vu$  transforms as a scalar (as any field does under translations, independently of its properties under rotations),
\be\label{transx}
\begin{array}{rl}
\vx&\ra\vx'=\vx\ +\va\,\\
\vu(t,\vx)&\ra \vu'(t,\vx')=\vu(t,\vx)\, .
\end{array}
\ee
This transformation is logically distinct from \eq{transu}, but it has the same effect on the variable $\vy$.

An important simplification takes place when the solid, beside being homogeneous, is also isotropic. In this case, $c_{ijkl}$ can only be build with combinations of products of Kronecker delta's, and (assuming that there is no contribution from vorticity) the most general form consistent with the symmetries 
(\ref{csym1}) and (\ref{csym2}) is
\be\label{cijklLame}
c_{ijkl}=\lambda \delta_{ij} \delta_{kl}+\mu ( \delta_{ik} \delta_{jl}+ \delta_{il} \delta_{jk})\, ,
\ee
where the constants $\lambda$ and $\mu$ are  the Lam\'e coefficients. Then, \eq{LDyson} becomes
\be\label{LDysonLame}
{\cal L}=\frac{1}{2}  \[  \rho\dot{u}_i\dot{u}_i -\lambda (\pa_iu_i)^2-2\mu\, u_{ij}u_{ij}
\]\, .
\ee
Note that $\pa_iu_i$ can be written in terms of $u_{ij}$ as $\pa_iu_i=u_{ii}$.
The Lam\'e coefficients parametrize the two independent quadratic terms that can be formed with $u_{ij}$, namely
$u_{ii}^2$ and $u_{ij}u_{ij}$.

We next want to relax the assumption that the elastic body is homogeneous.
Indeed, as found in  refs.~\cite{Dyson:1969zgf,Hudelist:2022ixo}, and as we will confirm below with our formalism, in the TT gauge the coupling of  GWs to the bulk of an elastic homogeneous medium vanishes, and the only coupling to GWs comes from internal inhomogeneities, or from the boundary conditions at the surface of the elastic body. It is therefore interesting to generalize the formalism to non-homogeneous materials. As we will see later, this will be non-trivial when we move to the relativistic formalism;
for the non-relativistic Lagrangian (\ref{LDyson}), however, the extension is obvious, and we 
just  add a spatial dependence to $\rho$ and $c_{ijkl}$, writing them as  $\rho(\vx)$ and $c_{ijkl}(\vx)$. The simplest generalization of \eqs{LfiniteV}{LDyson} is then
\be\label{LDysoninhom}
 L= \int_V d^3x\, \[ \frac{1}{2} \rho(\vx)\,\dot{u}_i\dot{u}_i- \frac{1}{2} c_{ijkl}(\vx)\pa_i u_j\pa_k u_l\] \, ,
\ee
where the integration is over the volume $V$ of the body. In the isotropic limit,
\be\label{cijklLamevx}
c_{ijkl}(\vx)=\lambda(\vx) \delta_{ij} \delta_{kl}+\mu(\vx) ( \delta_{ik} \delta_{jl}+ \delta_{il} \delta_{jk})\, ,
\ee
so
\be\label{LDysoninhomisotr}
 L= \frac{1}{2} \int_V d^3x\, \[  \rho(\vx) \dot{u}_i\dot{u}_i -\lambda(\vx) (\pa_iu_i)^2-2\mu(\vx)\, u_{ij}u_{ij} \]\, .
\ee
We observe that the transformation (\ref{transu}) is still a symmetry of the Lagrangian (\ref{LDysoninhom}). In contrast, if we interprete \eq{LDysoninhom} as a field theory for the field $\vu(t,\vx)$ in a  fixed external background given by $\rho(\vx)$ and $c_{ijkl}(\vx)$, the invariance under spatial translations (\ref{transx}) is lost.\footnote{One could further include an explicit time dependence writing $\rho(t,\vx)$ and $c_{ijkl}(t,\vx)$. However, in the following we will be mostly interested to situations in which, on the timescales of interest, the time dependence of $\rho(t,\vx)$ and $c_{ijkl}(t,\vx)$ can be neglected, and, in this case, time translations are a symmetry.}
There are, however, other ways of generalizing \eqs{LfiniteV}{LDyson} to the inhomogeneous case, which are equivalent at the quadratic level; in particular, one could write
\be\label{LDysoninhomxplusu}
 L= \int_V d^3x\, \[ \frac{1}{2} \rho(\vx+\vu)\,\dot{u}_i\dot{u}_i- \frac{1}{2} c_{ijkl}(\vx+\vu)\pa_i u_j\pa_k u_l\] \, ,
\ee
so that $\rho$ and $c_{ijkl}$ are now evaluated at the perturbed point $\vy$.
Note that now the Lagrangian depends explicitly on $\vu$ and not only on its derivatives.
At the quadratic level \eqs{LDysoninhom}{LDysoninhomxplusu} are obviously equivalent since, expanding
$ \rho(\vx+\vu)= \rho(\vx)+u_k\pa_k\rho+\ldots $, the extra term $\dot{u}_i\dot{u}_i u_k\pa_k\rho$ is cubic in $u$, and similarly for the expansion of $c_{ijkl}(\vx+\vu)$. 
However, the symmetries of the Lagrangian (\ref{LDysoninhomxplusu}) are different; now, neither \eq{transu} nor \eq{transx} are, separately, invariances of the Lagrangian; instead, the combination of a 
spatial translation (\ref{transx}) with parameter $\va$ and a
transformation  in field space (\ref{transu}) with parameter $-\va$ leaves $\vx+\vu$ invariant, so the Lagrangian (\ref{LDysoninhomxplusu}) is invariant under the transformation
\be\label{transxu}
\begin{array}{rl}
\vx&\ra\vx'=\vx\ +\va\,\\
\vu(t,\vx)&\ra \vu'(t,\vx')=\vu(t,\vx)-\va\, .
\end{array}
\ee
This point of view on the symmetries will be useful when we proceed to the relativistic generalization.

\subsection{Equations of motion}

From the Lagrangian, we can now derive the equations of motion.  For later use, we find useful to go through these results explicitly, even if they are quite standard. We begin with the  general Lagrangian (\ref{LDysoninhom}), without assuming immediately homogeneity and isotropy, and we consider
in general an  elastic body with finite extension, so that 
in \eq{LDysoninhom} the integration is over a finite volume $V$; in this case, we must  be careful with the boundary terms in the variational procedure. The action $S$ is defined by
\be
S=\int_{-\infty}^{\infty} dt\int_V d^3x\, {\cal L}\, ,
\ee
and its variation  is
\be\label{deltaSboundary}
\delta S=\int_{-\infty}^{\infty} dt \int_V d^3x\, \[ \frac{\d {\cal L}}{\d u_i} \d u_i+\frac{\d {\cal L}}{\d \dot{u}_i}\d \dot{u}_i
+\frac{\d {\cal L}}{\d (\pa_ju_i)}\delta (\pa_ju_i)\]\, .
\ee
The term $\d \dot{u}_i=\d (du_i/dt)=
d( \d u_i)/dt$ can be integrated by parts as usual, and does not leave a boundary term, since the integral over $t$ in \eq{deltaSboundary} runs from $-\infty$ to $+\infty$ and we assume the usual boundary conditions that the perturbation $u_i(t,\vx)$ vanishes sufficiently fast as $t\ra\pm\infty$. In contrast, using 
$\delta (\pa_ju_i)=\pa_j \delta u_i$ and integrating by parts,
\be
\int_V d^3x \frac{\d {\cal L}}{\d (\pa_ju_i)}\delta (\pa_ju_i)= -\int_V d^3x \(\pa_j \frac{\d {\cal L}}{\d (\pa_ju_i)}\) \delta u_i
+\int_{\pa V}d^2s\,  n_j \frac{\d {\cal L}}{\d (\pa_ju_i)} \delta u_i\, ,
\ee
where $n_j$ is the outer normal to the boundary $\pa V$, with surface element $d{\bf s} =d^2s\,  \hatn$.
Therefore, overall, the variation of the action is
\be\label{deltaSconBound}
\delta S= -\int_{-\infty}^{\infty} dt \int_V d^3x\, \[ -\frac{\d {\cal L}}{\d u_i}+\frac{d}{dt}\frac{\d {\cal L}}{\d \dot{u}_i}+\pa_j\frac{\d {\cal L}}{\d (\pa_ju_i)}\] \d u_i
+ \int dt \int_{\pa V}d^2s\,  n_j \frac{\d {\cal L}}{\d (\pa_ju_i)} \delta u_i
\, .
\ee
The vanishing of the volume term in \eq{deltaSconBound} gives the Euler-Lagrange equation 
\be\label{EulLag}
-\frac{\d {\cal L}}{\d u_i}+\frac{d}{dt}\frac{\d {\cal L}}{\d \dot{u}_i}+\pa_j\frac{\d {\cal L}}{\d (\pa_ju_i)}=0\, ,
\ee
while the vanishing of the boundary term imposes the boundary condition
\be
\label{boundarycond}
n_j\frac{\d {\cal L}}{\d (\pa_ju_i)}=0\, .
\ee
It is convenient to define the stress tensor,
\be\label{defsigmaij}
\sigma_{ij}\equiv -\frac{\d {\cal L}}{\d (\pa_ju_i)}\, .
\ee
For the Lagrangian (\ref{LDysoninhom}),
the term ${\d {\cal L}}/{\d u_i}$ that appears in \eq{EulLag} vanishes, and  the equation of motion (\ref{EulLag}) becomes
\be\label{EulLagNRexplicit}
\frac{d}{dt}\( \rho \dot{u}_i\) -\pa_j \sigma_{ij}=0\, ,
\ee
while
the boundary condition reads
\be\label{njsigmaij}
n_j \sigma_{ij}=0\, ,
\ee
and
$\sigma_{ij}$ has the explicit form
\be \label{Sijexpl}
\sigma_{ij}(\vx)= c_{ijkl}(\vx)\pa_k u_l\, .
\ee
The above equations are valid independently of whether  \eq{csym2} holds or not. If it does, i.e. in the absence of vorticity, we also have
\be\label{sigmasym}
\sigma_{ij}=\sigma_{ji}\, .
\ee
If we now further restrict to the isotropic case, using \eq{cijklLamevx}, we get
\be\label{sigmaijLame}
\sigma_{ij}=\lambda(\vx)\delta_{ij}\pa_ku_k +\mu(\vx) (\pa_i u_j+\pa_j u_i)\, .
\ee
\Eq{EulLagNRexplicit}  holds for a generic time-dependent density $\rho(t,\vx)$;  taking  now $\rho$ independent of time, and inserting \eq{sigmaijLame} into \eq{EulLagNRexplicit} 
we get
\be\label{NavierCauchycomponentsnof}
\rho(\vx) \ddot{u}_i=\pa_i\[\lambda(\vx)\pa_ju_j\]+\partial_j\[ \mu(\vx)  (\partial_i u_j+\partial_j u_i)\]\, ,
\ee
while, using again \eq{sigmaijLame}, the boundary condition  (\ref{njsigmaij}) becomes
\be\label{bclamevx}
\lambda(\vx) n_i \pa_ju_j +\mu(\vx)  n_j(\pa_iu_j+\pa_ju_i)=0\, .
\ee
If we further assume homogeneity, \eq{NavierCauchycomponentsnof} becomes
\be\label{NavierCauchy0comp}
\rho \ddot{u}_i=(\lambda+\mu) \pa_i\pa_ju_j+\mu \n^2 u_i\, ,
\ee
while the boundary condition  (\ref{bclamevx}) becomes
\be\label{2bodyboundcomp}
\lambda \,n_i \pa_ju_j +\mu\,  n_j (\pa_iu_j+\pa_ju_i)=0\, .
\ee
In vector form, \eqs{NavierCauchy0comp}{2bodyboundcomp} read
\be\label{NavierCauchy0}
\rho \ddot{\vu}=(\lambda+\mu) \n ( \n{\bf\cdot} \vu)  +\mu\n^2\vu\, ,
\ee
and
\be\label{2bodybound}
\lambda (\n\cdot {\bf u} )\hatn +2\mu (\hatn\cdot \n) {\bf u}+
\mu \hatn\times (\n\times {\bf u})=0\, .
\ee
\Eq{NavierCauchy0} becomes more clear performing the Helmholtz decomposition
\be
\vu=\vu_{\perp}+\vu_{\parallel}\, ,
\ee
where
\be
\n{\bf\cdot}\vu_{\perp}=0\, ,\qquad
\n\times\vu_{\parallel}=0\, ,
\ee
and writing $\vu_{\parallel}=\n\psi$. Then \eq{NavierCauchy0} separates into
\bees
\rho \ddot{\vu}_{\perp}&=&\mu\n^2\vu_{\perp}\, ,\label{equperp}\\ 
\rho \ddot{\psi}&=&(\lambda+2\mu)\n^2\psi\, ,\label{equpar}
\ees
showing that the transverse modes $\vu_{\perp}$ propagate with a speed of sound 
\be\label{vperp}
v_{\perp}=(\mu/\rho)^{1/2}\, ,
\ee 
while the longitudinal modes $\vu_{\parallel}$ propagate with a speed of sound 
\be\label{vparallel}
v_{\parallel}=[(\lambda+2\mu)/\rho]^{1/2}\, .
\ee
The whole formalism is immediately extended to the presence of an external force per unit volume, 
${\bf f}(t,\vx)$, adding a term $u_i f_i$ to the Lagrangian density, so that now the variation $\d {\cal L}/\d u_i$ in \eq{EulLag} no longer vanishes, and \eq{NavierCauchycomponentsnof} becomes
\be\label{NavierCauchycomponentswithf}
\rho(\vx) \ddot{u}_i=\pa_i\[\lambda(\vx)\pa_ju_j\]+\partial_j\[ \mu(\vx)  (\partial_i u_j+\partial_j u_i)\] +f_i\, ,
\ee
while
\eq{NavierCauchy0} becomes 
 \be\label{NavierCauchy1}
\rho \ddot{\vu}=(\lambda+\mu)\n (\n{\bf\cdot} \vu) +\mu\n^2\vu +{\bf f}\, ,
\ee
which is called the Navier-Cauchy equation. 

\subsection{Symmetries and conserved quantities}\label{sect:symmetriesNR}


We next turn to the consequences of the symmetries of the theory, using Noether's theorem, and applying it first  to  the Lagrangian   (\ref{LDysoninhom}).
 Let us recall the basic formalism of  Noether's theorem
(we follow the notation in Section~3.2 of  \cite{Maggiore:2005qv}). 
Consider a field theory with 
fields  $\phi_i(x)$, where $x$ denotes collectively  $(t,\vx)$ and the index $i$ is a generic label for the fields; in our case we are interested in the three components $u_i$ of a vector field $\vu$, but the notation is completely general. We use for convenience a relativistic formalism, with a four-vector notation $x^{\mu}=(t,\vx)$,  and summation over repeated upper and lower  Lorentz indices, with metric $\emn=(-,+,+,+)$. This, however, does not necessarily imply that the Lagrangian is Lorentz-invariant and, in fact, the Lagrangian (\ref{LDysoninhom}) that we are considering in this section is not.

We then consider 
an infinitesimal transformation of the coordinates and of the fields,
parametrized by a set of infinitesimal parameters $\epsilon^a$,
with $a=1, \ldots , N$, of the  form
\be\label{3variaz}
\begin{array}{rl}
x^{\mu}&\ra {x'}^{\mu}= x^{\mu}+\epsilon^a A_a^{\mu}(x)\\
\phi_i(x)&\ra\phi_i'(x')=\phi_i(x) +\epsilon^a F_{i,a}(\f ,\pa\f )\, ,
\end{array}
\ee
with $A_a^{\mu}(x)$ and $F_{i,a}(\f ,\pa\f )$ given.
The transformation (\ref{3variaz}) is
a symmetry of the theory if it leaves the action $S(\phi)=\int d^4x\, {\cal L}$ invariant.
Noether's theorem states that, if \eq{3variaz} is a global, but not a local, symmetry of
the theory, on a classical
solution of the equations of motion there are  $N$ conserved currents $j^{\mu}_a$,
\be\label{3Noe}
\pa_{\mu}j^{\mu}_a=0\, ,\qquad (a=1,\ldots, N)\, .
\ee
The currents can be obtained explicitly in terms of the Lagrangian density and the parameters $A_a$ and $F_{i,a}$ that define the transformation (\ref{3variaz}), as
\be\label{3current}
j_a^{\mu}=\frac{\delta \cal L}{\delta (\pa_{\mu}\phi_i)}
\[ A_a^{\nu}(x)\pa_{\nu}\phi_i-  F_{i,a}(\f, \pa\f ) \]
-A_a^{\mu}(x) {\cal L}\, .
\ee
We can now apply this to the transformation (\ref{transu}), which is indeed a global (but not a local)  symmetry of the action corresponding to the Lagrangian density (\ref{LDysoninhom}) (given that this is an internal symmetry that does not touch the coordinates, the invariance of the action is equivalent to the invariance of the Lagrangian density). For the transformation (\ref{transu}) the index denoted by $a$
in \eq{3variaz} is just a spatial index $i$, and the parameters $\epsilon^a$ are   the components $a^i$ of the vector $\va$ (taken now to be infinitesimal) that defines the translation in field space.  The coefficients $A_a^{\mu}(x)$ vanish since the coordinates do not change, while the transformation  of the fields $u_i$ in \eq{transu} can be written as
$u_i\ra u_i+a_j \delta_{ij}$,
so $F_{i,a}$ is now written as $F_{i,j}$, and is given by $F_{i,j}=\delta_{ij}$. Then, there are three conserved currents $j^{\mu}_i$, 
\be\label{pamjimu}
\pam  j_i^{\mu}=0\, ,
\ee
where the index $i=1,2,3$ is in correspondence with the three parameters $a^i$ of the transformation (\ref{transu}) and is therefore a spatial index, while $\mu$ is a Lorentz index, and \eq{3current} gives 
\be\label{jimu}
j_i^{\mu}=-\frac{\delta \cal L}{\delta (\pa_{\mu}u_j)}\delta_{ij} = -\frac{\delta \cal L}{\delta (\pa_{\mu}u_i)}\, .
\ee
Explicitly,
\bees
j_i^{0}&=& -\frac{\delta \cal L}{\delta \dot{u}_i}=-\rho \dot{u}_i\, ,\label{ji0}\\
j_i^{k}&=& -\frac{\delta \cal L}{\delta (\pa_k u_i)}=\sigma_{ik}\, ,\label{jij}
\ees
where $\sigma_{ij}$ was defined in \eq{defsigmaij} and, for the Lagrangian (\ref{LDysoninhom}), is given explicitly in \eq{Sijexpl}. Inserting \eqs{ji0}{jij} into \eq{pamjimu}, we get
\be\label{ConservNRexplicit}
\frac{d}{dt}\( \rho \dot{u}_i\) -\pa_j \sigma_{ij}=0\, ,
\ee
and we see that this is the same as the Euler-Lagrange equation (\ref{EulLagNRexplicit}). 
The fact that the conservation equation associated to the symmetry (\ref{transu}) is the same as the Euler-Lagrange equation of the theory can be understood, in our  setting, observing that the transformation (\ref{transu}) is a special case  of the broader set of variations $u_i\ra u_i+\delta u_i$, at fixed $\vx$,  used to derive the equations of motion by requiring that the variation of the action vanishes.

Consider next  space-time translations. Time translations are symmetries of the Lagrangian (\ref{LDysoninhom}) only if $\rho$ and $c_{ijkl}$ are independent of $t$, and spatial translations only if   they  are independent of $\vx$. The latter condition, in particular, implies that the elastic body has an infinite spatial extent, and therefore is never realized in a real system. Let us however assume for the moment 
that $\rho$ and $c_{ijkl}$ are constants in space and time, so that the Lagrangian reduces to  \eq{LDyson}. 
Then, the theory is invariant under space-time translations, 
\be\label{3spacetime}
\begin{array}{rl}
x^{\mu}&\ra {x'}^{\mu}= x^{\mu}+\epsilon^{\mu}\, ,\\
u_i(x)&\ra u_i'(x')=u_i(x) \, .
\end{array}
\ee
Comparing with \eq{3variaz} we see that the index $a$ is now a Lorentz index, that we denote by $\nu$, and we have $A_{\nu}^{\mu}=\delta^{\mu}_{\nu}$ while $F_{i,\nu}=0$.
Noether's theorem then gives four conserved currents 
$\theta_{(\nu)}^{\mu}$, labeled by an index $\nu$, that (on the solutions of the equations of motion) satisfy 
$\pam \theta_{(\nu)}^{\mu}=0$.
Of course, since $\nu$ is a Lorentz index, $\theta^{\mu}_{(\nu)}$ is actually a Lorentz tensor;
we then write it simply as ${\theta^{\mu}}_{\nu}$ and we define $\theta^{\mu\nu}=\eta^{\nu\rho} {\theta^{\mu}}_{\rho}$. This is   the energy--momentum tensor of the theory.  
The conservation laws  $\pam  \theta_{(\nu)}^{\mu}=0$ then gives
\be\label{pamTMNzero}
\pam \theta^{\mu\nu}=0\, .
\ee
From \eq{3current},  we have\footnote{Of course, the overall sign and overall multiplicative coefficients in
the expression for the conserved current in  \eq{3current}
are  arbitrary since, if $j^{\mu}$ is conserved, $\kappa j^{\mu}$ is also conserved, for an arbitrary constant $\kappa$. The normalization of the currents obtained from Noether's theorem is in general fixed so to reproduce the standard normalizations of the corresponding physical quantities. Here we have fixed the overall sign so that, with our metric signature, $\theta^{00}$ gives the energy density. Note that in  \cite{Maggiore:2005qv} the opposite metric signature is used.}
\be\label{3Tmunu}
\theta^{\mu\nu}=
-\frac{\delta {\cal L}}{\delta (\pa_{\mu}u_i)} \pa^{\nu}u_i
+\eMN{\cal L}
\, ,
\ee
and, using the Lagrangian density (\ref{LDyson}), we get 
\bees
\theta^{00}&=&\frac{1}{2}  \rho\dot{u}_i\dot{u}_i+\frac{1}{2} c_{ijkl}\pa_i u_j\pa_k u_l\, ,\label{T00timetransl}\\
\theta^{i0}&=&-c_{ijkl}\dot{u}_j\pa_k u_l \, ,\label{Tio}\\
\theta^{0i}&=&-\rho \dot{u}_k\pa_i u_k\, ,\label{Toi}\\
\theta^{ij}&=&c_{iklm}\pa_ju_k\pa_lu_m+\delta_{ij}\frac{1}{2}\(  \rho\dot{u}_k\dot{u}_k-c_{klmn}\pa_k u_l\pa_m u_n\)\, .\label{T00timetranslij}
\ees
Some comments are useful here:

\vspace{1mm}\noindent
{\bf (1)} The tensor $\theta^{\mu\nu}$ is quadratic in $\pa u$ (where by $\pa u$ we denote here generically a spatial or a time derivative of $u_i$), contrary to the conserved current (\ref{ji0}, \ref{jij}), which is linear in $\pa u$.

\vspace{1mm}\noindent
{\bf (2)}
If  $\rho$ and $c_{ijkl}$ are time-independent, so that
time translations are a symmetry, the $\nu=0$ component of the conservation law (\ref{pamTMNzero}) holds, and gives
\be\label{pa0T00paiTi0}
\pa_0 \theta^{00}+\pa_i \theta^{i0}=0\, .
\ee
Since $\theta^{00}$ is the energy density, this shows that $\theta^{i0}$ is the energy flux. From \eqs{Tio}{Sijexpl}, we see that it can be rewritten in terms of the stress tensor as 
\be
\theta^{i0}=-\sigma_{ij}\dot{u}_j\, .
\ee 
Then, if $\hatn$ is the normal to the surface of the boundary of the body, $n_i\theta^{i0}=-\sigma_{ij}n_i\dot{u}_j$ and, because of the boundary condition (\ref{njsigmaij}), together with \eq{sigmasym},
this vanishes. Therefore, the physical meaning of the boundary condition (\ref{njsigmaij}) is that there is no flow of elastic energy outside the body.
Spatial translations would rather imply 
$\pa_0 \theta^{0j}+\pa_i \theta^{ij}=0$ but, as we have discussed,  this assumes a  homogeneous body of infinite extent.

\vspace{1mm}\noindent
{\bf (3)} Note that \label{point3}
$\theta^{\mu\nu}$ is not symmetric, $\theta^{0i}\neq \theta^{i0}$ (and, for $c_{iklm}$ generic, also
$\theta^{ij}\neq \theta^{ji}$). 
The fact that $\theta^{0i}\neq \theta^{i0}$ is a consequence of the fact that this tensor is derived from a Lagrangian that is not relativistic. Indeed,  $\theta^{i0}$ is the energy flux [as we see from \eq{pa0T00paiTi0}], while $\theta^{0i}$ is the momentum density. In a relativistic theory, where energy and momentum form a four-vector, the two quantities are related and 
$\theta^{0i}= \theta^{i0}$. However, in a non-relativistic theory, this is no longer true, as we see from this explicit example.
As we see from \eq{pamTMNzero}, a non-relativistic Lagrangian can still generate a covariantly conserved energy--momentum tensor, if the theory is invariant under space and time translations. However, the sign that this energy--momentum tensor comes from a non-relativistic theory is in the fact that $\theta^{0i}\neq \theta^{i0}$.\footnote{Actually, even in a  relativistically--invariant  classical field theory the energy--momentum tensor derived from Noether's theorem can be non-symmetric, but in this case it can be made symmetric by adding total derivative terms that do not affect the corresponding conserved charges, i.e.  the total energy and momentum, see e.g. Section~3.2.1 of \cite{Maggiore:2005qv}. This is not the case here, where 
$\theta^{0i}$ and $\theta^{i0}$ have nothing to do with each other, to the extent that one is proportional 
to $\rho$ and the other to $c_{ijkl}$. In contrast, the energy--momentum tensor computed in General Relativity as a variation of the action with respect to $\gmn$ (see \eq{TmnGR} below) is automatically symmetric.}

\section{Relativistic effective field theory of elasticity}\label{sect:EFT_Rel}

We now discuss the relativistic formulation of the EFT of elasticity, following 
refs.~\cite{Dubovsky:2005xd,Endlich:2010hf,Endlich:2012pz} and elaborating on them.  In Section~\ref{sect:coupling_to_GWs} we will generalize this construction to a full diffeomorphism invariant theory, which will allow us to determine the coupling to GWs, but in this section we restrict to special relativity. In this section we use units $c=1$.
 
\subsection{Fundamental field variables and spontaneous symmetry breaking}\label{sect:defxi}

Let us begin with the ``kinematics" of the problem, i.e. the introduction of the most appropriate variables, before studying their dynamics. 
The  first fundamental  idea, which in the context of relativistic elasticity goes back to \cite{Carter:1977qf}, is to make a distinction between  four-dimensional  space-time,  denoted by ${\cal M}$, and an abstract three-dimensional spatial manifold ${\cal B}$, called the ``material manifold" or the ``body manifold"; in this section  the manifold ${\cal M}$ is taken to be Minkowski space-time,   with  metric $\emn=(-,+,+,+)$, and $\mu,\nu=0,1,2,3$. The manifold
${\cal B}$, instead, can be seen as an abstract three-dimensional space made by the  collection of all points of the elastic body. We can for instance visualize ${\cal B}$ imagining that at each point of the body there is an idealized point-like constituent; we will refer to it as a ``molecule".\footnote{We are assuming that the elementary constituents do not have internal degrees of freedom or, more precisely, that they do not affect the long-distance dynamics to lowest order; otherwise, such a description will have to be enlarged  to take into account the internal degrees of freedom.}$^{,}$\footnote{Note that also non-relativistic elasticity can be formulated  making an explicit distinction between the space ${\cal B}$, that represents a reference configuration (or, equivalently, a set of labels of the material ``molecules"), and the actual three-dimensional space $\realnum^3_x$, and introducing a mapping from  $\realnum_t\times \realnum^3_x$ to ${\cal B}$, see e.g. Chapter~2 of \cite{Holzapfel}.  We did not introduce this slightly more complex formalism in Section~\ref{sect:EFT_NR}, where it was not essential for our purposes, while it will be crucial here to implement Lorentz invariance.}
 
 For the moment, we consider an   elastic medium that fills all of space, so that ${\cal B}$ is an $\realnum^3$ space, and we also take the medium to be homogeneous. However, similarly to what we discussed in the previous section,  we will later need to relax this assumption   since we will see that,  in the TT gauge, GWs  only couple to the inhomogeneities of the body, or to the discontinuities at the body's boundary. We introduce coordinates $\xm$  ($\mu=0,1,2,3$) in ${\cal M}$, and $X^a$ ($a=1,2,3$) in ${\cal B}$,  and we take $\delta_{ab}$ as the metric in ${\cal B}$.\footnote{See, however, footnote~\ref{foot:diffB} below
 for a possible different strategy involving  the construction of a theory invariant also under diffeomorphism in ${\cal B}$. }
 We can think of $X^a$ as the labels assigned to the ``molecules" of the elastic body.  The space-time configuration of the body can be specified by assigning, to each  space-time point $\xm=(t,\vx)$,  the molecule of the material that passes through the spatial point $\vx$ at time $t$. This is obtained by specifying a map ${\cal M}\ra {\cal B}$.\footnote{Observe that this assumes that the elastic body has infinite extent and also that it is topologically trivial, e.g. that it has no empty cavities since, to define this map,  we are implicitly assuming that, at each point $\vx$ of real space (and for each $t$) there is a molecule of the body that passes through it.} Given the above coordinate systems in ${\cal M}$ and in  ${\cal B}$, such a map can be written as
\be\label{mapXxi}
X^a=\xi^a(x)\, , \qquad (a=1,2,3)\, ,
\ee
where we use $x$ as a shorthand for $(t,\vx)$. 
The fundamental idea of this approach is to use the functions $\xi^a(x)$ as our fundamental fields, and construct an effective field theory describing their  long-distance dynamics.  
An equivalent way of thinking to this mapping 
is to attach a  label $\xi^a$ to each ``molecule" of the body, and follow its evolution through space, which will be given by a function $\vx(t;\xi^a)$. We assume that this mapping is invertible for all $t$ (once again, this assumes that the body has infinite spatial extent, so 
at each point $\vx$ of real space and for each $t$ there is a molecule of the body  passing through it), so this can be inverted to give $\xi^a(t,\vx)$.
 The description using $\vx(t;\xi^a)$ corresponds to the Lagrangian perspective in fluid mechanics, where we follow the evolution in time of a particle labeled by $\xi^a$. The description using $\xi^a(t,\vx)$ corresponds instead to the Eulerian perspective, and is the more natural if we want to build a field-theoretical description, in particular when we want to build a Lorentz-invariant formalism, since it treats $t$ and $\vx$ on the same footing.
Observe that the three fields $\xi^a(x)$ are  {\em scalars} from the point of view of rotations in ${\cal M}$, and in fact more generally under Lorentz or Poincar\'e transformations  (or even diffeomorphisms) in ${\cal M}$, since $X^a$ identifies abstractly the molecule in ${\cal B}$, irrespectively of the coordinate system used in ${\cal M}$. In contrast, $\vxi=(\xi^1,\xi^2,\xi^3)$ is a vector from the point of view of rotations in ${\cal B}$.

Given  an elastic body in a static equilibrium configuration,   there is a particularly convenient choice of coordinates $X^a=(X^1,X^2,X^3)$ in ${\cal B}$ and $x^{\mu}=(t,x,y,z)$ in ${\cal M}$, obtained setting 
$X^1=x$, $X^2=y$ and $X^3=z$, i.e. choosing the functions $\xi^a(x)$ as
\be\label{xiequilexplicit}
\xi^1(t,x,y,z)=x\, ,\qquad   \xi^2(t,x,y,z)=y\,, \qquad \xi^3(t,x,y,z)=z\, ,
\ee 
that we write compactly as
\be\label{xiaequilibrium}
\xi^a(x)=\delta_i^a x^i\, .
\ee
We will see below that, for an infinite and homogeneous body (but, crucially, only in that case), \eq{xiaequilibrium} is a solution of the equations of motion derived from the action that describes relativistic elasticity.
To understand the meaning of \eq{xiaequilibrium}, and the different roles of the $a$ and $i$ indices, 
it is important to realize that, in an equation such as $\xi^1(t,x,y,z)=x$, the left- and right-hand sides transform differently under translations and under  Lorentz transformations in ${\cal M}$: under these transformations $\xi^1(t,x,y,z)$ is a scalar field, and its label $a=1$ just identifies it as one of the three scalar fields $\xi^a(t,x,y,z)$. In contrast, $x$ is the first  component of the vector $\vx=(x,y,z)$. Therefore $\xi^1(t,x,y,z)$ and $x$ transform differently under translations and Lorentz transformations in ${\cal M}$, and an equation such as $\xi^1(t,x,y,z)=x$ can only hold in a specific reference frame in ${\cal M}$, or at most in a family of frames; conversely, from the point of view of rotations in ${\cal B}$ space, $\xi^a$ are the three components of   a vector  $\vxi=(\xi^1,\xi^2,\xi^3)$, while $x,y,z$ are scalars, since they are not affected by any transformation in ${\cal B}$.
Therefore, \eq{xiaequilibrium} selects a frame, both in ${\cal M}$ and in ${\cal B}$ (or, more precisely, still a family of frames, see below). In the language of  field theory, it 
corresponds to a choice of ``vacuum", and this choice  breaks spontaneously Lorentz and spatial translation  invariance in ${\cal M}$, as well as rotation and translation invariance in ${\cal B}$. 

However, consider the combined transformation 
\be\label{diagtrans}
X^a\ra X^a+C^a\, ,\qquad
x^i \ra x^i+c^i\, ,
\ee
where ${\bf C}=(C^1,C^2,C^3)$ is a constant vector in ${\cal B}$, while  ${\bf c}=(c^1,c^2,c^3)$  is a constant vector in ${\cal M}$, and the two are related by the condition that,  in a frame where
\eq{xiaequilibrium} holds, 
$C^1=c^1, C^2=c^2$ and $C^3=c^3$, i.e. $C^a=\delta^a_i c^i$. \Eq{diagtrans} is a combination of a translation  in ${\cal B}$ and a translation  in ${\cal M}$, and leaves \eq{xiaequilibrium} invariant. 
Similarly, we can perform a spatial rotation  in ${\cal B}$ on the $X^a$, and the corresponding rotation in ${\cal M}$ on the $x^i$
\be\label{diagrot}
X^a\ra {R^a}_{b}X^b\, ,\qquad
x^i  \ra {r^i}_j x^j\, ,
\ee
with ${r^i}_j=\delta^i_a\delta^b_j {R^a}_{b}$, and again \eq{xiaequilibrium} is invariant. Therefore, the choice of coordinates leading to \eq{xiaequilibrium} breaks the Poincar\'e group $ISO(3,1)$ of ${\cal M}$ (i.e., the group made by the Lorentz group $SO(3,1)$ together with space-time translations)  except for  the time translation symmetry, which is not broken by \eq{xiaequilibrium}, and it also breaks
 the group $ISO (3)$ of rotations and translations in ${\cal B}$. However, it leaves unbroken  the transformation made by   a simultaneous translation  in ${\cal M}$ and in ${\cal B}$, given in \eq{diagtrans}, as well as the simultaneous spatial rotations  in ${\cal M}$ and in ${\cal B}$  
given by \eq{diagrot}. In the group-theory language these are called the ``diagonal" combination, and we denote the corresponding  group of combined translations and rotations by $ISO(3)_{\rm diag}$.
So, keeping into account that the time translations of  $ISO(3,1)$ (that we denote by $T_t$) are not broken by \eq{xiaequilibrium},   the symmetry breaking pattern is~\cite{Esposito:2020wsn}
\be\label{symbreakpattern}
ISO(3,1)\times ISO(3)\ra  T_t\times ISO(3)_{\rm diag}\, .
\ee
To make contact with  the formalism of Section~\ref{sect:EFT_NR}, we now expand $\xi^a(x)$ around the unperturbed configuration (\ref{xiaequilibrium}), writing
\be\label{xiapi}
\xi^a(x)=\delta_i^a x^i +\pi^a(x)\, .
\ee
We have seen that the choice of the background configuration (\ref{xiaequilibrium}) breaks spontaneously the Lorentz and translation  invariance of the theory; then, in the field-theoretical language, the three scalar fields $\pi^a(x)$ are the Goldstone bosons associated to the spontaneous breaking  pattern (\ref{symbreakpattern}).\footnote{If we denote the symmetry generators in ${\cal M}$ by $P^0$ (time translations), $P^i$ (space translation), $J^i$ (rotations) and  $K^i$ (boosts), and those in ${\cal B}$ by $\bar{P}^i$ (space translations) and $\bar{J}^i$ (rotations), the unbroken generators of  $T_t\times ISO(3)_{\rm diag}$ are $P^0$, $P^i+\bar{P}^i$ and $J^i+\bar{J}^i$, while the broken generators are $K^i$, $P^i-\bar{P}^i$ and $J^i-\bar{J}^i$. Given that there are 9 broken generators one might naively think that Goldstone's theorem requires nine Goldstone bosons, while \eq{xiapi} shows that there are only three. However, the textbook version of Goldstone's theorem, that asserts the existence of one Goldstone boson for each broken symmetry generator, only holds for internal symmetries. When space-time symmetries are spontaneously broken, some of the Goldstone bosons can be expressed in terms of the others, through conditions known as ``inverse Higgs constraints." See, e.g., Section~3.5 of \cite{Penco:2020kvy} for a recent review.}
We now define  a vector field $\vu(x)$ in ${\cal M}$, with components $u^i(x)$ ($i=1,2,3$)  requiring that, {\em in the frame where \eq{xiaequilibrium} holds}, its components take the values $u^1(x)=-\pi^1(x)$, $u^2(x)=-\pi^2(x)$ and $u^3(x)=-\pi^3(x)$, which we write compactly as 
\be\label{piaui}
\pi^a(x)=-\delta^a_i u^i(x)\, .
\ee 
Once again, given a choice of coordinate $X^a$ in ${\cal B}$, an equation such as $u^1(x)=-\pi^1(x)$ can only hold in some frames in ${\cal M}$,   since, from the point of view of Lorentz transformations in  ${\cal M}$,  $\pi^1(x)$ is just one of the three scalar fields $\pi^a(x)$, while $u^1(x)$ is the first component of the vector field 
$\vu(x)=(u^1(x),u^2(x),u^3(x))$.\footnote{More precisely, denoting by ${\cal F}_{\cal B}$ a generic  frame in ${\cal B}$ and by ${\cal F}_{\cal M}$ a generic  frame in ${\cal M}$, \eq{piaui} only holds for pairs of frames $({\cal F}_{\cal B},{\cal F}_{\cal M})$ related by the unbroken group $T_t\times ISO(3)_{\rm diag}$, see \eq{symbreakpattern}.}
In such  frames,
\eq{xiapi} can then be written as 
\be\label{xiaandui}
\xi^a(x)=\delta_i^a [x^i -u^i(x)]\, ,
\ee  
or $\xi^a(x)=\delta_i^a \xi^i(x)$, where
\be\label{defxii}
\xi^i(x)=x^i -u^i(x)\, .
\ee
As we will see, the fields $u^i(x)$ defined in this way [including the choice of sign in front of $u^i$ in \eq{xiaandui}]
are just the same  as the fields $u^i(x)$ that were the basic field variables in Section~\ref{sect:EFT_NR}. The crucial difference is that in Section~\ref{sect:EFT_NR} we started directly from these fields; since they are spatial vectors, with them we can build a  Lagrangian invariant under spatial rotations, but not  under the full group of Lorentz transformations. In contrast, the fields $\xi^a(x)$ are scalars from the point of view of  Lorentz transformations in  ${\cal M}$ and, with them, we can build a Lorentz-invariant Lagrangian. The non-Lorentz invariant Lagrangian (\ref{LDyson}) will then emerge as a consequence of the fact that we choose to expand $\xi^a(x)$ around  the background configuration (\ref{xiaequilibrium}), which is not  Lorentz-invariant, i.e. as a consequence of the spontaneous symmetry breaking (\ref{symbreakpattern}). However the fundamental theory, at the level of the Lagrangian, will now be Lorentz invariant.

\subsection{Relativistic effective actions for elastic  bodies. General formalism}\label{sect:isoandhom}

We now want to construct an  action $S$, involving the fields $\xi^a$ and they first derivatives $\pam\xi^a$,  and invariant under Lorentz transformations in  ${\cal M}$.
We write it in the form
\be\label{Sxipaxi}
S=\int_{\cal M} d^4x\, {\cal L}(\xi,\pam\xi)\, ,
\ee
and we look for a Lorentz-invariant Lagrangian density ${\cal L}$, following \cite{Dubovsky:2005xd,Endlich:2010hf}.
We begin in this section with the simpler case of a homogeneous body, filling all of space. Recall, in particular, that the whole formalism based on the introduction of the mapping (\ref{mapXxi}) assumes  that the elastic body fills all of space, otherwise this mapping is not even well defined. A finite-size body can of course be modeled using a density $\rho(\vx)$  that has a compact support, so that it goes to a negligibly small value outside the body, but this will anyhow require to deal with an inhomogeneous situation, whose treatment  we defer to Section~\ref{sect:notisoandhom}.

In the logic of EFT, the dynamics of the Goldstone bosons must be invariant under the residual symmetry group after symmetric breaking, i.e. the group $T_t\times ISO(3)_{\rm diag}$ given in \eq{symbreakpattern}. In particular, the dynamics  must be invariant under the combined transformation (\ref{diagtrans}). 
For a homogeneous body of infinite extent, however, spatial translations in ${\cal M}$ are separately a symmetry.  If the system is invariant under the combined transformation (\ref{diagtrans}), as well under separate  spatial translations  $x^i \ra x^i+c^i$ in ${\cal M}$,  it follows that also the translations in ${\cal B}$, $X^a\ra X^a+C^a$, are separately a symmetry.
Therefore, 
in this case the action for the $\xi^a$ fields must be separately invariant also under 
$\xi^a(x)\ra\xi^a(x) +C^a$.
This  forces the Lagrangian density ${\cal L}$ in \eq{Sxipaxi}
to depend only on the derivatives $\pam\xi^a$ of the field, and not on the field  $\xi^a$ itself; in order to saturate the Lorentz index in  $\pam\xi^a$, we then need to contract it with another factor  $\pan\xi^b$, and Lorentz invariance  requires that the contraction is performed with the metric $\eMN$. Therefore, ${\cal L}$ can depend on the field $\xi^a$ and its derivatives only through the quantity 
\be\label{defBab}
B^{ab} \equiv \eMN \pam\xi^a\pan\xi^b = \pam\xi^a\paM\xi^b \, .
\ee
Inserting (\ref{xiapi}) into \eq{defBab} we get 
\be\label{Babpi}
B_{ab}=\delta_{ab}+\pa_a\pi_b+\pa_b\pi_a +\pam\pi_a\paM\pi_b\, .
\ee
Note that, since the metric in material space is $\delta_{ab}$, we can write the indices $a, b$ as upper or lower indices, at will. Similarly to \eq{piaui}, we can also 
write $B_{ab}=\delta_a^i \delta_b^j B_{ij}$,  and
we then have
\be\label{Bijui}
B_{ij}=\delta_{ij}-\pa_i u_j -\pa_j u_i +\pam u_i\paM u_j\, .
\ee
In the logic of EFT, we now want to write a long-distance action as an expansion in powers of the derivatives 
of $u$ (which we generically denote as $\pa u$ when counting the order of the expansion). Note, however, that $B_{ij}$ starts from $\delta_{ij}$, so it is of order one, i.e. $(\pa u)^0$, and higher powers of $B_{ij}$ are therefore still of order one, and cannot be considered as corrections. Therefore, the effective action will be fully non-linear in $B_{ij}$.

At this stage, the most general Lagrangian density is an arbitrary function of $B_{ij}$, which still leaves a huge freedom.  However, 
if we further assume that the body is isotropic, rotational symmetry  limits significantly the possible independent terms that can be written. The argument is the same as that used above for translations: we have seen that \eq{diagrot} is a symmetry of the system; however, if the body is isotropic, the spatial rotations $x^i  \ra {r^i}_j x^j$ are separately a symmetry operation, so also rotations in ${\cal B}$, of the form $X^a\ra {R^a}_{b}X^b$ must  separately be  an invariance of the Lagrangian. This means that 
the Lagrangian can only depend on quantities constructed from $B^{ab}$ which are invariant under rotations  in ${\cal B}$, i.e. with respect to the $(a,b)$ indices. For a $3\times 3$ matrix, there are only three independent invariant that can be constructed, which can be taken to be $\det B$, ${\rm Tr}\,  B$ and ${\rm Tr} (B^2)$.\footnote{In general, the principal invariants of a matrix $B$ are the coefficients of the characteristic polynomial $p(\lambda)=\det (B-\lambda I)$, where $I$ is the identity matrix and 
 the roots of $p(\lambda)$ are the (possibly complex) eigenvalues. For a $3\times 3$ matrix this gives the three invariants $I_1\equiv {\rm Tr}\, B$,
$I_2\equiv (1/2) \[ ({\rm Tr}\, B)^2- {\rm Tr} (B^2)\]$ and $I_3=\det B$. These invariants have a simple expression in terms of the eigenvalues, $I_1=\lambda_1+\lambda_2+\lambda_3$, $I_2=\lambda_1\lambda_2+\lambda_1\lambda_3+\lambda_2\lambda_3$ and $I_3=\lambda_1\lambda_2\lambda_3$. In our case, however, we find simpler to use ${\rm Tr} (B^2)$ instead of the combination $I_2$, as one of the independent invariants.
Another possibility, used in \cite{Endlich:2012pz}, is to use ${\rm Tr}\,  B$, ${\rm Tr} (B^2)$
and ${\rm Tr} (B^3)$ as a basis. In term of these,
\be
6\det B=({\rm Tr}\,  B)^3-3\, {\rm Tr}\,  B\, {\rm Tr}  (B^2) +2\, {\rm Tr} (B^3)\, .
\ee
}
Among these invariants, a special role is played by $\det B$.  As explained in \cite{Dubovsky:2005xd,Endlich:2010hf}, this is the only quantity that can be formed that, beside being rotationally invariant, is also invariant  under diffeomorphisms of the material manifold, 
$\xi^a\ra {\xi'}^a(\xi)$, that preserve the volume in material space, i.e. such that $\det (\pa{\xi'}^a/\pa\xi^b)=1$. This is a symmetry that is expected for a relativistic fluid (see the discussion in Section~5.1 of  \cite{Dubovsky:2005xd}).  We find convenient to use as variable $\sqrt{\det B}$, instead of $\det B$.
Therefore, the most general effective action for a relativistic fluid has the form~\cite{Dubovsky:2005xd} 
\be
S=\int_{\cal M} d^4x\, {\cal L}(\sqrt{\det B}\, )\, ,
\ee
for some arbitrary function ${\cal L}$.
For a solid, however, this extra symmetry is not present, and the most general action of an isotropic and homogeneous solid has the form
\be\label{SdetBTrBTrB2}
S=\int_{\cal M} d^4x\, {\cal L}\(\sqrt{\det B}, {\rm Tr}\,  B, {\rm Tr}\,  B^2 \)\, ,
\ee
again for some arbitrary function ${\cal L}$. In terms of the field $\xi^a$,
\bees
\det B
&=& \frac{1}{3!}\eps_{abc}\eps_{a'b'c'}\pam\xi^a\paM\xi^{a'} \pan\xi^b\paN\xi^{b'} \parho\xi^c\paR\xi^{c'}\, , \label{detB} \\
{\rm Tr}\,  B&=& \pam\xi^a\paM\xi^a\, ,  \label{TrB}\\
{\rm Tr}\,  B^2&=& \pam\xi^a\paM\xi^b \pan\xi^a\paN\xi^b\, .  \label{TrB2} 
\ees
The action (\ref{SdetBTrBTrB2}) is still very general, and includes all possible non-linear couplings consistent with homogeneity and isotropy. If we are interested only in linear elasticity (linear in $\pa u$ in the equations of motion, so quadratic in  $\pa u$ in the action)   
we can write $\xi^a(x)$ as in \eq{xiaandui} and expand the action to quadratic order in $\pa u$. We find convenient to introduce the notation
\be\label{defnpq}
n(x)= \sqrt{\det B}\, ,\qquad
p(x)=\frac{1}{3} {\rm Tr} \, B\, ,\qquad
q(x)=\frac{1}{3} {\rm Tr}\,  B^2\, ,
\ee
and work in term of these quantities. Note that $n$, $p$ and $q$,   as $B^{ab}$, are dimensionless. Then, the most general relativistic action of an isotropic and homogeneous solid can be written as
\be\label{SFnm}
S=-\int_{\cal M} d^4x\, F(n,p,q)\, ,
\ee
for a generic function $F(n,p,q)$ (we have extracted a minus sign for later convenience). We next expand $n,p$ and $q$ in powers of $\pa u$. 
For $n(x)$, the  expansion to quadratic order is given by
\be\label{nvsu}
n(x)=1-\pa_iu_i -\frac{1}{2}\dot{u}_i^2 +\frac{1}{2}\[(\pa_iu_i)^2-\pa_i u_j\pa_ju_i\]+{\cal O}
(\pa u)^3
\, .
\ee
We prove this result in App.~\ref{app:n}, where we also provide  a more geometric definition of $n(x)$, which makes contact with the formalism used in \cite{Magli:1992,Beig:2002pk}.
For $p(x)$, from \eq{Bijui} we get the exact expression 
\be\label{pvsu}
p(x)=1 -\frac{2}{3}\pa_iu_i -\frac{1}{3}\dot{u}_i^2 +\frac{1}{3}\pa_iu_j\pa_iu_j\, ,
\ee
and for $q(x)$ we get
\be\label{qvsu}
q(x)=1 -\frac{4}{3}\pa_iu_i -\frac{2}{3}\dot{u}_i^2 +\frac{2}{3}\pa_iu_j\pa_iu_j
+\frac{4}{3} u_{ij}u_{ij}+  {\cal O}(\pa u)^3 \, .
\ee
We now expand the action (\ref{SFnm})  up to quadratic order, writing $n=1+\delta n$, 
$p=1+\delta p$, $q=1+\delta p$ and Taylor-expanding $F(n,p,q)$ around $n=p=q=1$,
\bees\label{Sdeltanm}
S&=&-\int_{\cal M} d^4x\,  \[ F_0+ \(\frac{\pa F}{\pa n}\)_0 \delta n
+\(\frac{\pa F}{\pa p}\)_0 \delta p
+\(\frac{\pa F}{\pa q}\)_0 \delta q
\right. \nn\\
&&\hspace*{15mm}+
\frac{1}{2}\(\frac{\pa^2 F}{\pa n^2}\)_0 (\delta n)^2
+\frac{1}{2}\(\frac{\pa^2 F}{\pa p^2}\)_0 (\delta p)^2
+\frac{1}{2}\(\frac{\pa^2 F}{\pa q^2}\)_0 (\delta q)^2\nn\\
&&\hspace*{15mm} \left. 
+\(\frac{\pa^2 F}{\pa n\pa p}\)_0 \delta n\delta p
+\(\frac{\pa^2 F}{\pa n\pa q}\)_0 \delta n\delta q
+\(\frac{\pa^2 F}{\pa p\pa q}\)_0 \delta p\delta q
\]\, ,
\ees
where the subscript ``0" on $F$ and its  derivatives means that they are evaluated at the equilibrium configuration $n=p=q=1$. 
We then read $\delta n, \delta p$ and $\delta q$ from \eqss{nvsu}{pvsu}{qvsu}. The term $F_0$ gives an irrelevant constant, while the terms linear in $\pa u$ are  proportional to $\pa_i u_i$ and therefore are a total derivative, that gives zero after integration by parts with the boundary condition that $u_i$ vanishes at spatial infinity.\footnote{The coefficient of $\pa_i u_i$  is in fact the combination on the right-hand side of \eq{rhopaF} below, that we will identify with the constant density $\rho$.} This shows, as anticipated, that the background configuration (\ref{xiaequilibrium}) is a solution of the equations of motion; observe that this is true for any function $F$.\footnote{More generally, following the same steps, we can check that  the configuration
\be\label{xiequilexplicitlambda}
\xi^1(t,x,y,z)=\lambda_1 x\, ,\qquad   \xi^2(t,x,y,z)=\lambda_2 y\,, \qquad \xi^3(t,x,y,z)=\lambda_3 z\, ,
\ee
with arbitrary constants $\lambda_1$, $\lambda_2$, $\lambda_3$, is also a solution. }
We are then left with the quadratic and higher-order terms,
\be\label{ScalL2u3}
S= \int_{\cal M} d^4x\,  \[{\cal L}_2+{\cal O}(\pa u)^3\]\, .
\ee
For the quadratic part of the Lagrangian density, ${\cal L}_2$, we get, after some integrations by parts,
\be\label{LDysonLameagain}
{\cal L}_2=\frac{1}{2}  \[  \rho\dot{u}_i\dot{u}_i -\lambda (\pa_iu_i)^2-2\mu\, u_{ij}u_{ij}
\]\, ,
\ee
where
\bees
\rho &=& \(\frac{\pa F}{\pa n}\)_0+\frac{2}{3} \(\frac{\pa F}{\pa p}\)_0
+\frac{4}{3} \(\frac{\pa F}{\pa q}\)_0\, ,\label{rhopaF}\\
\lambda &=&- \frac{2}{3} \(\frac{\pa F}{\pa p}\)_0- \frac{4}{3} \(\frac{\pa F}{\pa q}\)_0
+ \(\frac{\pa^2 F}{\pa n^2}\)_0
+ \frac{4}{9} \(\frac{\pa^2 F}{\pa p^2}\)_0
+ \frac{16}{9} \(\frac{\pa^2 F}{\pa q^2}\)_0\nn\\
&&+\frac{4}{3} \(\frac{\pa^2 F}{\pa n\pa p}\)_0
+\frac{8}{3} \(\frac{\pa^2 F}{\pa n\pa q}\)_0
+\frac{16}{9} \(\frac{\pa^2 F}{\pa p\pa q}\)_0
\, ,\\
\mu &=&\frac{2}{3}  \(\frac{\pa F}{\pa p}\)_0
+\frac{8}{3}  \(\frac{\pa F}{\pa q}\)_0\, .\label{mufromF}
\ees
At quadratic order, we have therefore recovered the non-relativistic Lagrangian (\ref{LDysonLame}). The specific form of the function $F(n,p,q)$ determines the constants $\rho,\lambda$ and $\mu$, as well as all the non-linear couplings.

An interesting limiting case of the above formulas is obtained when $F$ is independent of $p$ and $q$, which, as discussed in \cite{Dubovsky:2005xd,Endlich:2010hf}, corresponds to the limit of a relativistic fluid. Then $\rho$ and $\lambda$ are determined by the first and second derivatives of $F(n)$, 
\be\label{rholambdaFn}
\rho=\(\frac{dF}{dn}\)_0\, ,\qquad \lambda = \(\frac{d^2F}{dn^2}\)_0\, ,
\ee
while $\mu=0$.  In this limit, from \eqs{vperp}{vparallel}, we see that the transverse modes no longer propagates, and the propagating waves become purely  longitudinal. 

An even simpler case is given by the choice $F(n)=k n$, where $k$ is a constant. Then, from \eq{rholambdaFn}, $\rho=k$ while $\lambda=\mu=0$. Therefore, the action
\be\label{Sd4xn}
S=-\rho \int_{\cal M} d^4x\, n(x)
=-\rho \int_{\cal M} d^4x\, \sqrt{\det B}\, ,
\ee
at quadratic order, describes an ensemble of free relativistic particles, with no elastic forces. Elastic forces would however manifest beyond linear order, expanding $n(x)$ up to cubic order in $\pa u$.\footnote{The fact that, up to  quadratic order, $\int d^4x\, n(x)$ only produces a kinetic term can also be directly seen from \eq{nvsu}, observing that the terms $\pa_i u_i$ and $(\pa_iu_i)^2-\pa_i u_j\pa_ju_i =\pa_i (u_i\pa_j u_j-u_j\pa_j u_i)$ are total derivatives, so, apart from an irrelevant constant,  only $(-1/2)\dot{u}_i^2$ remains.}

\subsection{Action for homogeneous and anisotropic elastic bodies}\label{sect:homanis}

We now consider the more general case of a body which is not isotropic, while still assuming for the moment homogeneity. When looking for the most general action, the lack of isotropy now allows us to contract $B^{ab}$ with arbitrary tensors in ${\cal B}$ space. Of course, there is an infinity of  Lagrangians that, at quadratic order, reduce to \eq{LDyson}, and differ in the non-linear terms and, if we are eventually  interested just in   linearized elasticity (although derived from a fully Lorentz-invariant formalism), we can just pick up any convenient choice.\footnote{Actually, as discussed in \cite{Alberte:2018doe}, in principle the form of the effective Lagrangian can be fully reconstructed by just measuring the response to time-independent and homogeneous deformations.}
Given that the kinetic term in \eq{LDysoninhom} can be  obtained from \eq{Sd4xn}, a simple choice is
\be\label{Sd4xnandBabhom}
S=- \int_{\cal M} d^4x\, \[ \alpha\,  \sqrt{\det B} +\frac{1}{2} w_{abcd} B^{ab}B^{cd}\]
\, ,
\ee
where $\alpha$ is a constant   and 
$w_{abcd} $ is a  constant tensor in ${\cal B}$ space.   Since $B^{ab}B^{cd}$ is symmetric under $a\leftrightarrow b$, under $c\leftrightarrow d$, and under the simultaneous exchange $(a\leftrightarrow c, b\leftrightarrow d)$,  $w_{abcd}$ can be taken to satisfy
\be\label{wabcdsym}
w_{abcd}=w_{cdab}\,,\qquad
w_{abcd}=w_{bacd}=w_{abdc}\, ,
\ee 
since anyway the antisymmetric parts do not contribute to the contraction. This is similar to \eqs{csym1}{csym2}.
Note, however, that \eq{csym1} was a mathematical condition coming from the structure of the contractions, while \eq{csym2} was a physical assumption, which implies that we neglect the effect of vorticity. 
In contrast, all conditions in \eq{wabcdsym} are  mathematical conditions coming from the structure of the contractions.

We now expand the action (\ref{Sd4xnandBabhom}) around the background configuration $\bar{\xi}^a=\delta^a_i x^i$, writing $\xi^a(x)=\delta_i^a [x^i -u^i(x)]$ as in
\eq{xiaandui}. 
We  write,  as usual,  $B_{ij}=\delta_i^a \delta_j^b B_{ab}$ and 
\be\label{wijjkdeltaswabcd}
w_{ijkl}=\delta_i^a \delta_j^b\delta_k^c \delta_l^d\, w_{abcd}\, ,
\ee 
so that 
\be\label{wijklsym}
w_{ijkl}=w_{klij}\,,\qquad
w_{ijkl}=w_{jikl}=w_{ijlk}\, ,
\ee 
and we use  \eq{Bijui} for $B_{ij}$.
It is also useful to define
\be
w_{ij}\equiv w_{ijkk}= w_{kkij}\, ,
\ee
and
\be\label{deftracewij}
w\equiv w_{ii}\, .
\ee
Note that
\be\label{wijsym}
w_{ij}=w_{ji}\, .
\ee 
Then, we get
\be\label{SL0L1L2}
S= \int_{\cal M} d^4x\, \[ {\cal L}_0 +{\cal L}_1+{\cal L}_2 +{\cal O}(\pa u)^3\]\, ,
\ee
where 
\bees
\hspace*{-5mm}{\cal L}_0 &=&-\alpha -\frac{w}{2} \, ,\label{calL0}\\
\hspace*{-5mm}{\cal L}_1 &=& \pa_i \( \alpha u_i+2 w_{ij}u_j  \) \, ,\label{calL1}\\
\hspace*{-5mm}{\cal L}_2 &=& \dot{u}_i\dot{u}_j \( \frac{\alpha}{2}\delta_{ij}+ w_{ij} \)
-\frac{1}{2}  ( 4w_{ijkl}+ 2w_{jl}\delta_{ik}) \pa_i u_j\pa_k u_l
 -\frac{ \alpha}{2} \pa_i (u_i\pa_j u_j-u_j\pa_j u_i)
 \, .
 \label{L2wijkwij}
\ees
The fact that ${\cal L}_1$ is a total derivative, and therefore integrates to zero, shows that, when expanding around  $\bar{\xi}^a=\delta^a_i x^i$, there are no  terms linear in $u_i$ in the action, and therefore
$\bar{\xi}^a=\delta^a_i x^i$ is a solution of the equations of motion also for the anisotropic action (\ref{Sd4xnandBabhom}).  The last term in ${\cal L}_2$ is also a total derivative. We now define
\be\label{cijklwijkl}
C_{ijkl}=4w_{ijkl}+2 w_{jl}\delta_{ik}\, .
\ee
We observe that, because of \eqs{wijklsym}{wijsym}, $C_{ijkl}$ automatically obeys
$C_{ijkl}=C_{klij}$.
However, it is in general not symmetric under the separate exchange  $i\leftrightarrow j$, or  
$k\leftrightarrow l$.
In order to write the action in a form as close as possible to \eq{LDysonuij} we then proceed as follows.
We  first use \eq{defuutilde} to write
\be\label{Cpaupau}
C_{ijkl} \pa_i u_j\pa_k u_l=C_{ijkl}u_{ij}u_{kl}
+C_{ijkl}(u_{ij}\tilde{u}_{kl}+\tilde{u}_{ij}u_{kl})+C_{ijkl}\tilde{u}_{ij}\tilde{u}_{kl}\, .
\ee
In the first term, we  use the symmetries of $u_{ij}u_{kl}$ to rewrite it as
\bees
C_{ijkl}u_{ij}u_{kl}&=&
\frac{1}{4} \( C_{ijkl}+C_{jikl}+C_{ijlk}+C_{jilk}\) u_{ij}u_{kl}\nn\\
&=& \[ 4w_{ijkl}
+\frac{1}{2}\( w_{jl}\delta_{ik}+w_{il}\delta_{jk}+w_{jk}\delta_{il}+w_{ik}\delta_{jl}\)
\] u_{ij}u_{kl} \, .
\label{cijklvsw}
\ees
The second term is rewritten as 
\bees
C_{ijkl}(u_{ij}\tilde{u}_{kl}+\tilde{u}_{ij}u_{kl})&=&2 C_{ijkl} u_{ij}\tilde{u}_{kl}\nn\\
&=&4 w_{jl}\delta_{ik}u_{ij}\tilde{u}_{kl}\nn\\
&=& 2\( w_{jl}\delta_{ik}-w_{ik}\delta_{jl}\)u_{ij}\tilde{u}_{kl}
\, ,
\ees
where we used $w_{ijkl}\tilde{u}_{kl}=0$, since $w_{ijkl}$ is symmetric in $(k,l)$ 
while $\tilde{u}_{kl}$ is antisymmetric.
The last term in \eq{Cpaupau} is instead manipulated as follows:
\bees
C_{ijkl}\tilde{u}_{ij}\tilde{u}_{kl}&=&2w_{jl}\delta_{ik}\tilde{u}_{ij}\tilde{u}_{kl}\nn\\
&=&2\tilde{w}_{jl}\delta_{ik}\tilde{u}_{ij}\tilde{u}_{kl}
+\frac{2w}{3} \tilde{u}_{ij}\tilde{u}_{ij}\, ,
\ees
where we defined the traceless combination
\be\label{deftildew}
\tilde{w}_{ij}=w_{ij}-\frac{w}{3}\delta_{ij}\, .
\ee
We next  use the identity
\be\label{identityuu}
\tilde{u}_{ij}\tilde{u}_{ij}=u_{ij}u_{ij}-(\pa_i u_i)^2 +\pa_i (u_i\pa_ju_j-u_j\pa_j u_i)\, ,
\ee
which is easily verified writing all factors explicitly in terms of $\pa_i u_j$. We also use $\pa_i u_i=u_{ii}$, so $u_{ij}u_{ij}-(\pa_i u_i)^2=(\delta_{ik}\delta_{jl}-\delta_{ij}\delta_{kl})u_{ij}u_{kl}$ which can  be rewritten, in terms of a fully symmetric tensor, as
\be
 u_{ij}u_{ij}-(\pa_i u_i)^2=\frac{1}{2}\(\delta_{ik}\delta_{jl}
 + \delta_{jk}\delta_{il}
 -2\delta_{ij}\delta_{kl}\)u_{ij}u_{kl}\, .
\ee
Then
\be\label{Ctildetilde}
C_{ijkl}\tilde{u}_{ij}\tilde{u}_{kl}=\frac{w}{3}
(\delta_{ik}\delta_{jl}+\delta_{jk}\delta_{il}-2\delta_{ij}\delta_{kl})u_{ij}u_{kl}
+2\tilde{w}_{jl} \delta_{ik}\tilde{u}_{ij}\tilde{u}_{kl}
+\frac{2w}{3}\pa_i (u_i\pa_ju_j-u_j\pa_j u_i)\, .
\ee
Observe that the last term 
in \eq{Ctildetilde} is a total derivative, with the same structure as that appearing in \eq{L2wijkwij}.
Then, putting everything together, \eq{L2wijkwij} can be rewritten as
\bees
{\cal L}_2&=& \dot{u}_i\dot{u}_j \( \frac{\alpha}{2}\delta_{ij}+ w_{ij} \)
-\frac{1}{2}  c_{ijkl} u_{ij}u_{kl}
-d_{ijkl} u_{ij}\tilde{u}_{kl}
-\frac{1}{2}e_{ijkl}\tilde{u}_{ij}\tilde{u}_{kl}\nn\\
&&
-\frac{1}{2}\(\alpha+\frac{2w}{3}\)  \pa_i (u_i\pa_j u_j-u_j\pa_j u_i)
\, ,\label{L2wijkwij2}
\ees
where
\bees
\hspace*{-4mm}c_{ijkl}&=&4w_{ijkl}
+\frac{1}{2}\( \tilde{w}_{jl}\delta_{ik}+\tilde{w}_{il}\delta_{jk}+\tilde{w}_{jk}\delta_{il}+\tilde{w}_{ik}\delta_{jl}\)+\frac{2w}{3}
\(\delta_{ik}\delta_{jl}
 + \delta_{jk}\delta_{il} 
 -\delta_{ij}\delta_{kl}\) , \label{cijklw}\\
\hspace*{-4mm}d_{ijkl}&=& \tilde{w}_{jl}\delta_{ik}-\tilde{w}_{ik}\delta_{jl}\, ,\label{dijklw}\\
\hspace*{-4mm}e_{ijkl}&=&2\tilde{w}_{jl} \delta_{ik}\label{eijklw}\, .
\ees
The term $w_{ij} \dot{u}_i\dot{u}_j$, in \eq{L2wijkwij2} is smaller by a factor ${\cal O}(v_s^2/c^2)$ compared to 
$\alpha\delta_{ij} \dot{u}_i\dot{u}_j$, and we will neglect it.\footnote{Indeed, from \eq{cijklw}, we see that $w_{ijkl}$ is of the same order as $c_{ijkl}$. The latter, comparing to the isotropic limit 
(\ref{cijklLame}) and to \eqs{vperp}{vparallel}, is of order $\rho v_s^2$, where we denote generically by $v_s$ the order of magnitude of the speed of sound (either $v_{\perp}$ or  $v_{\parallel}$). 
The term $w_{ij}\dot{u}_i\dot{u}_j$
is therefore a correction of order $v_s^2$ (or, reinstating $c$, of order $v_s^2/c^2$) to the kinetic term
$(1/2)\alpha\dot{u}_i^2$ which, in the same approximation, is equal to $(1/2)\rho \dot{u}_i^2$.
Note, however, that in the general anisotropic case the term produces an anisotropic correction to the kinetic term.} Dropping also the total derivatives, the action can then be written as
\be\label{SdetBwexpanded}
S= \int_{\cal M} d^4x\, 
\[
-\rho
+\frac{1}{2}\rho \dot{u}^2_i
-\frac{1}{2}  c_{ijkl} u_{ij}u_{kl}
-d_{ijkl} u_{ij}\tilde{u}_{kl}
-\frac{1}{2}e_{ijkl}\tilde{u}_{ij}\tilde{u}_{kl}
+{\cal O}(\pa u)^3 
\]\, ,
\ee
where, to lowest order in $v_s^2/c^2$,  $\rho=\alpha$.
The term $-\rho$ is a zero-th order term  associated to the background rather than to the fluctuations, and produces a term $+\rho$ in the corresponding energy density (it is the rest energy density of the fluid, which automatically emerges in a relativistic formalism, just as the $mc^2$ rest mass for a relativistic particle). 
The tensor $c_{ijkl}$ has the symmetries (\ref{csym1}) and (\ref{csym2}) and reproduces 
the corresponding term in Dyson's Lagrangian density (\ref{LDyson}). However, for a generic choice of the tensor $w_{abcd}$ in the action (\ref{Sd4xnandBabhom}),  
there is also a contribution to the Lagrangian from the vorticity,  encoded in the  tensors $d_{ijkl}$
and $e_{ijkl}$.
This can be eventually traced to the fact that, as we see from \eq{Bijui}, $B_{ij}$ depends not only on $u_{ij}$ but also on $\pa_ku_i\pa_ku_j$, which can be written as a combination of $u_{ij}$ and $\tilde{u}_{ij}$. A dependence on $\tilde{u}_{ij}$ therefore generically appears in a relativistic formulation, constructed using $B_{ij}$; 
however, the dependence on vorticity  disappears if we choose $w_{abcd}$, and therefore $w_{ijkl}$, such that $w_{ij}$ is proportional to $\delta_{ij}$. Indeed, in this case, using $w=w_{ii}$, we have  
$w_{ij}=(w/3)\delta_{ij}$ and therefore $\tilde{w}_{ij}=0$, so
\eqs{dijklw}{eijklw} give $d_{ijkl}=0$ and $e_{ijkl}=0$.

In the isotropic limit $w_{abcd}$  takes the  general form
\be\label{wijklLamehom}
w_{abcd}=2\beta \delta_{ab} \delta_{cd}+\gamma 
( \delta_{ac} \delta_{bd}+ \delta_{ad} \delta_{bc})
\, ,
\ee
for some constants  $\beta$ and $\gamma$.  Inserting here \eq{wijklLamehom}  we get the corresponding action
\be\label{Swabcdisotropichom}
S=- \int_{\cal M} d^4x\, \[ \alpha\, \sqrt{\det B} +\beta  ({\rm Tr}\,  B)^2 
  +\gamma\,  {\rm Tr} (B^2) \]
\, .
\ee
The expansion to quadratic order can be obtained specializing to this action the general results of Section~\ref{sect:homanis}. In this case
\be
w_{ijkl}=2\beta \delta_{ij} \delta_{kl}
+\gamma ( \delta_{ik} \delta_{jl}+ \delta_{il} \delta_{jk})\, ,
\ee
and therefore
\be\label{wijisotrobg}
w_{ij}=(6\beta+2\gamma)\delta_{ij}\, ,
\ee 
so $w_{ij}$ is proportional to $\delta_{ij}$ and $\tilde{w}_{ij}=0$. Therefore all contributions associated to vorticity vanish,
\be
d_{ijkl}=e_{ijkl}=0\, ,
\ee 
while  \eq{cijklw} gives
\be\label{cijklmulambda2}
c_{ijkl}=\lambda\delta_{ij} \delta_{kl}
+\mu
(\delta_{ik} \delta_{jl}+\delta_{il} \delta_{jk})\, ,
\ee
with
\be\label{lambdamubetagamma}
\lambda=-4\beta-4\gamma\, ,\qquad
\mu =12\beta+8\gamma\, .
\ee
We have therefore recovered the form  (\ref{cijklLame}) for $c_{ijkl}$, and we have found the explicit expression for 
the Lam\'e coefficients $\lambda$ and $\mu$ in terms of the parameters $\beta$ and $\gamma$ that enter  the  specific  relativistic action (\ref{Swabcdisotropichom}).
We can also obtain the expression for the density, including the corrections to the lowest-order result $\rho=\alpha$, from the kinetic term in \eq{L2wijkwij2}, which in the isotropic case becomes
\be
\dot{u}_i\dot{u}_j \( \frac{\alpha}{2}\delta_{ij}+ w_{ij} \)=
\frac{1}{2} (\alpha+12\beta+4\gamma) \dot{u}_i\dot{u}_i\, ,
\ee
showing that 
\be\label{rhobetagamma}
\rho=\alpha+12\beta+4\gamma\, .
\ee
We can double-check these results comparing \eq{Swabcdisotropichom} with \eqs{SFnm}{defnpq}. We see that  the action  (\ref{Swabcdisotropichom}) corresponds to the choice
\be\label{Fnpqiso}
F(n,p,q)=\alpha n +9\beta p^2+3\gamma q\, ,
\ee
and, expanding to quadratic order,  this gives a Lagrangian density of the form (\ref{LDysonLameagain}) with $\rho$, $\lambda$ and $\mu$ given by \eqst{rhopaF}{mufromF}. Inserting \eq{Fnpqiso}
into \eqst{rhopaF}{mufromF} we get again \eqs{lambdamubetagamma}{rhobetagamma}.

From \eqs{vperp}{vparallel}, we then also have
\be
v^2_{\perp}=\frac{12\beta+8\gamma}{\alpha+12\beta+4\gamma}\, ,\qquad
v^2_{\parallel}=\frac{20\beta+12\gamma}{\alpha+12\beta+4\gamma}\, .
\ee
Inverting these relations, and denoting generically the order of $v_{\perp}$ and $v_{\parallel}$ by $v_s$, we have
\be\label{alpharho}
\alpha =\rho \[1+{\cal O}(v_s^2)\]\, ,\qquad \beta,\gamma ={\cal O}(\rho v_s^2)\, .
\ee
Therefore, apart from corrections proportional to the square of the speed of sound (or, restoring $c$, to $v_s^2/c^2$), we see again that $\alpha$ is the same as the density of the body, while $\beta$ and $\gamma$ are combinations of the Lam\'e coefficients. 

As already mentioned, 
there are of course an infinity of fully non-linear relativistic Lagrangians constructed with $B_{ab}$ that, after expanding over the background solution $\bar{\xi}^a$, reduce to the  Lagrangian (\ref{LDyson}) at quadratic order, but differ in the non-linear terms. Since eventually we are only interested in the linearized theory (however, derived from a  fully Lorentz-invariant action, so written in terms of $B_{ab}$ before expanding around a symmetry-breaking background), we can just use any simple choice. In the following we will use the action (\ref{Sd4xnandBabhom}), or its isotropic limit (\ref{Swabcdisotropichom}). We will next generalize them to the non-homogeneous case.\footnote{An alternative could be to factorize $\sqrt{\det B}$ and use an action of the form
\be\label{Swtilde}
S=- \int_{\cal M} d^4x\, \sqrt{\det B}  \[ \alpha\,  +\frac{1}{2} \tilde{w}_{abcd} B^{ab}B^{cd}\]\, .
\ee
Such a choice would be closer in spirit to the formalism in refs.~\cite{Magli:1992,Beig:2002pk,Hudelist:2022ixo}. At the quadratic level in the perturbations the actions (\ref{Sd4xnandBabhom}) and (\ref{Swtilde}) 
are equivalent but, when in Section~\ref{sect:notisoandhom} we will generalize to inhomogeneous bodies, promoting $\alpha$ and $w_{abcd}$ to functions, we will find that \eq{Sd4xnandBabhom} is slightly more convenient, because of a nice cancellation of a term involving the spatial derivatives of $\alpha$.}

\subsection{Inhomogeneous  bodies}\label{sect:notisoandhom}

We  now generalize to an elastic body which is not  homogeneous. This will be essential to the study of the coupling of elastic media to GWs because, as already found in~\cite{Dyson:1969zgf,Hudelist:2022ixo}, and as we will derive again below with our formalism, in the TT gauge the coupling to GWs vanishes for a homogeneous body of infinite extent, and the leading contribution comes from internal inhomogeneities, or from the discontinuity associated to the boundary of a body with finite extent. However, we will see that this introduces some extra technical complication, since the configuration $\bar{\xi}^a=\delta^a_i x^i$ will no longer be a solution of the equations of motion.

We still assume that the body formally has an infinity extent, which is necessary to introduce the mapping (\ref{mapXxi}). However, we now assume that  it  is characterized by a generic density $\rho(\vx)$; the case of a body of finite extent can then be recovered taking a sequence of functions that vanish faster and faster outside the body, approaching  a function $\rho(\vx)$ with  compact support. Similarly, the quantities that characterize the elastic properties of the body, such as  the tensor $c_{ijkl}$ in \eq{LDysoninhom}, can be taken to have the same compact support.

\subsubsection{Effective action}

As a first attempt, it might seem that a natural generalization of \eq{Sd4xnandBabhom} to the non-homogeneous case could be given by
\be\label{Sd4xnandBab}
S=- \int_{\cal M} d^4x\, \[ \alpha(\vx) \, \sqrt{\det B} +\frac{1}{2} w_{abcd}(\vx) B^{ab}B^{cd}\]
\, ,
\ee
where $\alpha(\vx)$ is a function  of $\vx$ and 
$w_{abcd}(\vx) $  a  tensor in ${\cal B}$ space, again function of $\vx$,  and  with the symmetries 
(\ref{wabcdsym}).

However, if we want to construct a Lorentz-invariant theory,  \eq{Sd4xnandBab} is not viable because
the explicit dependence of $\alpha$ and $ w_{abcd}$ on $\vx$ breaks Lorentz invariance. A Lorentz-invariant  (and in fact also Poincar\'e-invariant) action  is rather obtained writing them as functions of $\xi$, \be\label{Sd4xnandBabxi}
S[\xi,\pa\xi]= -\int_{\cal M} d^4x\, \[ \alpha(\xi) \, \sqrt{\det B} +\frac{1}{2} w_{abcd}(\xi) B^{ab}B^{cd}\]
\, ,
\ee
where $\alpha$ is taken to be a Lorentz scalar function  of the Lorentz scalar fields $\xi^a$, and similarly for $w_{abcd}$. This action is therefore  invariant under Poincar\'e transformations in ${\cal M}$. Note that the action now depends  both on $\xi$, through $\alpha$ and $ w_{abcd}$, and on its derivatives, through $B_{ab}$; indeed, for an inhomogeneous body, translations in $\vx$ space are no longer a symmetry, so the argument above \eq{defBab}, that led to the conclusion  that the action can depend on $\xi$ only through $\pam\xi^a$, is no longer valid.
\Eq{Sd4xnandBabxi} provides a Lorentz and Poincar\'e invariant theory at the full non-linear level, for an elastic body which is not homogeneous nor isotropic.

%
Similarly to \eq{wijklLamehom}, in the isotropic limit, $w_{abcd}(\xi)$  takes the  general form
\be\label{wijklLame}
w_{abcd}(\xi)=2\beta(\xi) \delta_{ab} \delta_{cd}+\gamma(\xi) ( \delta_{ac} \delta_{bd}+ \delta_{ad} \delta_{bc})\, ,
\ee
for some functions $\beta(\xi)$ and $\gamma(\xi)$.
Inserting \eq{wijklLame} into \eq{Sd4xnandBabxi},  in this limit we get
\be\label{Swabcdisotropic}
S=- \int_{\cal M} d^4x\, \[ \alpha(\xi) \, \sqrt{\det B} +\beta(\xi)  ({\rm Tr}\,  B)^2 
  +\gamma(\xi)\,  {\rm Tr} (B^2) \]
\, ,
\ee
which is the generalization of \eq{Swabcdisotropichom}.

\subsubsection{Equations of motion and background solutions}

We now write down the equations of motion and look for  background solutions. As we will see, \eq{xiaequilibrium} is no longer a solution of the equations of motion derived from the action (\ref{Sd4xnandBabxi}), not even in the isotropic limit (\ref{Swabcdisotropic}). Therefore, the first step is to find the correct background solution, around which we will then expand the action to quadratic order, to obtain the generalization of \eqst{LDysonLameagain}{mufromF} to inhomogeneous bodies.\footnote{An alternative route could be to promote the metric in ${\cal B}$ from $\delta_{ab}$ to a generic metric $\gamma_{ab}(\xi)$, and construct the theory so that it is also invariant  under diffeomorphisms in ${\cal B}$. Then, one could try to set $\xi^a(x)=\delta_i^a x^i$ as a choice of coordinates in  ${\cal B}$, at the price of dealing with a non-trivial metric $\gamma_{ab}(\xi)$. Note that each different physical setting (e.g., slowly varying inhomogeneities, one-dimensional inhomogeneities, etc.) will in general give rise to a different $\gamma_{ab}(\xi)$. If we do not want to have a different action for each situation, we must treat also $\gamma_{ab}(\xi)$ as a dynamical field, adding its dynamics through the action of three-dimensional Euclidean gravity. This alternative approach could give useful insights, but we will not pursue it here.\label{foot:diffB}}
 We restrict to the isotropic case, so we use the action (\ref{Swabcdisotropic}), and the Lagrangian is
\be\label{Lalphabetagamma}
{\cal L}= -\[ \alpha(\xi) \, \sqrt{\det B} +\beta(\xi)  ({\rm Tr}\,  B)^2 
  +\gamma(\xi)\,  {\rm Tr} (B^2)\]\, .
\ee
We stress that \eq{Lalphabetagamma} is just an example of a Lagrangian that describes, in a Lorentz--covariant manner, the  elasticity theory of  an inhomogeneous body. As we saw already in the homogeneous case, see  \eqst{SFnm}{mufromF}, an infinity of other possibilities, differing at the level of non-linear terms, can be written down. We will use the Lagrangian (\ref{Lalphabetagamma}) as an  example of the results that are obtained from an explicit relativistic Lagrangian, but we will then see how these explicit results allow us to understand the general structures that can be obtained, and how,  when it comes to determining the energy--momentum tensor of the theory and then the coupling to GWs, these generalize the ``naive" (i.e. non-covariant) approach based on \eq{LDysoninhom}.
The Euler-Lagrange equation is 
\be\label{eqmotionpam}
\frac{\d{\cal L}}{\delta\xi^a}-\pam \frac{\d{\cal L}}{\delta(\pam\xi^a)}=0\, .
\ee
We  look for static background solutions, so the time derivative $\pam$ with $\mu=0$ in \eq{eqmotionpam} vanishes, and the Euler-Lagrange equation reduces to
\be\label{eqmotionpaj}
\frac{\d{\cal L}}{\delta\xi^a}-\pa_j \frac{\d{\cal L}}{\delta(\pa_j\xi^a)}=0\, ,
\ee
where $\pa_j=\pa/\pa x_j$. As we show  in App.~\ref{app:contoeqm}, inserting the Lagrangian (\ref{Lalphabetagamma}) into \eq{eqmotionpaj} we get\footnote{Recall that the indices $a,b$ are raised and lowered with $\delta_{ab}$, so we can write them indifferently as lower or upper indices.} 
\bees
&&\alpha(\xi) \pa_j\[\sqrt{\det B}\, (B^{-1})_{ab}\pa_j\xi^b \] 
+4\beta(\xi)\pa_j\(  {\rm Tr}\,  B\, \pa_j\xi^a\) 
+4\gamma(\xi)\pa_j\(  B_{ab}\pa_j\xi^b\) \nn\\
&&=\frac{\pa\beta}{\pa\xi^b}   {\rm Tr}\,  B \[ \delta_ {ab}{\rm Tr}\,  B-4B_{ab}\]
+\frac{\pa\gamma}{\pa\xi^b}    \[\delta_{ab}{\rm Tr} (B^2) - 4 B_{ac}B_{cb}\]\, .
\label{eqmfinal}
\ees
As a check we can verify that, when $\alpha$, $\beta$ and $\gamma$ are constants, \eq{xiaequilibrium} is a solution. Indeed, in this case the terms on the right-hand side of \eq{eqmfinal} vanish. On the left-hand side we make use of the fact that, when $\xi^a(x)=\delta_i^a x^i$, we have  $\pa_j\xi^a=\delta_i^a$, and then
$B_{ab}=\delta_{ab}$; therefore,  on the left-hand side, all expressions on which $\pa_j$ acts are constants,  so we get again zero.\footnote{More generally, if $\xi^a(x)=\delta_i^a \lambda^a x^i$ (no sum over $a$), we get
$B_{ab}=\lambda_a\lambda_b\delta_{ab}$, which is still a constant,  and again we get a solution of \eq{eqmfinal}, compare with \eq{xiequilexplicitlambda}.\label{foot:lambdaalambdab}}
Note that, in fact, \eq{xiaequilibrium} is a solution of the equations of motion of the Lagrangian (\ref{Lalphabetagamma})  even for a generic function $\alpha(\xi)$, as long as $\beta$ and $\gamma$ are constants. This is due to the fact that the terms proportional to $\pa\alpha/\pa\xi^a$ do not appear in \eq{eqmfinal}, because of a cancelation that we discuss explicitly in App.~\ref{app:contoeqm}.

When  $\beta$ and $\gamma$ depend on $\xi$, however, $\xi^a(x)=\delta_i^a x^i$ is no longer a solution, and we must find a new background $\bar{\xi}^a$ around which will then define the fluctuations. In its full form, \eq{eqmfinal} is a highly non-linear differential equation for the field $\xi^a(\vx)$, because $B_{ab}$ depends  on $\pa_i\xi^a$ [see 
\eqst{detB}{TrB2}], and therefore an even more complicated and fully non-linear dependence on $\pa_i\xi^a$ is carried by $ (B^{-1})_{ab}$. 
However, there are two interesting limiting cases where
\eq{eqmfinal} becomes analytically treatable. The first is when the body is taken to be homogeneous and of infinite extent in the $(x,y)$ directions, and the only spatial dependence is through the $z$ variable.  The second case corresponds to a full three dimensional setting where, however, the inhomogeneities are taken to be small perturbations over a homogeneous background. We analyze these two cases in turn.

\vspace{1mm}
\noindent{\bf One-dimensional inhomogeneities}. We first consider a three-dimensional body which is   homogeneous and of infinite extent in the $(x,y)$ directions, and inhomogeneous with respect to the $z$ variable.  In this case,  generalizing \eq{xiequilexplicit}, we  make the ansatz
\be
\xi^1(x,y,z)=x\, ,\qquad
\xi^2(x,y,z)=y\, ,\qquad
\xi^3(x,y,z)=\xi^3(z)\, ,
\ee
for some function $\xi^3(z)$ to be determined. In this case 
$B={\rm diag}(1,1, B_{33}(z) )$, where 
\be\label{B33dxidz}
B_{33}(z)= \(\frac{d\xi^3}{dz}\)^2\, ,
\ee
so $\det B=B_{33}$, ${\rm Tr}\,  B=2+B_{33}$, ${\rm Tr}\,  B^2=2+B_{33}^2$
and $B^{-1}={\rm diag}(1,1, B^{-1}_{33}(z) )$.
In \eq{eqmfinal}  we take $\beta(\xi)=\beta(\xi^3)$ and $\gamma(\xi)=\gamma(\xi^3)$, so only the $a=3$ component  of the equation is non-trivial. Furthermore, the term proportional to $\alpha(\xi)$ vanishes, since
\bees
\pa_j\[ \sqrt{\det B}\, (B^{-1})_{3b}\pa_j\xi^b \]&=&\frac{d}{dz} \[ \sqrt{\det B}\, (B^{-1})_{33}\frac{d\xi^3}{dz}\] \nn\\
&=&\frac{d}{dz} \[ B^{1/2}_{33}\, B^{-1}_{33}\,  B^{1/2}_{33}\] =0\, .
\ees
Since $\xi^3=\xi^3(z)$, in the terms  $d\alpha/d\xi^3$ and $d\beta/d\xi^3$ the derivatives with respect to $\xi^3$ can be expressed
as derivatives with respect to $z$ using
\be
\frac{d}{dz}=\frac{d\xi^3}{dz}\frac{d}{d\xi^3}= B_{33}^{1/2}\frac{d}{d\xi^3}\, ,
\ee
so 
\be
\frac{d}{d\xi^3}=B_{33}^{-1/2}\frac{d}{dz}\, .
\ee
Then \eq{eqmfinal} gives
\be\label{dB33dz}
\frac{dB_{33}}{dz}=\frac{(2+B_{33})(2-3B_{33})\beta'+(2-3B_{33}^2)\gamma'}{\beta (4+6B_{33})+6\gamma B_{33}}\, ,
\ee
where $\beta'=d\beta/dz$, $\gamma'=d\gamma/dz$. In principle, given the functions $\beta(z)$ and $\gamma(z)$, 
\eq{dB33dz} allows us to determine $B_{33}(z)$ and therefore $\xi^{3}(z)$.

\vspace{2mm}
\noindent{\bf Small perturbations over a homogeneous body.}
Another situation that is both interesting and amenable to analytic treatment is when  $\alpha(\xi)$, 
$\beta(\xi)$ and $\gamma(\xi)$ describe small and slowly-varying perturbations of a homogeneous background. In this case  we write  
\be\label{alpha0plus1}
\alpha(\xi)=\alpha_0+\alpha_1(\xi)\, ,
\qquad
\beta(\xi)=\beta_0+\beta_1(\xi)\, ,
\qquad
\gamma(\xi)=\gamma_0+\gamma_1(\xi)\, ,
\ee
where $\alpha_0$, $\beta_0$ and $\gamma_0$ are constants, while $ \alpha_1(\xi)$, $\beta_1(\xi)$ and 
$\gamma_1(\xi)$ are small perturbations. 
Correspondingly, we look for a solution that is a small perturbation of the solution
(\ref{xiaequilibrium}) valid for a homogeneous body, i.e. we look for a solution of the form 
$\xi^a(x)=\delta_i^a \xi_i(\vx)$, where
\be\label{xiofxlin}
\xi_i(\vx)= x_i + \eps_i(\vx)\, .
\ee
Therefore $\pa_j\xi_i=\delta_{ij}+\pa_j\eps_i$ and the assumption that $\alpha(\xi)$,  $\beta(\xi)$ and $\gamma(\xi)$ are slowly-varying functions of $\xi$ implies that $|\pa_j\eps_i|\ll 1$. Then, the linearization of \eq{eqmfinal} is straightforward (see  App.~\ref{app:contoeqm} for details) and gives
\be\label{eqepslin0}
(8\beta_0+4\gamma_0)\pa_i\pa_j\eps_j+
(12\beta_0+8\gamma_0)\n^2\eps_i=-\frac{\pa}{\pa\xi_i} \[3\beta_1(\xi)+\gamma_1(\xi)\]\, .
\ee
Since the right-hand side is proportional to $\beta_1$ and $\gamma_1$ and is therefore already of first-order, there we can replace $\xi_i$ with $x_i$ and $\pa/\pa\xi_i$ with $\pa/\pa x_i\equiv \pa_i$, so we get 
\be\label{eqepslin1}
(8\beta_0+4\gamma_0)\pa_i\pa_j\eps_j+
(12\beta_0+8\gamma_0)\n^2\eps_i=-\pa_i \[3\beta_1(\vx)+\gamma_1(\vx)\]\, .
\ee
In general, an equation of this form can be solved separating the vector field $\eps_i(\vx)$ into its transverse and longitudinal parts, 
\be\label{espespTpapsi}
\eps_i(\vx)=\eps^T_i(\vx)+\pa_i\psi(\vx)\, ,
\ee
where $\pa_i \eps^T_i=0$. However,  to linear order the right-hand side of \eq{eqepslin1} is proportional to $\pa_i$, i.e. is purely longitudinal. Then, the solution of \eq{eqepslin1} is given by $\eps^T_i(\vx)=0$, while $\psi$ satisfies
\be\label{n2psiexplicit}
4(5\beta_0+3\gamma_0)\n^2\psi=-\[3\beta_1(\vx)+\gamma_1(\vx)\]\, .
\ee
Recalling that $\beta_0$ and $\gamma_0$ are constants, and using the Green's function of the Laplacian,
\be
G(\vx-\vx')=-\frac{1}{4\pi}\, \frac{1}{|\vx-\vx'|}\, ,
\ee
the solution for $\psi(\vx)$ is
\be\label{solpsi}
\psi(\vx)=\frac{1}{16\pi}\, \frac{1}{5\beta_0+3\gamma_0}\, 
\int d^3x'\, \frac{3\beta_1(\vx')+\gamma_1(\vx')}{|\vx-\vx'|}\, ,
\ee
where we assumed that the functions $\beta_1(\vx)$ and $\gamma_1(\vx)$ go to zero at infinity sufficiently fast, so  that the integral converges.

In conclusion, we have found a static background solution of the equations of motion of the Lagrangian (\ref{Lalphabetagamma}), in the limit where the functions $\beta$ and $\gamma$ are slowly-varying,  of the form
\be\label{xixpapsi}
\bar{\xi}^a(\vx)= \delta^a_i(x^i + \pa^i\psi)\, ,
\ee
with $\psi(\vx)$ given explicitly by \eq{solpsi}.

\subsubsection{Expansion to quadratic order}

To understand the physical content of the  inhomogeneous theory we now expand the action to quadratic order. We therefore write 
\be
\xi^a(x)=\bar{\xi}^a(x)+\pi^a(x)\, ,
\ee 
where $\bar{\xi}^a(x)$ is a generic  background field configuration and $\pi^a(x)$ is a  perturbation around it.  For generality, we work at first with an arbitrary background solution (possibly even time-dependent), rather than using from the beginning the explicit form  (\ref{xixpapsi}).
For the perturbation, we use at first the Goldstone fields $\pi^a(x)$, since some properties of the action are better understood in the language of Goldstone boson physics, but one can immediately rewrite the results  in terms of $u^i(x)$ using \eq{piaui}. We limit ourselves to isotropic (but inhomogeneous) media, using the action (\ref{Swabcdisotropic}).

The expansion of the action (\ref{Swabcdisotropic}) up to quadratic order in the perturbations $\pi^a(x)$ gives (after some integration by parts in the linear term)
\be\label{Spiainhom}
S=\int_{\cal M}d^4x\,\( \bar{\cal L}+\bar{\mathcal{E}}_a\pi^a
-\frac{1}{2}\mathcal{\bar{M}}_{ab}\pi^a\pi^b
+\mathcal{\bar{G}}^\mu_{ab}\pi^a \partial_\mu \pi^b
+\mathcal{\bar{H}}^{\mu\nu}_{ab}\partial_\mu\pi^a\partial_\nu\pi^b
 \)\,,
\ee
where we introduced the background quantities
\bees
\bar{\cal L}&=& -\[ \alpha(\bar{\xi})  (\det \bar{B})^{1/2} +\beta(\bar{\xi})  ({\rm Tr}\,  \bar{B})^2 +\gamma(\bar{\xi})\,  {\rm Tr} (\bar{B}^2)\]\,,\\
\bar{\mathcal{E}}_a&=&\alpha(\bar{\xi}) \pa_\mu\[ (\det \bar{B})^{1/2} (\bar{B}^{-1})_{ab}\pa^\mu\bar{\xi}^b \]
+4\beta(\bar{\xi})\pa_\mu\(  {\rm Tr}\,  \bar{B}\, \pa^\mu\bar{\xi}^a\) 
+4\gamma(\bar{\xi})\pa_\mu\(  \bar{B}_{ab}\pa^\mu\bar{\xi}^b\)\,\nonumber\\
&+&\(\frac{\pa\beta}{\pa\xi^b}\)_{ \bar{\xi} } {\rm Tr}\,  \bar{B} \[4\bar{B}_{ab} - \delta_ {ab}{\rm Tr}\,  \bar{B}\]
+\(\frac{\pa\gamma}{\pa\xi^b}\)_{\bar{\xi}} \[4 \bar{B}_{ac}\bar{B}_{cb} - \delta_{ab}{\rm Tr} (\bar{B}^2)\]\,,\label{eqcalE}
\ees
\bees
\hspace*{-5mm}\mathcal{\bar{M}}_{ab}&=& 
\(\frac{\partial^2\alpha}{\pa\xi^a\pa\xi^b}\)_{\bar{\xi}}  (\det \bar{B})^{1/2} +\( \frac{\pa^2\beta}{\pa\xi^a\pa\xi^b}\)_{\bar{\xi}}  ({\rm Tr}\,  \bar{B})^2 
+\(\frac{\pa^2\gamma}{\pa\xi^a\pa\xi^b}\)_{\bar{\xi}}\, {\rm Tr} (\bar{B}^2)\,,\label{calMpia}\\
\hspace*{-5mm}\mathcal{\bar{G}}^\mu_{ab}&=&-\partial^\mu\bar{\xi}^c\[
(\det \bar{B})^{1/2}\(\frac{\pa\alpha}{\pa\xi^a}\)_{\bar{\xi}}\(\bar{B}^{-1}\)_{bc}
+4\(\frac{\pa\beta}{\pa\xi^a}\)_{\bar{\xi}}\d_{bc}{\rm Tr}\, \bar{B}
+4\(\frac{\pa\gamma}{\pa\xi^a}\)_{\bar{\xi}}\bar{B}_{bc}\],\\
\hspace*{-5mm}\mathcal{\bar{H}}^{\mu\nu}_{ab}&=&
\frac12\alpha(\bar{\xi})\bar{C}_{abcd}\pa^{\mu}\bar{\xi}^c\,\pa^{\nu}\bar{\xi}^d
-\frac12\alpha(\bar{\xi})(\det \bar{B})^{1/2}\, \eta^{\mu \nu}\left(\bar{B}^{-1}\right)_{ab}\\
&&-\beta(\bar{\xi}) \[ 4\pa^{\mu}\bar{\xi}_a\,\pa^{\nu}\bar{\xi}_b+2\,\eta^{\mu\nu}\d_{ab}{\rm Tr}\, \bar{B}\]-2\gamma(\bar{\xi})\[ \pa^{\mu}\bar{\xi}_b\,\pa^{\nu}\bar{\xi}_a
+\d_{ab}\, \pa^{\mu}\bar{\xi}_c\,\pa^{\nu}\bar{\xi}^c+\eta^{\mu\nu}\bar{B}_{ab}\] \, ,\nn
\ees
and we defined
\bees
\bar{B}_{ab}&=&\partial_\mu\bar{\xi}_a\partial^\mu\bar{\xi}_b\, ,\\
\bar{C}_{abcd}&=&(\det \bar{B})^{1/2}\,  \[\(\bar{B}^{-1}\)_{ab}\(\bar{B}^{-1}\)_{cd}-\(\bar{B}^{-1}\)_{ac}\(\bar{B}^{-1}\)_{bd}+\(\bar{B}^{-1}\)_{ad}\(\bar{B}^{-1}\)_{bc}\] \, .
\ees
By definition, the  vanishing of the linear term gives the equation of motion, which therefore is
\be\label{eqmotionE}
\bar{\mathcal{E}}_a=0\, .
\ee
As a check, when looking for a static  solution $\bar{\xi}^a(\vx)$,   
\eqs{eqcalE}{eqmotionE}  are the same as \eq{eqmfinal}.

We also see from \eqs{Spiainhom}{calMpia} that, when $\alpha$, $\beta$ and $\gamma$ depend on $\xi$, a  non-vanishing mass term ${\cal \bar{M}}_{ab}$ is generated for the fields $\pi^a(x)$. In particular, the mass matrix ${\cal \bar{M}}_{ab}$ depends on the second derivatives of $\alpha(\xi)$, $\beta(\xi)$ and $\gamma(\xi)$.
In the field-theoretical language, the $\pi^a(x)$ are now pseudo-Goldstone bosons, because the translation symmetry is broken by the fact that $\alpha$, $\beta$ or $\gamma$ depend on $\xi$, and therefore become massive. The translation symmetry in field space $\pi^a(x)\ra \pi^a(x)+ C^a$, with $C^a$ a constant, is broken both by the mass term and by the term $\mathcal{\bar{G}}^\mu_{ab}\pi^a \partial_\mu \pi^b$.
Observe also that $\mathcal{\bar{M}}_{ab}$, $\mathcal{\bar{G}}^\mu_{ab}$ and $\mathcal{\bar{H}}^{\mu\nu}_{ab}$ are functions of $x$, through their dependence on $\bar{\xi}^a(x)$.

We can now specialize these general results to the background solution (\ref{xixpapsi}).
For slowly-varying functions we write  $\alpha(\xi)$, $\beta(\xi)$ and $\gamma(\xi)$ as in
\eq{alpha0plus1} 
Inserting now the explicit static background solution (\ref{xixpapsi}), determined by $\beta_1$ and 
$\gamma_1$ as in eq.~(\ref{solpsi}) and, working at first order in the small perturbations $\pa_i\psi$, $ \alpha_1$, $\beta_1$ and 
$\gamma_1$, we get
\bees
\bar{B}_{ab}&=& \delta_{ab} + 2 \partial_a \partial_b \psi \, ,\\
\bar{C}_{abcd}&=& (1+\n^2 \psi) [\delta_{ab}\delta_{cd}-\delta_{ac}\delta_{bd}+\delta_{ad}\delta_{bc}]\\
&&
-2\partial_e \partial_f \psi [\delta_{ab}\delta^e_c\delta^f_d 
+\delta_{cd}\delta^e_a\delta^f_b-\delta_{ac}\delta^e_b\delta^f_d-\delta_{bd}\delta^e_a\delta^f_c+\delta_{ad}\delta^e_b\delta^f_c+\delta_{bc}\delta^e_a\delta^f_d] \,,\nn
\ees
and then
\be
\mathcal{\bar{M}}_{ab}=  \frac{\partial^2}{\pa x^a\pa x^b}\[\alpha_1(\vx)+9\beta_1(\vx)+3\gamma_1(\vx)\] \, ,
\ee
\be
\mathcal{\bar{G}}^\mu_{ab} = -\delta^\mu_b \frac{\pa}{\pa x^a}\[\alpha_1(\vx)+12\beta_1(\vx)+4\gamma_1(\vx)\] \, ,
\ee
and
\bees
\mathcal{\bar{H}}^{\mu\nu}_{ab}&=&\d^\mu_0\d^\nu_0\left\{\[\frac12\alpha(\vx)-2\gamma(\vx)+\frac12\alpha_0\n^2\psi(\vx)\]\d_{ab}-\alpha_0\pa_a\pa_b\psi(\vx)\right\}\,\nonumber\\
&&-\[\frac12\alpha(\vx) +\frac12\alpha_0\n^2\psi(\vx)+4\beta(\vx)\]\d^\mu_a \d^\nu_b+\[\frac12\alpha(\vx) +\frac12\alpha_0\n^2\psi(\vx) -2\gamma(\vx)\]\d^\mu_b \d^\nu_a\,\nonumber\\
&&+\[\frac12\alpha_0-4\beta_0\]\[\d^\mu_a \pa^\nu \pa_b\psi(\vx)+\d^\nu_b \pa^\mu\pa_a\psi(\vx)\]\,\nonumber\\
&&-\[\frac12\alpha_0+2\gamma_0\]\[\d^\nu_a \pa^\mu \pa_b\psi(\vx)+\d^\mu_b \pa^\nu\pa_a\psi(\vx)\]\,\nonumber\\
&&-2\eta^{\mu\nu}\{\d_{ab}\[3\beta(\vx)+2\gamma(\vx)+2\beta_0\n^2\psi(\vx)\]+2\gamma_0\pa_a\pa_b\psi(\vx)\}\,\nonumber\\
&&-4\gamma_0\d_{ab}\pa^\mu\pa^\nu \psi(\vx)\,,
\label{Hmunuablin}
\ees
In  eq.~(\ref{Hmunuablin})  it is meant that 
\be \label{alphabetagamma}
\alpha(\vx)=\alpha_0+\alpha_1(\vx)\, ,\qquad
\beta(\vx)=\beta_0+\beta_1(\vx)\, ,\qquad
\gamma(\vx)=\gamma_0+\gamma_1(\vx)\, , 
\ee
since, to this order, we could replace $\xi$ by $\vx$ in their argument,
and $\n^2\psi$ is given explicitly, in terms of these quantities, in \eq{n2psiexplicit}.

Writing $\pi^a=-\delta^a_i u^i$, performing some integration by parts and using \eq{n2psiexplicit} to simplify the structure of a term, it is possible to write the quadratic part of the Lagrangian as
\bees
\label{Lagrangianquadraticinhom}
{\cal L}_2&=&\frac{1}{2}  \[  \rho(\vx)\dot{u}_i\dot{u}_i -\lambda(\vx) (\pa_i u_i)^2-2\mu(\vx)\, u_{ij}u_{ij}\]+\(4\gamma_0-\alpha_0\)\dot{u}_i\dot{u}_j\pa_i\pa_j\psi(\vx)+\,\nonumber\\
&+&\(\alpha_0+8\beta_0\)(\pa_i u_i) u_j \pa_j\n^2\psi(\vx)
+\frac12 (\alpha_0 - 12\beta_0 - 12\gamma_0) u_i u_j \pa_i\pa_j\n^2\psi(\vx)\,\\
&+&\pa_j\pa_k\psi(\vx)\left[\(\alpha_0-8\beta_0\) (\pa_i u_i)\pa_j u_k-\(\alpha_0+4\gamma_0\)\pa_i u_j\pa_k u_i-4\gamma_0\pa_i u_j\pa_i u_k-4\gamma_0\pa_j u_i\pa_k u_i\right]\,,\nn
\ees
with
\bees
\label{rhoinhom}
\rho(\vx)&=& \alpha(\vx)+12\beta(\vx)+4\gamma(\vx)+\(\alpha_0+8\beta_0\)\n^2\psi(\vx) \, ,\\
\label{lambdainhom}
\lambda(\vx)&=&-4\beta(\vx)-4\gamma(\vx)-8\beta_0\n^2\psi(\vx)\, ,\\
\label{muinhom}
\mu(\vx)&=&12\beta(\vx)+8\gamma(\vx)+8\beta_0\n^2\psi(\vx)\, .
\ees
Observe that, in $\lambda(\vx)$ and $\mu(\vx)$, the dependence on $\alpha(\vx)$ actually canceled.
The equations of motion derived from the Lagrangian (\ref{Lagrangianquadraticinhom}) are
\bees
&\rho(\vx)\ddot{u}_i& + 2(4\gamma_0-\alpha_0) \ddot{u}_j \partial_j \partial_i \psi(\vx) = \partial_i\Bigr[\lambda(\vx)\partial_j u_j\Bigr] + \partial_j\Bigr[\mu(\vx)(\partial_i u_j+\partial_j u_i)\Bigr] \nn\\
&&+8\beta_0\partial_i\Bigr[\partial_j\partial_k\psi(\vx)\partial_j u_k+\partial_k u_k\n^2\psi(\vx)\Bigr]+8\beta_0\partial_j\Bigr[\partial_i\partial_j\psi(\vx)\partial_k u_k-\partial_iu_j\n^2\psi(\vx)\Bigr]\nn\\
&&+4\gamma_0\partial_j\Bigr[\partial_i\partial_k\psi(\vx)(\partial_k u_j+2\partial_j u_k)+\partial_j\partial_k\psi(\vx)(\partial_i u_k +2\partial_k u_i)\Bigr]\nn\\
&&
-(20\beta_0+12\gamma_0)u_j\partial_j\partial_i\n^2\psi(\vx)\, .\label{eqmotiondiLquadraticinhom}
\ees
Two comments are in order:

\vspace{1mm}\noindent
{\bf (1)} The  terms $\lambda(\vx) (\pa_i u_i)^2$ and $\mu(\vx)\, u_{ij}u_{ij}$, in the first line of \eq{Lagrangianquadraticinhom}, contains two spatial derivatives acting on $u_i$, while all terms in the second and third  line contain four spatial derivatives. Nevertheless, if we want to keep the spatial dependence of $\lambda(\vx)$ and $\mu(\vx)$,  for consistency we must keep also all other terms. For instance, from \eqs{lambdainhom}{alphabetagamma}, we see that, for slowly varying backgrounds,  $\lambda(\vx)$ has the form $\lambda(\vx)=\lambda_0+\lambda_1(\vx)$, with 
$\lambda_1(\vx)$ of  order  $\eps_1(\vx)$, where, for the purpose of estimating the order of the various terms,  we denote here by $\eps_1(\vx)$ a parameter such 
$\eps_1(\vx)\sim \alpha_1(\vx)\sim \beta_1(\vx)\sim \gamma_1(\vx)$. So, the spatially-dependent part of $\lambda(\vx) (\pa_i u^i)^2$ is, schematically, of order $\eps_1\pa u\pa u$, i.e. one power of the small quantity $\eps_1$, two derivatives, and two powers of $u$. Compare this, for instance, with the term $\eps_0 u^i u^j \pa_i\pa_j\n^2\psi$ in the second line of \eq{Lagrangianquadraticinhom}
(where, again, we define $\eps_0$ as a parameter such that $\eps_0\sim\alpha_0\sim\beta_0\sim\gamma_0$).
Since, from \eq{n2psiexplicit}, $\eps_0\n^2\psi$ is of order $\eps_1$, this term is actually of order
$u_i u_j\pa_i\pa_j\eps_1$ or, integrating by parts,  $\eps_1\pa_i\pa_j  (u_i u_j)$,  so again one power of  $\eps_1$, two derivatives, and two powers of $u$, just as 
$\lambda_1(\vx) (\pa_i u^i)^2$. Similarly for all other terms with four spatial derivatives. Therefore, either one works at zero-th order, neglecting all spatial dependences and replacing $\lambda(\vx)$ and $\mu(\vx)$ by two constants $\lambda_0$ and $\mu_0$ or else, if one wants to keep the spatial dependence in $\lambda(\vx)$ and $\mu(\vx)$, by consistency one must keep  also all other terms in \eq{Lagrangianquadraticinhom}. Similarly, writing $\rho(\vx)=\rho_0+\rho_1(\vx)$, the same argument holds for the term proportional to $\dot{u}^i\dot{u}^j\pa_i\pa_j\psi$, compared to  $\rho_1(\vx)\dot{u}_i\dot{u}_i$.

\vspace{1mm}\noindent
{\bf (2)} If one  uses the non-relativistic Lagrangian (\ref{LDysoninhomisotr}), the 
 equation of motion is
\be\label{eqmotinhomiisot}
\rho\ddot{u}_i=\pa_i\[\lambda(\vx)\pa_ju_j\]
+\pa_j\[ \mu(\vx) (\pa_i u_j+\pa_j u_i)\]\, ,
\ee
which is the same as that obtained setting $\psi=0$ in \eq{eqmotiondiLquadraticinhom}. The difference  between these results can be traced to the fact that  the   Lagrangian (\ref{LDysoninhomisotr}) is not Lorentz invariant;  in contrast, the Lagrangian (\ref{Lalphabetagamma}) is Lorentz-invariant, but Lorentz-invariance is spontaneously  broken by the background solution and, in the inhomogeneous case, the latter is not given simply by \eq{xiaequilibrium},
but by  \eq{xixpapsi}, which involves a non-vanishing function $\psi(\vx)$.
As long as one is not interested in the coupling of elasticity to GWs, the use of a Lagrangian that is not covariant is perfectly fine (at least, for non-relativistic oscillations); elasticity is anyhow a phenomenological theory, and the functions that appear in a given Lagrangian, such as $\lambda(\vx)$ and $\mu(\vx)$ in \eq{LDysoninhomisotr}, are determined by comparison with observation, within the context of that theory. Note also that $\psi(\vx)$ is in principle determined in terms of  $\beta(\vx)$ and $\gamma(\vx)$ [and therefore in terms of
$\lambda(\vx)$, $\mu(\vx)$ and $\rho(\vx)$] by \eq{solpsi}, so both \eqs{eqmotiondiLquadraticinhom}{eqmotinhomiisot} will fit the observations using the same number of independent functions.
From this point of view, since \eq{eqmotinhomiisot} is so much simpler than \eq{eqmotiondiLquadraticinhom}, it is naturally to be preferred, and there is no point in using 
\eq{eqmotiondiLquadraticinhom}, that is phenomenologically similar, but much more complicated.
As we will see in  Section~\ref{sect:Coupling}, however, when we couple elasticity to GWs the extra terms induced by $\psi(\vx)$ can be important; their effect is that the coupling to GWs is no longer determined by the same functions, such as $\mu(\vx)$, that appear in the part of the equation that do not involve GWs, as was the case in ref.~\cite{Dyson:1969zgf}.


\section{Coupling elasticity to GWs}\label{sect:coupling_to_GWs}

We now discuss how to couple the elasticity theory to GWs. We begin, in Section~\ref{sect:PDFvsTT}, by recalling the difference between proper detector frame and TT frame. In Section~\ref{sect:normal modes} we will recall the standard picture in the proper detector frame, where the effect of GWs on an elastic medium can be described  as an effective Newtonian force, while space-time can be taken as flat. Within this approach, the field-theoretical machinery developed above is not needed; however,  as we will stress, this approach eventually breaks down at sufficiently large frequencies. 
In Section~\ref{sect:Coupling} we will then see how to couple elasticity theory to GWs in the TT gauge. This approach has no intrinsic validity limit in frequency, and here we will need the field-theoretical formalism developed in Section~\ref{sect:EFT_Rel}. Finally, in Section~\ref{sect:Comparison}, we will see how to compare the two approaches in the low-frequency regime, where they are both valid.

\subsection{Proper detector frame vs. TT frame: a reminder}\label{sect:PDFvsTT}

In this section we recall the definitions and properties  of the proper detector frame and of the TT frame. This is standard textbook material, and we will follow closely the presentation in Section~1.3.3 of \cite{Maggiore:2007ulw}. However, a clear understanding of the meaning of these two frames is essential in the following, so we find useful to provide here this short reminder.

In General Relativity (GR) the choice of coordinates is arbitrary, and different choices have different physical meaning. When coupling an elastic medium to GWs, one must therefore be aware of the coordinate system that one is using, since the definition of the displacement vector $\vu$, that determines the position of a given volume element with respect to an unperturbed position,  also depends on this choice.  
An important tool for understanding the physical meaning of a given coordinate choice is the
equation of the geodesic deviation: consider a space-time with a generic metric $\gmn(x)$, and let $x^{\mu}(\tau)$ be a time-like geodesic, parametrized by its proper time $\tau$, and  $x^{\mu}(\tau)+\zeta^{\mu}(\tau)$ an infinitesimally close geodesics. Then, in a generic coordinate system,   
$\zeta^{\mu}(\tau)$ satisfies the equation of the geodesic deviation,
\be\label{1geodev1}
\frac{d^2\zeta^{\mu}}{d\tau^2} +2\Gamma^{\mu}_{\nu\rho}(x)
\frac{d\xn}{d\tau}\frac{d\zeta^{\rho}}{d\tau}
+\zeta^{\sigma}\, \pas\GMnr (x) \,\frac{d\xn}{d\tau}\frac{dx^{\rho}}{d\tau}
=0\, .
\ee
The TT frame is defined as the coordinate system associated to the TT gauge: one considers a linearized metric of the form $\gmn=\emn+\hmn$ and shows, with standard arguments (e.g., Section 1.2 of \cite{Maggiore:2007ulw}) that we can choose the coordinate system so that $\hmn$ satisfies
\be\label{1TT}
h^{0\mu}=0\, ,\hspace{5mm} {h^i}_{i}=0\, ,\hspace{5mm}\pa^j\hij =0\, .
\ee
This defines the {\em transverse-traceless gauge}, or TT~gauge. When we wish to stress that we are writing $\hmn$ in this gauge, we will write the corresponding metric perturbation as $h_{ij}^{\rm TT}$.
Therefore, in the TT frame, the space-time interval reads (writing here $c$ explicitly)
\be\label{ds2TT}
ds^2= -c^2dt^2 +(\d_{ij}+\hTTij ) dx^i dx^j\, .
\ee
The physical meaning of this coordinate choice is obtained looking at   the geodesic deviation equation (\ref{1geodev1}), that in this metric becomes
\be\label{ddzetaTT}
\ddot{\zeta}^i=-\dot{h}^{\rm TT}_{ij}\dot{\zeta}^j\, ,
\ee
where the dot denotes the derivative with respect to proper time $\tau$.
We see that, if at $\tau =0$ we have $\dot{\zeta}^i =0$, then also
$\ddot{\zeta}^i=0$, so the coordinate separation $\zeta^i$ of two test masses initially at rest relative to each other remains constant at all
times, even when a GW passes. The TT gauge clearly illustrates the fact that, in GR, the physical effects are not expressed by what happens to the coordinates, since the theory is invariant under coordinate
transformations. In this gauge, indeed, as a GW passes, the coordinate distance between two test masses initially at rest  does not change.
Of course, this does not mean that the GW has no physical effect, but only that we used the freedom of choosing the coordinate system to {\em define} the 
coordinates by  using the position of test masses; e.g. we can say that a given test mass  defines the origin of the coordinate system, $\vx=(0,0,0)$ and  another test mass (or another volume element) defines the point with coordinates, say, $\vx=(1,0,0)$ (in some units), and this holds by definition even when the GW passes.
So, in this frame,  by definition the coordinates of test masses, or of volume elements of an elastic body, are not affected by the passage of a GW.
Physical effects, however,  can  be found monitoring  proper distances (or  proper times) rather than coordinate distances. The proper  distance $s$ between two events is obtained integrating 
$ds =\( \gmn dx^{\mu} dx^{\nu} \)^{1/2}$ along a space-like trajectory $x^{\mu}(\lambda)$, parametrized by a parameter $\lambda$, so that the proper distance between two space-time points
depends on the metric $\gmn$,  and therefore in the TT gauge it is affected by the  GW, that enters through \eq{ds2TT}. The proper distance between two volume elements of an elastic medium, or between the two mirrors of an interferometer arm, is therefore affected by a passing GW, and this is a gauge-invariant statement, since proper distances are invariant under coordinate transformations.

The proper detector frame, in contrast,  is defined by making use of the fact that, in a sufficiently small spatial neighborhood of a given space-time point P, we can choose the coordinates so that the metric is flat, 
$\gmn=\emn$, which is the content of the equivalence principle. The corresponding frame is called a locally inertial frame. In a locally inertial frame the metric is flat only at a given point in space and a given moment in time, but this can be extended by constructing a frame where a test mass is in free fall all along its  geodesic, rather than just at one space-time point, and this defines the ``freely falling frame" (the corresponding coordinates are known as Fermi normal coordinates). Then, in a freely falling frame, 
we can choose the
coordinates $(t,\vx )$ so that,  even in the presence of a gravitational field,  for any point P on the geodesic and sufficiently close to it, we have 
\be\label{eq:flat}
ds^2 \simeq -c^2dt^2 + \d_{ij}dx^idx^j\, .
\ee
To linear order in $|x^i|$ there are no corrections to this metric, 
since in a freely falling frame
the derivatives of $\gmn$ vanish at the point P on the geodesic around which we
expand. Pursuing the expansion to second order, one finds~\cite{Ni:1978zz}
\be\label{1suffR}
ds^2 \simeq -  c^2dt^2  \( 1+ R_{0i0j}x^ix^j \)- 2 cdt\, dx^i \(\frac{2}{3}R_{0jik}x^jx^k \)
+ dx^idx^j\( \d_{ij}-\frac{1}{3}R_{ikjl}x^kx^l \)\, ,
\ee
where the Riemann tensor is evaluated at the given point P along the geodesic.
We see that, if $L_B$ is the typical variation scale of the metric, so that 
$\Rmnrs =O(1/L_B^2)$, the corrections to the flat metric are
$O(r^2/L_B^2)$, where $r^2=x^ix^i$. In particular, if the Riemann tensor is due to a GW with reduced wavelength $\lbar$, the scale $L_B$ is $\lbar$, and the corrections to the flat metric, at a distance $r$ from the point P, are ${\cal O}(r^2/\lbar^2)$. 

To understand the effect of GWs in this frame, we can consider again the equation of the geodesic deviation, \eq{1geodev1}, that with this metric becomes
\be\label{1geodev}
\ddot{\zeta}^i=-c^2\, {R^i}_{0j0}\zeta^j\, ,
\ee
where  now the dot denotes the derivative with respect to the  coordinate time
$t$ of the proper detector frame. To compute the Riemann tensor ${R^i}_{0j0}$ 
due to the GWs in the proper detector frame, it is convenient to 
observe that, in linearized theory, the Riemann tensor is invariant,
rather than just 
covariant as in full GR, so we can compute
it in the frame that we prefer. The most convenient  choice is then to compute it in
the TT frame, since in this frame GWs have the simplest form. One then obtains
\be\label{1Riimme}
{R^i}_{0j0}=-\frac{1}{2c^2}\,\ddot{h}_{ij}^{\rm TT}\, ,
\ee
and therefore the equation of the geodesic deviation 
in the proper detector frame reads
\be\label{ddotzetalab}
\ddot{\zeta}^i=\frac{1}{2}\, \ddot{h}_{ij}^{\rm TT}\zeta^j\, .
\ee
From this we see that,  in the proper detector frame, measuring the position of  test masses as the geodesic deviation  with respect to an origin defined by another given test mass,  the effect of 
GWs on a point particle of mass $m$ can be described in terms of 
a Newtonian force
\be\label{1GWforce}
F_i=\frac{m}{2}\, \ddot{h}_{ij}^{\rm TT}\zeta^j\, ,
\ee
while at the same time, for distances $|\zeta^i|$ much smaller than $\lbar$, space-time can be approximated as flat, see \eq{eq:flat}.  Therefore 
the response of the detector to GWs can be analyzed in a purely Newtonian
language, without any further reference to General Relativity.

To summarize, in the TT gauge the coordinate separation is not affected by GWs. Then, below, we will not be surprised to find that, in the TT gauge, the  coordinates of  infinitesimal volume elements inside a (homogeneous) elastic medium are not affected by the passing of a GW. However, physical effects, in the TT gauge, are obtained looking at invariant quantities such as proper distances and proper times, and these are affected by the GW, since the metric is given by \eq{ds2TT}, and is not flat.
In the proper detector frame, in contrast, to lowest order in $r/\lbar$, where $r$ is the distance between two test masses and $\lbar$ the reduced wavelength of the GW, space-time is flat, \eq{eq:flat}. We can then forget about GR, and use our Newtonian intuition. At the same time, the action of a GW on a test mass can be described in terms of a Newtonian force, given by \eq{1GWforce}. Let us stress again that \eq{1GWforce} only holds in the proper detector frame, and not in the TT frame, despite the fact that $\hTTij$ appears there; this just came out because the Riemann tensor, in the linearized theory, is invariant, and can be computed in any frame, and then it is convenient to compute it in the TT frame, where $\hmn$ takes the simplest form.

A crucial difference between the TT frame and the proper detector frame is that the TT frame description is exact (in the context of linearized theory, i.e. writing $\gmn=\emn+\hmn$ and keeping only the linear orders in $\hmn$), in the sense that it  is valid independently of the distance $r$ between the bodies that are monitored. In contrast, the proper detector frame can  only be used in the limit $r\ll \lbar$. For instance, resonant-mass detectors  had a typical length $L$ of order a few meters, and their fundamental mode had a frequency $f_0$ of the order of the kHz, so they were searching GWs with $\lbar\sim 50$~km. In this case, therefore, $L\ll\lbar$ [the small ratio between these quantities  is in fact just  the factor $\pi v_s/c$ \eq{Lsulbar}], so they can be analyzed in the proper detector frame. Current ground-based GW interferometers operate at frequencies between tens of Hz and a few kHz, so approximately $\lbar$ is between 100 and 5000~km, while their arm length $L$ is 3~km for Virgo and 4~km for LIGO, so again $L\ll\lbar$; to lowest order in $L/\lbar$, or equivalently in $\omega L/c$, where $\omega =2\pi f$, it is then possible to analyze them in the proper detector frame. However, if one wants their response for higher frequencies, only the TT frame gives the exact answer, 
while the proper detector frame computation only reproduces the leading-order term in an expansion in $\omega L/c$   (see Section~9.1 of \cite{Maggiore:2007ulw}). As another example,  the space interferometer LISA is sensitive to GWs with $f$ approximately in the range  $10^{-3}-10^{-2}\, {\rm Hz}$, corresponding to $\lbar\sim (5\times 10^6) - (5\times 10^7)\, {\rm km}$,  while the arm-length, provided by the distance between the spacecrafts, is $L\sim 2.5\times 10^6\, {\rm km}$. In this case, the condition $\lbar\gg L$ is not really satisfied, and the response of LISA can only be computed in the TT frame.

\subsection{Proper detector frame description:  GWs as a force driving the normal modes}\label{sect:normal modes}

We now recall the standard computation of the response of an  elastic body to a GW in the proper detector frame,  in terms of a Newtonian force  acting on the normal modes of the body; we follow   Section~8.4.1 of \cite{Maggiore:2007ulw} (see also \cite{Lobo:1995sc}, and \cite{Brunetti:1998cc,Maggiore:1999wm} for the extension to scalar modes of the gravitational field), keeping however the formalism more general.
In this section we limit ourselves for simplicity  to isotropic elastic media, considering however both the homogeneous and inhomogeneous cases.
We use the proper detector frame which, as we have discussed above, implies that we restrict to GWs whose reduced wavelength $\lbar$ is much larger  than the typical linear size $L$ of the elastic body. We have seen 
in Section~\ref{sect:PDFvsTT} that in the proper detector frame, when $L\ll \lbar$ (and only in this limit!), the effect of the GW is described in terms of a Newtonian force $F_i$ given by \eq{1GWforce}, and therefore by a force per unit volume
\be\label{forcedensity}
f_i(t,\vx)=\frac{1}{2}\, \rho(\vx) \ddot{h}_{ij}^{\rm TT}(t,\vx) x_j\, .
\ee
Here we have set an origin inside the body (e.g., in the center of mass of the body) and we measure the geodesic deviation with respect to this point, so that $\zeta^j$ in \eq{1GWforce} is just the same as the coordinate $x^j$ of the volume element at position $\vx$ (note that spatial indices can be written equivalently as upper or lower); we also assumed that the density does not change appreciably on the time-scale of the interaction with the GW, so we took it time-independent. In the proper detector frame, therefore, we can describe 
the dynamics using \eq{NavierCauchycomponentswithf} with the force density given by \eq{forcedensity}. This gives
\be\label{rhoddotu}
\rho(\vx) \ddot{u}_i=\pa_i\[\lambda(\vx)\pa_ju_j\]+\partial_j\[ \mu(\vx)  (\partial_i u_j+\partial_j u_i)\] 
+\frac{1}{2}\rho(\vx) \ddot{h}_{ij}^{\rm TT}x_j\, .
\ee
We also observe that, in the same limit $L\ll \lbar$ in which the proper detector frame description holds, the spatial variation of the GW over the body is negligible, so we can  neglect the $\vx$ dependence in $h_{ij}^{\rm TT}(t,\vx)$ and we just write it as
 $h_{ij}^{\rm TT}(t)$. The force density (\ref{forcedensity}) can therefore be written in the separable form,
\be\label{fiseparable}
f_i(t,\vx)=\ddot{h}_{ij}^{\rm TT}(t) f_j(\vx) \, ,
\ee
(more precisely, as a sum of separable terms, because of the contraction on the $j$ index),
where
\be\label{fjrhoxj}
f_j(\vx) =\frac{1}{2}\, \rho(\vx) x_j\, .
\ee
The normal modes $\vu_N(\vx)$ of the body are defined by setting  ${\bf f}(t,\vx)=0$ and looking 
for  solutions of \eq{rhoddotu} and of the boundary condition  (\ref{bclamevx}), of the form
\be
\vu(t,\vx)=\vu_N(\vx) e^{-i\omega_N t}+ {\rm c.c.}\, .
\ee
By definition, therefore, the normal modes satisfy
\be\label{NavierCauchyuNinhom}
\pa_i\[\lambda(\vx)\pa_ju_N^j\]+\pa_j\[ \mu(\vx)  \(\pa_i u_N^j+\pa_j u_N^i\)\]=-\omega_N^2 \rho(\vx) u_N^i\, ,
\ee
together with the boundary condition
\be\label{2bodybounduNinhom}
n_i\pa_j \[ \lambda(\vx) u_N^j \] 
+  n_j\left\{ \pa_i \[ \mu(\vx)u_N^j\]+\pa_j \[ \mu(\vx) u_N^i\] \right\}=0\, ,
\ee
where $u_N^i\equiv (\vu_N)^i$.
In the homogeneous case these become (in vector notation)
\be\label{NavierCauchyuN}
(\lambda+\mu)\n (\n{\bf\cdot} \vu_N) +\mu\n^2\vu_N =-\rho(\vx)\omega_N^2\vu_N
\, ,
\ee
and
\be\label{2bodybounduN}
\lambda (\n\cdot \vu_N )\hatn +2\mu (\hatn\cdot \n) \vu_N+
\mu \hatn\times (\n\times \vu_N)=0\, .
\ee
The index $N$ denotes generically all the  indices needed to label the independent solutions of
\eq{NavierCauchyuN}, that satisfy the  boundary condition (\ref{2bodybounduN}). Observe that the normal modes
$\vu_N(\vx)$ can in general be  complex functions.
The normal modes are orthogonal with respect to the scalar product
\be
\langle \vu_{N'} | \vu_N\rangle =\int d^3x\, \rho(\vx) \vu^*_{N'}(\vx){\bf\cdot} \vu_N(\vx)\, .
\ee
A convenient choice of normalization is obtained setting
\be\label{othonorm}
\int d^3x\, \rho(\vx) \vu^*_{N'}(\vx) {\bf\cdot}\vu_N(\vx)=M\delta_{N N'}\, ,
\ee
where $M$ is the total mass of the body. The normal modes, so normalized, 
form an orthonormal complete set, so the most general solution for $\vu(t,\vx)$ can be written in the form
\be\label{vusumNuN}
\vu(t,\vx)=\sum_N\xi_N(t)\vu_N(\vx)\, ,
\ee
where $\xi_N(t)$ is the amplitude associated to the normal mode $\vu_N(\vx)$.  Inserting
\eq{vusumNuN} into \eq{rhoddotu} and using \eq{NavierCauchyuNinhom} we get
\be
\sum_N\rho(\vx)\vu_N(\vx)\( \ddot{\xi}_N+\omega_N^2\xi_N\)={\bf f}(t,\vx)\, ,
\ee
where ${\bf f}(t,\vx)$ is given by \eq{fiseparable}.
Taking the scalar product with $ \vu^*_{N'}(\vx) $, integrating over $d^3x$ using \eq{othonorm}, and renaming $N'\ra N$, we get 
\be
M\( \ddot{\xi}_N+\omega_N^2\xi_N\) =\int d^3x \, \vu^*_{N}(\vx){\bf\cdot}{\bf f}(t,\vx)\, .
\ee
Finally, using \eqs{fiseparable}{fjrhoxj}, we get
\be\label{eqforxiN}
\ddot{\xi}_N+\omega_N^2\xi_N=\kappa_{N,ij}\ddot{h}_{ij}^{\rm TT}(t) \, ,
\ee
where
\bees
\kappa_{N,ij}&=&\frac{1}{M}\int d^3x\, [\vu^*_{N} (\vx)]_i f_j(\vx)\nn\\
&=&\frac{1}{2M}\int d^3x\,\rho(\vx)  [\vu^*_{N} (\vx)]_i \, x_j\, .\label{kappaNij}
\ees
The constants $\kappa_{N,ij}$ determine the coupling of the $N$-th mode to the GW.

\Eq{eqforxiN} neglects dissipation. This can be included, at a phenomenological level, by adding a friction term to \eq{eqforxiN}, that becomes
\be\label{eqforxiNdiss}
\ddot{\xi}_N+\gamma_N\dot{\xi}_N+\omega_N^2\xi_N=\kappa_{N,ij}\ddot{h}_{ij}^{\rm TT}(t) \, .
\ee
For completeness, in App.~\ref{app:NMsolutions} we discuss in more detail the solution of these equations, in particular when a large number of normal modes contribute.

\subsection{Coupling the field theory of elasticity to GWs. Covariant approach}\label{sect:Coupling}

We now discuss how to couple elasticity to GWs using the covariant formulation of the field theory of elasticity. As discussed in Section~\ref{sect:PDFvsTT}, in the limit where the GW frequency $f$ is much larger than the frequencies of the lowest normal modes of the system, this will have two important advantages over the normal-mode description discussed in Section~\ref{sect:normal modes}: (1)  in the homogeneous case the description will depend only on few phenomenological quantities such as the density and the Lam\'e coefficients, that are in principle directly observable, rather than on the unknown frequencies and damping times of an infinity of highly excited  normal modes (in the inhomogeneous case the corresponding functions will however enter); (2) we will be able to formulate the coupling in the TT gauge, which does not suffer from the restriction $\lbar\gg L$ of the proper detector frame. In App.~\ref{sect:Dyson} we will then compare the results of our first-principle computation with those presented in the original Dyson's paper~\cite{Dyson:1969zgf}.

The first step is to provide a correct derivation of 
the  energy--momentum tensor of relativistic elasticity. To this purpose we covariantize the action (\ref{Sd4xnandBabxi}), which is obtained simply writing
\be\label{Sd4xnandBabxiCov}
S[\xi,\pa\xi]= -\int_{\cal M} d^4x\sqrt{-g}\, \[ \alpha(\xi) \, \sqrt{\det B} +\frac{1}{2} w_{abcd}(\xi) B^{ab}B^{cd}\]
\, ,
\ee
where now \eq{defBab} is replaced by
\be\label{defBabCov}
B^{ab} = \gMN \pam\xi^a\pan\xi^b  \, .
\ee
The energy--momentum tensor is then given by the standard general-relativistic expression
\be\label{TmnGR}
\Tmn=-\frac{2}{\sqrt{-g}}\, \frac{\d S}{\d\gMN}\, ,
\ee
and is automatically conserved, on the solutions of the equations of motion, because of the diffeomorphism invariance of the action (\ref{Sd4xnandBabxiCov}). Defining the linearized perturbations over flat space by $\gmn=\emn+\hmn$, we have $\gMN=\eMN-\hMN+{\cal O}(h^2)$ and therefore, to linear order in $\hmn$, the coupling of elasticity with metric perturbations is given by
\be\label{calLinthT}
{\cal L}_{\rm int}=+\frac{1}{2} \TMN\hmn\, .
\ee
The functional derivatives are easily computed using
\be
\frac{\d B^{ab}}{\d\gMN}=\pam\xi^a\pan\xi^b\, ,
\ee
and
\bees
\frac{\d\det B}{\delta \gMN}&=&
\frac{\d\det B}{\delta B^{ab}}\, \frac{\delta B^{ab}}{\delta \gMN}\nn\\
&=& (\det B) (B^{-1})_{ab} \, \pam\xi^a\pan\xi^b\, ,
\ees
where we used  \eq{vardetB}.
This gives
\bees \label{emtensor}
\Tmn &=& -\gmn \[ \alpha(\xi)  \,\sqrt{\det B} +\frac{1}{2} w_{abcd}(\xi) B^{ab}B^{cd}\]\nn\\
&&+\pam\xi^a\pan\xi^b\, 
\[ \alpha(\xi)   \,\sqrt{\det B}\, (B^{-1})_{ab}+2w_{abcd}(\xi) B^{cd}\]\, .
\label{TmnfullNL}
\ees
The flat-space expression is then obtained replacing $\gmn\ra \emn$ into this expression and, by construction, satisfies
\be
\pam\TMN=0\, .
\ee
\Eq{TmnfullNL} gives the full non-linear expression for $\TMN$, with the non-linearities that correspond to the specific non-linear action (\ref{Sd4xnandBabxi}). We are only interested in its expression to quadratic order in the fluctuations, which correspond to linearized (relativistic) elasticity theory.

\subsubsection{Homogeneous case}\label{sect:EFThom}

We begin by considering the homogeneous case, so that we can expand around the background solution $\bar{\xi}^a(x)=\delta_i^a x^i$. We therefore introduce $u^i$ as in \eq{xiaandui} and expand \eq{TmnfullNL} to second order in $u^i$, setting $\gmn=\emn$. 
We write  as usual $B_{ij}=\delta_i^a \delta_j^b B_{ab}$, and 
$w_{abcd}=\delta_i^a \delta_j^b\delta_k^c \delta_l^d w_{ijkl}$. The exact expression for $B_{ij}$ is given by
\eq{Bijui}, i.e.
\be
B_{ij}=\delta_{ij}-2u_{ij}-\dot{u}_i\dot{u}_j+\pa_k u_i\pa_k u_j\, ,
\ee
while the expansion
of  $\sqrt{\det B}$ to second order was already given in \eq{nvsu}, and the expansion  of  $(B^{-1})_{ij}$ to second order is given by
\be
(B^{-1})_{ij} =\delta_{ij}+2u_{ij}+\dot{u}_i\dot{u}_j-\pa_k u_i\pa_k u_j
+4u_{ik}u_{kj}+  {\cal O}(\pa u)^3\, .
\ee
Using  \eqst{calL0}{cijklwijkl}, we  then get [neglecting again the correction ${\cal O}(v_s^2/c^2)$, as in \eq{SdetBwexpanded}]
\be
T_{00}= \rho +\pa_iK_i
+\frac{1}{2}\rho \dot{u}_i^2 
+ \frac{1}{2} C_{ijkl}\pa_i u_j\pa_k u_l +{\cal O}(\pa u)^3 \, ,
\ee
where\footnote{If one includes the ${\cal O}(v_s^2/c^2)$ corrections, \eq{defKi} reads
$K_i\equiv  -\alpha u_i-2 w_{ij}u_j +(1/2) \alpha (u_i\pa_j u_j-u_j\pa_j u_i)$, and $\alpha$ is equal to $\rho$ times terms $1+{\cal O}(v_s^2/c^2)$, as we see from  \eq{alpharho} and, for the inhomogeneous case, \eqst{rhoinhom}{muinhom}.}
\be\label{defKi}
K_i\equiv  -\rho u_i+\frac{1}{2} \rho (u_i\pa_j u_j-u_j\pa_j u_i)\, .
\ee
and $C_{ijkl}$ was defined in \eq{cijklwijkl}.
The zero-th order term $T_{00}=\rho$ is the energy density associated to the equilibrium configuration, where $u^i=0$ and, if we are interested in the energy density associated to the fluctuations, we can just drop it.
The term $\pa_i K_i$ is a total derivative (observe that it has both a part linear and a part quadratic in $u_i$). As such,  it does not contribute to the energy of the fluctuations 
\be
E=\int d^3x\, T^{00}\, ,
\ee
with the boundary conditions that  $u_i$ goes to zero at infinity.  It also does not contribute to the coupling to GWs in the TT gauge, trivially because in the TT gauge $h_{00}=0$ and therefore the whole expression for $T_{00}$ is irrelevant for the coupling to GWs.   Note, however, that this term must be kept if we are interested in the coupling of the fluctuations of the elastic medium with a generic  external gravitational field with $h_{00}\neq 0$, such as a Newtonian potential, since, after integration by parts, it contributes to the interaction Lagrangian (\ref{calLinthT}) with a term $-(1/2)K_i\pa_i h_{00}$ or, to linear order in the fluctuations, $(1/2)\rho u_i\pa_i h_{00}$. So, in the end,  up to ${\cal O}(\pa u)^2$ the general expression for $T^{00}$ associated to the fluctuations is
\be\label{T00hom}
T_{00}=\pa_iK_i
+\frac{1}{2}\rho \dot{u}_i^2 
+ \frac{1}{2} C_{ijkl}\pa_i u_j\pa_k u_l +{\cal O}(\pa u)^3 \, .
\ee
Equivalently, in terms of $c_{ijkl}$ defined in \eq{cijklw}, we get
\be\label{T00hom2}
T_{00}=\pa_iK_i +\frac{1}{2}\rho \dot{u}_i^2
+ \frac{1}{2} c_{ijkl}u_{ij}u_{kl} + \tilde{w}_{jk} \tilde{u}_{ik}(2u_{ij}+\tilde{u}_{ij})
+{\cal O}(\pa u)^3\, .
\ee
For $T_{0i}$, instead, there is no zero-th order term and the leading term is ${\cal O}(\pa u)$,
\be\label{T0ihom}
T_{0i}=-\rho \dot{u}_i +{\cal O}(\pa u)^2\, .
\ee
where again we neglect corrections ${\cal O}(v_s^2/c^2)$.
From this expression of $T_{0i}$ we also further understand the need for the $\pa_i K_i$ term in \eq{T00hom}. Indeed, to linear order in $u$, the conservation equation $\pa_0 T^{00}+\pa_i T^{0i}=0$ is satisfied precisely because of the $\pa_i K_i$ term in $T^{00}$. In fact,  to linear order in $u$ and using $T^{0i}=-T_{0i}=+\rho \dot{u}_i$, we get   
$\pa_0 T^{00}+\pa_i T^{0i}=\pa_0\pa_i(-\rho u_i)+\pa_i (\rho \dot{u}_i)$, which indeed vanishes (also for a generic spatially-dependent $\rho$).

Note that $T_{0i}=T_{i0}$ since the covariant energy--momentum tensor (\ref{emtensor}), derived from the variation with respect to $\gmn$, is automatically symmetric.

Finally, for $T_{ij}$ \eq{TmnfullNL} gives
\be \label{Tij}
T_{ij}=-\[ 4w_{ijkl}\partial_k u_l -2\delta_{ij}w_{kl}\partial_k u_l
+2 \( w_{ik}\partial_j u_k+w_{jk}\partial_i u_k \) \]+{\cal O}(\pa u)^2\, ,
\ee
where we have dropped a term $2w_{ij}-(1/2)w\delta_{ij}$, which is associated to the equilibrium configuration rather than to the fluctuations and we have kept only the leading term, which is  ${\cal O}(\pa u)$. Observe that $T_{ij}$ is automatically symmetric, $T_{ij}=T_{ji}$. In terms of $c_{ijkl}$, defined in \eq{cijklw}, and $\tilde{w}_{ij}$ defined in \eq{deftildew}, we can rewrite this as
\be
T_{ij}=-c_{ijkl}\pa_ku_l-\tilde{w}_{ik}(u_{jk}+2\tilde{u}_{jk})-\tilde{w}_{jk}(u_{ik}+2\tilde{u}_{ik})+2\delta_{ij}\tilde{w}_{kl}\pa_k u_l\, .
\ee
To sum up, our final expression for the energy--momentum tensor of the fluctuations, to the lowest non-trivial order in $\pa u$ and neglecting corrections ${\cal O}(v_s^2/c^2)$ to the leading terms, is
\bees
T_{00}&=&\pa_iK_i +\frac{1}{2}\rho \dot{u}_i^2
+ \frac{1}{2} c_{ijkl}u_{ij}u_{kl} + \tilde{w}_{jk} \tilde{u}_{ik}(2u_{ij}+\tilde{u}_{ij})
+{\cal O}(\pa u)^3 \, ,\label{T00final} \\
T_{0i}&=&T_{i0}= -\rho \dot{u}_i +{\cal O}(\pa u)^2\, ,\label{T0ifinal}\\
T_{ij}&=&-c_{ijkl}\pa_ku_l-\tilde{w}_{ik}(u_{jk}+2\tilde{u}_{jk})-\tilde{w}_{jk}(u_{ik}+2\tilde{u}_{ik})+2\delta_{ij}\tilde{w}_{kl}\pa_k u_l\, ,
\label{Tijfinal}
\ees
where $K_i$ is defined in \eq{defKi}.
In the isotropic limit, where  $c_{ijkl}$ takes the form (\ref{cijklmulambda2}) and
$\tilde{w}_{ij}=0$, we get
\bees
T_{00}&=&\pa_iK_i
+\frac{1}{2}\rho \dot{u}_i^2 
+ \frac{1}{2} \[ \lambda (\pa_i u_i)^2+\mu \pa_iu_j (\pa_i u_j+\pa_j u_i)\]
+{\cal O}(\pa u)^3 \, ,\label{T00finaliso}\\
T_{0i}&=&T_{i0}= -\rho \dot{u}_i +{\cal O}(\pa u)^2\, ,\label{T0ifinaliso}\\
T_{ij}&=&-\[ \lambda\delta_{ij}\pa_ku_k+\mu (\pa_i u_j+\pa_j u_i)\]\, .\label{Tijlambdamu}
\ees
From these expressions it follows  that, for an infinite homogeneous elastic body, and to linear order in $u_i$ in the equations of motion (i.e., including terms up to quadratic order in $u_i$ in the Lagrangian), in the TT gauge the coupling  to GW vanishes. Indeed, using 
\eqs{LDysonLameagain}{calLinthT}, 
the total action  is
\bees\label{Stotcoupled}
S&=&S_2+S_{\rm int}\nn\\
&=&\frac{1}{2}  \int d^4x\,  \[  \rho\dot{u}_i\dot{u}_i -\lambda (\pa_iu_i)^2-2\mu\, u_{ij}u_{ij}
\]
+\frac{1}{2}  \int d^4x\, \TMN\hmn
\, .
\ees
Using the TT gauge condition 
$h^{0\mu}=0$, see \eq{1TT}, this becomes
\be\label{ShomhijTij}
S=\frac{1}{2}  \int d^4x\,  \[  \rho\dot{u}_i\dot{u}_i -\lambda (\pa_iu_i)^2-2\mu\, u_{ij}u_{ij} +h_{ij}^{\rm TT} T_{ij}
\]\, .
\ee
However, when contracted with $h_{ij}$, the term proportional to $\lambda\delta_{ij}$ in \eq{Tijlambdamu} vanishes because of the TT gauge condition $h_{ii}=0$, while the term proportional to $\mu (\pa_i u_j+\pa_j u_i)$ vanishes, upon integration by parts, because of  the TT gauge condition $\pa_ih_{ij}=0$. The equation of motion is therefore not affected by $h_{ij}^{\rm TT}$, and is still 
given by \eq{NavierCauchy0comp}, that we repeat here in the form
\be\label{eqmoticasohomogeneo}
\rho\ddot{u}_i=\lambda\pa_i\pa_ju_j
+\mu\pa_j(\pa_i u_j+\pa_j u_i)\, .
\ee
However, for a body of finite extent, the interaction with the GW comes from a boundary term. Indeed, 
using \eq{defsigmaij}, we now have\footnote{This result was already obtained, within the geometric formalism, in ref.~\cite{Hudelist:2022ixo}, see their eqs.~(15) and (20). If one does not specialize to the TT gauge, it reads $\sigma_{ij}=\lambda \delta_{ij} e_{kk} +2\mu e_{ij}$, where $e_{ij}=(1/2)(\pa_iu_j+\pa_ju_i+h_{ij})$.}
\be\label{sigmaijhijTT}
\sigma_{ij}=\lambda\delta_{ij}\pa_ku_k +\mu (\pa_i u_j+\pa_j u_i+ h_{ij}^{\rm TT})\, ,
\ee
so the boundary condition is now
\be\label{bcDyson}
n_j \[ \lambda\delta_{ij}\pa_ku_k +\mu (\pa_i u_j+\pa_j u_i+ h_{ij}^{\rm TT})\] =0\, ,
\ee
and depends on $h_{ij}^{\rm TT}$.
Observe that the  fact that \eq{eqmoticasohomogeneo} is independent of $h_{ij}^{\rm TT}$
crucially depends on the fact that here we are considering a homogeneous body with $\mu$ constant. otherwise, the integration by part would produce a non-vanishing coupling  to $h_{ij}^{\rm TT}$. As we will discuss in App.~\ref{sect:Dyson}, in the inhomogeneous case
Dyson~\cite{Dyson:1969zgf} uses the energy--momentum tensor
\be\label{TijlambdamuDysonx}
T_{ij}=-\[ \lambda(\vx)\delta_{ij}\pa_ku_k+\mu(\vx) (\pa_i u_j+\pa_j u_i)\] \, .
\ee
As we will see in Section~\ref{sect:inhomcasecoupling}, this is not the full answer. However, if  for the moment we use \eq{TijlambdamuDysonx} then, in the TT gauge, upon integration by parts, in the inhomogeneous case the action (\ref{ShomhijTij}) becomes
\be\label{SinhomhijTijincomplete}
S=\frac{1}{2}  \int d^4x\,  \[  \rho(\vx)\dot{u}_i\dot{u}_i -\lambda(\vx) (\pa_iu_i)^2-2\mu(\vx)\, u_{ij}u_{ij} 
-2  h_{ij}^{\rm TT} \mu(\vx)\pa_i u_j\]\, .
\ee
The corresponding equation of motion is\footnote{To compare with Dyson \cite{Dyson:1969zgf}, observe that we define
$\hmn$ from $\gmn=\emn+\hmn$, while Dyson uses
$\gMN=\eMN+\hMN$, see his eq.~(2.14), so for him $\gmn=\emn-\hmn+{\cal O}(h^2)$. Therefore, at linear order in $h$, his definition of $\hmn$ differs from ours by the sign. See App.~\ref{sect:Dyson} for a more detailed comparison with ref.~\cite{Dyson:1969zgf}.}
\be\label{eqmotinhomincomplete}
\rho(\vx)\ddot{u}_i=\pa_i[\lambda(\vx)\pa_ju_j]
+\pa_j\[ \mu(\vx) (\pa_i u_j+\pa_j u_i)\]
+h_{ij}^{\rm TT}\pa_j\mu\, ,
\ee
which, making use of $\pa_j  h_{ij}^{\rm TT}=0$,  can be rewritten as
\be\label{eqmotinhomhijinside}
\rho(\vx)\ddot{u}_i=\pa_i[\lambda(\vx)\pa_ju_j]
+\pa_j\[ \mu(\vx) (\pa_i u_j+\pa_j u_i+h_{ij}^{\rm TT})\]\, ,
\ee
while the boundary condition, computed as in \eq{boundarycond}, is
\be\label{bcDysoninhom}
n_j \[ \lambda(\vx)\delta_{ij}\pa_ku_k +\mu(\vx) (\pa_i u_j+\pa_j u_i+ h_{ij}^{\rm TT})\] =0\, .
\ee
The physical meaning of this result  can be understood as follows. In Section~\ref{sect:PDFvsTT} we recalled that, in the TT gauge, the coordinates of free particles are not affected by a passing GW simply because, in the TT gauge, we use the diffeomorphism invariance of GR to introduce  coordinates  that  are {\em defined} by the position of free particles; therefore, they are unaffected by a passing GW  by definition. However, we stressed that physical effects are contained in invariant quantities, such as proper distances or proper time intervals, and these are affected by a passing GW. 
We see that this result carries over from the TT-gauge coordinates of free particles to the TT-gauge coordinates defining the position of volume elements in an elastic body, as long as the elastic forces are uniform across the body and the body has infinite extent, so there are no boundary conditions to be imposed at the body's surface: in this case, in the TT gauge, the relative coordinate positions of volume elements are unaffected by a passing GW. Once again, this does not mean that GWs have no effect on uniform elastic bodies, but rather than we should look for them using invariant quantities, such as the proper distances between volume elements.

Even if, as we will see below, \eq{eqmotinhomincomplete} is not yet the full equation of motion for inhomogeneous bodies, still it already allows us to understand that the fact that coordinate distances between volume elements, in the TT gauge, are unaffected by a passing GW, only holds for homogeneous body. If the elastic restoring forces felts by two volume elements are different, not surprisingly, the result valid for free particle no longer goes through.

\subsubsection{Inhomogeneous case}\label{sect:inhomcasecoupling}

We now consider the inhomogeneous case, restricting ourselves to isotropic elastic bodies and small deviation from homogeneity. We then consider the background solution $\bar{\xi^a}(x)=\delta^a_i[x^i+\partial^i \psi(x)]$, with $\psi(x)$ given by eq. (\ref{solpsi}), and we expand \eq{emtensor} around this solution to second order in $u^i$; we also restrict to first order in the small quantities $\alpha_1,\beta_1,\gamma_1$ and $\psi$,  and we set $g_{\mu\nu}=\eta_{\mu\nu}$.
Neglecting again ${\cal O}(v_s^2)$ corrections,  for $T_{00}$ we get 
\bees
T_{00} &=& \partial_i D_i + \frac12 \rho(\vx) {\dot{u}_i}^2 -\alpha_0 \dot{u}_i\dot{u}_j \partial_i \partial_j \psi(\vx) + \frac12\lambda(\vx)(\partial_i u_i)^2 + \mu(\vx)u_{ij}u_{ij}\,\label{T00inhomo} \\
&&+ \alpha_0 \partial_i \partial_j \psi(\vx) (\partial_i u_k \partial_k u_j - \partial_i u_j \partial_k u_k) - \alpha_0 u_i \partial_i \n^2 \psi(\vx) \partial_j u_j  - \frac12 \alpha_0 u_i u_j \partial_i \partial_j \n^2 \psi(\vx)\,\nonumber
\ees
where we discarded the energy density associated to the equilibrium configuration. The functions $\rho(\vx)$ and $\psi(\vx)$ are given in terms of the functions $\alpha(\vx)$, $\beta(\vx)$ and $\gamma(\vx)$ by
\eqs{rhoinhom}{n2psiexplicit} and, to the order at which we are working,
$\alpha(\vx)$, $\beta(\vx)$ and $\gamma(\vx)$ are given by \eq{alphabetagamma}.
Neglecting ${\cal O}(v_s^2)$ corrections, the quantity $D_i$, in the first line of \eq{T00inhomo},  is given by
\be
D_i \equiv -\rho(\vx)u_i+\alpha_0 u_j \partial_i \partial_j \psi (\vx) + \frac12\rho(\vx)(u_i \partial_j u_j-u_j \partial_j u_i)+\frac12 u_i u_j \partial_j \rho(\vx)\, ,
\ee
and reduces to $K_i$, given in \eq{defKi}, if we set $\psi=0$, i.e. to zero-th order in the inhomogeneities.
For $T_{0i}$, there is no term associated to the equilibrium configuration, and  the leading order is linear in $u_i$, 
\bees\label{T0iinhompapapsi}
T_{0i}=-\rho(\vx) \dot{u}_i+\alpha_0\dot{u}_j\partial_i\partial_j \psi(\vx)\,,
\ees
where again ${\cal O}(v_s^2)$ corrections have been neglected. 
This generalizes  \eq{T0ihom} to the $\vx$-dependent case.
Finally, for $T_{ij}$ we get (again, at leading order ${\cal O}(\pa u)$ and discarding the contribution of the equilibrium configuration)
\bees \label{Tijinhom}
-T_{ij}&=&\lambda(\vx)\delta_{ij}\partial_k u_k 
+ \mu (\vx)(\partial_i u_j+\partial_j u_i) 
+ \delta_{ij} u_k \partial_k [3 \beta_1 (\vx)+\gamma_1(\vx)]\nn\\
&&- (4\beta_0+12\gamma_0)\delta_{ij}\partial_k \partial_l \psi(\vx) \partial_k u_l + 16\gamma_0 [\partial_i \partial_k \psi(\vx)u_{kj}+\partial_j\partial_k \psi(\vx)u_{ki}]\nn\\
&&+(12\beta_0+8\gamma_0)[\partial_i\partial_k\psi(\vx)\partial_j u_k+\partial_j\partial_k \psi(\vx)\partial_i u_k] + 16\beta_0 \partial_i \partial_j \psi(\vx) \partial_k u_k\, ,
\ees
where 
$\lambda(\vx)$ and $\mu(\vx)$ are the functions   given in \eqst{lambdainhom}{muinhom}. Note that $T_{ij}$ does not depend on $\alpha$. Therefore all the terms in eq. (\ref{Tijinhom}) are of the same order ${\cal O}(v_s^2)$ and cannot be neglected. Again, this reduces to \eq{Tijlambdamu} in the homogeneous case. We observe, however, that the result for the inhomogeneous case is not simply obtained from \eq{Tijlambdamu} by replacing $\lambda\ra\lambda(\vx)$ and $\mu\ra\mu(\vx)$; rather, extra terms appear, which depend on the first-order perturbations $\beta_1(\vx)$ and $\gamma_1(\vx)$, and on the second derivatives of the function $\psi(\vx)$ [which are of the same order, recall \eqs{n2psiexplicit}{solpsi}]. Note that these terms are of the same order as the $\vx$ dependent part of $\lambda(\vx)$ and $\mu(\vx)$, 
see \eqss{alpha0plus1}{lambdainhom}{muinhom},
and therefore cannot be neglected. 

We can now couple this energy--momentum tensor to GWs in the TT gauge. Again, at the quadratic level in the fluctuations, the action has the form (\ref{ShomhijTij}), but now $T_{ij}$ is given by
\eq{Tijinhom}. However, all terms in $T_{ij}$  proportional to $\delta_{ij}$ again do not contribute, because in the TT gauge $h_{ii}=0$, so the coupling simplifies, and we get
\bees
S&=&S_2+S_{\rm int}\nn\\
&=& \int d^4x\,\Big\{\frac{1}{2}  \[  \rho(\vx)\dot{u}_i\dot{u}_i -\lambda(\vx) (\pa_i u_i)^2-2\mu(\vx)\, u_{ij}u_{ij}\]+\(4\gamma_0-\alpha_0\)\dot{u}_i\dot{u}_j\pa_i\pa_j\psi(\vx)+\,\nonumber\\
&+&\(\alpha_0+8\beta_0\)(\pa_i u_i) u_j \pa_j\n^2\psi(\vx)
+\frac12 (\alpha_0 - 12\beta_0 - 12\gamma_0) u_i u_j \pa_i\pa_j\n^2\psi(\vx)\,\\
&+&\pa_j\pa_k\psi(\vx)\left[\(\alpha_0-8\beta_0\) (\pa_i u_i)\pa_j u_k-\(\alpha_0+4\gamma_0\)\pa_i u_j\pa_k u_i-4\gamma_0\pa_i u_j\pa_i u_k-4\gamma_0\pa_j u_i\pa_k u_i\right]\,\nonumber \\
&-& h_{ij}^{\rm TT} \[ \mu(\vx) \pa_j u_i
+16\gamma_0 \pa_i\pa_k \psi(\vx) u_{kj}
+(12\beta_0+8\gamma_0)\partial_i\partial_k\psi(\vx)\pa_j u_k
+8\beta_0\pa_i\pa_j\psi(\vx)\pa_k u_k
\]\Big\}
\, .\nn
\ees
The corresponding equation of motion is
\bees 
&\rho(\vx)\ddot{u}_i& + 2(4\gamma_0-\alpha_0) \ddot{u}_j \partial_j \partial_i \psi(\vx) = \partial_i\Bigr[\lambda(\vx)\partial_j u_j\Bigr] + \partial_j\Bigr[\mu(\vx)(\partial_i u_j+\partial_j u_i)\Bigr] \nn\\
&&+8\beta_0\partial_i\Bigr[\partial_j\partial_k\psi(\vx)\partial_j u_k+\partial_k u_k\n^2\psi(\vx)\Bigr]+8\beta_0\partial_j\Bigr[\partial_i\partial_j\psi(\vx)\partial_k u_k-\partial_iu_j\n^2\psi(\vx)\Bigr]\nn\\
&&+4\gamma_0\partial_j\Bigr[\partial_i\partial_k\psi(\vx)(\partial_k u_j+2\partial_j u_k)+\partial_j\partial_k\psi(\vx)(\partial_i u_k +2\partial_k u_i)\Bigr]\nn\\
&&
-(20\beta_0+12\gamma_0)u_j\partial_j\partial_i\n^2\psi(\vx)
+h_{ij}^{\rm TT}\partial_j \mu(\vx)
+(12\beta_0+16\gamma_0)h_{jk}^{\rm TT}\partial_i\partial_j\partial_k\psi(\vx)\nn\\
&& +8\beta_0 \partial_i \Bigr[h_{jk}^{\rm TT}\partial_j\partial_k\psi(\vx)\Bigr]
+8\gamma_0\partial_j\Bigr[h_{ik}^{\rm TT}\partial_j\partial_k\psi(\vx)\Bigr] \, .
\label{equmotioninhom}
\ees
We now use eqs. (\ref{n2psiexplicit}), (\ref{lambdainhom}) and (\ref{muinhom}) to rewrite $\n^2\psi(\vx)$ as function of $\lambda(\vx)$ and $\mu(\vx)$. We get
\bees \label{nablapsimulambda}
\n^2\psi(\vx)=\frac{1}{2}\frac{3\beta_0+\gamma_0}{11\beta_0+6\gamma_0}-\frac{1}{8}\frac{3\lambda(\vx)+2\mu(\vx)}{11\beta_0+6\gamma_0} \, .
\ees
Using the Green's function of the Laplacian, \eq{nablapsimulambda} gives $\psi[\lambda(\vx),\mu(\vx)]$. By inserting this explicit form into \eq{equmotioninhom}, the equations of motion can then be expressed in terms of $\rho(\vx)$, $\lambda(\vx)$ and $\mu(\vx)$. Following the strategy outlined in \eqs{deltaSconBound}{boundarycond}, the boundary condition at the surface of the body reads
\bees
&&\hspace*{-8mm}n_j\left[\lambda_0 \delta_{ij} \partial_k u_k 
+ \mu_0 \(\partial_i u_j +\partial_j u_i\)
+\mu(\vx)h_{ij}^{\rm TT}\right.\,\label{boundarywithGW}\\
&&+8\gamma_0 h_{ik}^{\rm TT} \left.\partial_j\partial_k \psi
+(12\beta_0+16\gamma_0)h_{jk}^{\rm TT}\partial_i\partial_k\psi
+8\beta_0\delta_{ij} h_{kl}^{\rm TT}\partial_k\partial_l\psi 
\right]=0 \,,\nonumber
\ees
where $\vx$ is the position of a point at the surface
and $n_j$ the normal vector to the surface of the body, and reduces to \eq{bcDyson} in the homogeneous case. 

\Eq{equmotioninhom}, as it stands, is quite complicated, and for practical  applications one might want to simplify it as much as possible. One possibility is to neglect the $\vx$ dependence in all terms on the right-hand side that are not coupled to GWs, since this already gives a non-trivial leading term. However, in the coupling with $h_{ij}^{\rm TT}$ the leading term is obtained from the $\vx$-dependent terms, so here these must be kept. This leads to the simpler form
\bees 
\rho\ddot{u}_i& \simeq &\lambda_0 \partial_i\partial_j u_j + \mu_0\partial_j(\partial_i u_j+\partial_j u_i) +h_{ij}^{\rm TT}\partial_j \mu \label{equmotioninhomsimplified}\\
&&
+(12\beta_0+16\gamma_0)h_{jk}^{\rm TT}\partial_i\partial_j\partial_k\psi
 +8\beta_0 \partial_i \( h_{jk}^{\rm TT}\partial_j\partial_k\psi\)
 +8\gamma_0\partial_j\(h_{ik}^{\rm TT}\partial_j\partial_k\psi\)\, ,\nn
\ees
where, in our approximation,  $\pa_j\mu$ can also be written as $\pa_j\mu_1$. Note that, in any case, the coupling to GWs is more complicated than that always used in the literature, following \cite{Dyson:1969zgf}, that only included the term $h_{ij}^{\rm TT}\partial_j \mu$. In the same approximation, the boundary condition (\ref{boundarywithGW}) simplifies to
\be
\label{boundarywithGWsimplified}
n_j \[ \lambda_0 \delta_{ij}\partial_k u_k + \mu_0 \(\partial_i u_j +\partial_j u_i+ h_{ij}^{\rm TT}\)\]=0\, .
\ee
A further simplification is obtained considering GWs whose reduced wavelength $\lbar$ is much larger than the scale $\ell$ of the inhomogeneities. Then, each spatial derivative applied on $h_{ij}^{\rm TT}$ brings a factor of order $1/\lbar$, while a derivative applied on $\psi$ brings a factor $1/\ell$.  So, for instance, in the second line of \eq{equmotioninhomsimplified},
\be
\partial_i \(h_{jk}^{\rm TT}\partial_j\partial_k\psi\)=
(\pa_ih_{jk}^{\rm TT})\partial_j\partial_k\psi+
h_{jk}^{\rm TT}\pa_i\partial_j\partial_k\psi\, ,
\ee
and the first term on the right-hand side is smaller than the second by a factor of order $\ell/\lbar$, and can be neglected. Then, \eq{equmotioninhomsimplified} becomes
\be\label{equmotioninhomsimplified2}
\rho\ddot{u}_i \simeq \lambda_0 \partial_i\partial_j u_j + \mu_0\partial_j(\partial_i u_j+\partial_j u_i) 
+h_{ij}^{\rm TT}\partial_j (\mu+8\gamma_0 \n^2\psi)
+(20\beta_0+16\gamma_0)h_{jk}^{\rm TT}\partial_i\partial_j\partial_k\psi\, .
\ee
\Eq{equmotioninhomsimplified2}  shows that, on top of the term $h_{ij}^{\rm TT}\partial_j \mu$  already found by Dyson, there are extra terms, that depends on the function $\psi$, that can be used to construct two more independent structures, $h_{jk}^{\rm TT}\partial_i\partial_j\partial_k\psi$ and
$h_{ij}^{\rm TT} \n^2\pa_j\psi$. While the specific coefficients of these terms in \eq{equmotioninhomsimplified2}, as well as \eq{solpsi},  that fixes $\psi$ in terms of  $\beta(\vx)$ and $\gamma(\vx)$ [and therefore in terms of
$\lambda(\vx)$, $\mu(\vx)$ and $\rho(\vx)$], are specific to the Lagrangian (\ref{Lalphabetagamma}) that we are using as an explicit example, the more general lesson that one learns is that the coupling of an elastic medium to GWs is more complicated, compared to the $h_{ij}^{\rm TT}\partial_j \mu$ term found by Dyson, and depends on at least one more function $\psi(\vx)$.\footnote{For more complex actions one could imagine that also the transverse vector field $\eps^T_i(\vx)$ could be non-zero, see \eq{espespTpapsi}.}
This function   can in principle be determined in terms of $\lambda(\vx)$, $\mu(\vx)$ and $\rho(\vx)$, given a specific relativistic theory, or else must be kept as a new phenomenological function, if we do not want to commit  ourselves to a specific relativistic action. In this more general form,
\eq{equmotioninhomsimplified2} can be written as
\be\label{equmotioninhomsimplified3}
\rho\ddot{u}_i \simeq \lambda_0 \partial_i\partial_j u_j + \mu_0\partial_j(\partial_i u_j+\partial_j u_i) 
+h_{ij}^{\rm TT}\partial_j ( \mu+c_1\n^2\psi) +c_2 h_{jk}^{\rm TT}\partial_i\partial_j\partial_k\psi\, ,
\ee
for some phenomenological coefficients $c_1$, $c_2$ and a phenomenological function $\psi(\vx)$ that, together with 
$\rho(\vx)$, $\lambda(\vx)$ and $\mu(\vx)$, describe the elastic medium.\footnote{Note that one of the two coefficients $c_1, c_2$ could be reabsorbed into the normalization of $\psi(\vx)$.} Equivalently, defining 
\be\label{defnu1nu2}
\nu_1(\vx)= \mu(\vx)+c_1\n^2\psi(\vx)\, ,\qquad
\nu_2(\vx)=c_2\psi(\vx)\, ,
\ee
we get 
\be\label{equmotioninhomsimplified4}
\rho\ddot{u}_i \simeq \lambda_0 \partial_i\partial_j u_j + \mu_0\partial_j(\partial_i u_j+\partial_j u_i) 
+h_{ij}^{\rm TT}\partial_j \nu_1 +h_{jk}^{\rm TT}\partial_i\partial_j\partial_k\nu_2\, .
\ee
The function $\psi(\vx)$ and the constants $c_1, c_2$, and therefore  $\nu_1(\vx)$ and $\nu_2(\vx)$, can in principle be computed in terms of $\lambda(\vx)$, $\mu(\vx)$ and $\rho(\vx)$ if we assume a specific form for the relativistic action. However,   if we do not want to commit  ourselves to a specific relativistic action, we must keep $\psi(\vx)$ as new phenomenological function and $c_1, c_2$ as new phenomenological parameters.

\subsection{Matching the TT frame and proper detector frame computations}\label{sect:Comparison}

We now want to compare the computations in the TT frame and in the proper detector frame. In this section we denote by $x^{\mu}$ the coordinates in the TT frame and by $x^{\mu}_{\rm lab}$ the coordinates in the proper detector frame (which are the natural coordinates from the point of view of the laboratory, where one naturally think in terms of Newtonian forces in flat space). Similarly, we denote by
$u_i(t,\vx)$ the displacement from equilibrium of a volume element of the elastic body in the TT frame, defined in \eq{xiaandui},  and by
$u^{\rm lab}_i(t,\vx_{\rm lab})$ the displacement in the proper detector frame.
As we have discussed in Section~\ref{sect:PDFvsTT}, in both frames the effect of the GW can be described using $h_{ij}^{\rm TT}$, because in the  linearized theory the Riemann tensor is invariant under linearized diffeomorphisms (rather than just covariant, as in the full theory), see \eq{1Riimme} and the discussion above it. Therefore, both in the  equations of the TT frame and in the equations of the proper detector frame, $h_{ij}^{\rm TT}$ appears. In this section, to make the notation lighter, we will then denote $h_{ij}^{\rm TT}$ simply as $h_{ij}$, but still it satisfies $\pa_i h_{ij}=0$ and $h_{ii}=0$.

\vspace{2mm}\noindent
{\bf Relation between $x^{\mu}$ and  $x^{\mu}_{\rm lab}$.}
Let us first discuss the relation between $x^{\mu}$ and  $x^{\mu}_{\rm lab}$. In the TT frame the interval is given by \eq{ds2TT}, that in the present notation reads
\be\label{ds2TTnew}
ds^2= -c^2dt^2 +\[\d_{ij}+h_{ij}(t,\vx) \] dx^i dx^j\, .
\ee
In the proper detector frame, and sufficiently close to the point P (or to the geodesic) where we could write the metric as flat, it is instead given by \eq{eq:flat}, so
\be\label{eq:flatnew}
ds^2 = -c^2dt_{\rm lab}^2 + \d_{ij}dx_{\rm lab}^idx_{\rm lab}^j\, .
\ee
Since the interval is invariant, we must have
\be
 -c^2dt_{\rm lab}^2 + \d_{ij}dx_{\rm lab}^idx_{\rm lab}^j=
 -c^2dt^2 +\[\d_{ij}+h_{ij}(t,\vx)\]  dx^i dx^j\, .
 \ee
This shows that
\be
 t_{\rm lab}=t\, ,
\ee
and
\be\label{dxTTdxlab1}
\d_{ij}dx_{\rm lab}^idx_{\rm lab}^j= \[\d_{ij}+h_{ij}(t,\vx)\]   dx^i dx^j\, .
\ee
Looking for a solution $dx_{\rm lab}^i= dx^i+\eps_{ij}dx^j$, with $\eps_{ij}$ of order $h_{ij}$, and substituting this into \eq{dxTTdxlab1}, to first order in $h_{ij}$ we get $2\eps_{ij}=h_{ij}$, and therefore
\be\label{dxTTdxlab2}
dx_{\rm lab}^i= dx^i+\frac{1}{2} h_{ij}(t,\vx) dx^j +{\cal O}(h^2)\, .
\ee
To integrate this equation and get the relation between finite displacements
we now, crucially, assume that the wavelength of the GW is much larger than the region over which we want to integrate this relation, and we therefore neglect the spatial dependence in $h_{ij}$, that we therefore write simply as $h_{ij}(t)$. Then, \eq{dxTTdxlab2} integrates to
\be\label{xTTxlab}
x_{\rm lab}^i= x^i+\frac{1}{2} h_{ij}(t) x^j +{\cal O}(h^2)\, .
\ee
Since we have neglected the spatial variation of $h_{ij}$, this relation holds only over a distance $r\ll\lbar$ and can be used over the whole scale of a detector of linear size $L$ only if $L\ll\lbar$. So, even if the results in the TT frame are valid for arbitrary frequencies of the GW, they can be translated to the proper detector frame only for $r\ll\lbar$, which is the regime where  computations can also  be performed directly in the proper detector frame.\footnote{Of course, one can in principle improve \eq{xTTxlab} perturbatively, including higher and higher orders in $r/\lbar$. In any case, as $r/\lbar$ becomes of order one or larger, the expansion breaks down.}

As a check, consider a test mass in the TT frame, and denote by $x_a^i(t)$ its trajectory in the TT frame coordinates [measured with respect to an origin defined by another test mass, so that $x_a^i(t)$ is the same as the geodesic deviation $\zeta(t)$]. The index $a$ labels the particle, and allows us to distinguish between a trajectory $x_a^i(t)$ and a generic spatial coordinate $x^i$. According to \eq{ddzetaTT}, if the particle was initially at rest before the arrival of the GW, it will remain at rest, i.e.
\be
\ddot{x}_a^i(t)=0\, .
\ee
Then, from \eq{xTTxlab}, in the proper detector frame the same test mass will be described by a trajectory
$x_{a,\rm lab}^i(t)$ such that
\be
\ddot{x}_{a,\rm lab}^i(t)=\frac{1}{2} \ddot{h}_{ij}(t) x_a^j(t)+{\cal O}(h^2)\, .
\ee
Since, from  \eq{xTTxlab}, $x_a^j(t)=x_{a,\rm lab}^j(t)+{\cal O}(h)$, in the term 
$ \ddot{h}_{ij}(t) x_a^j(t)$ we can just replace $x_a^j(t)$ by  $x_{a,\rm lab}^j(t)$, and therefore
\be
\ddot{x}_{a,\rm lab}^i(t)=\frac{1}{2} \ddot{h}_{ij}(t) x_{a,\rm lab}^j(t)+{\cal O}(h^2)\, ,
\ee
in agreement with \eq{ddotzetalab}. We stress again that this relation is valid to first order in $h$, and only across  distances much smaller than the GW wavelength.

\vspace{2mm}\noindent
{\bf Relation between $u_i$ and $u_i^{\rm lab}$: homogeneous case.}
We next  discuss the transformation of the variables describing the position of a volume element of an elastic medium,  in the TT gauge and in the proper detector frame.  We work directly in the long-wavelength limit, where  spatial derivatives of  $h_{ij}$ can be neglected and therefore $h_{ij}$ can be taken independent of $\vx$.
We begin with the homogeneous case. 
The equation of motion for $u^i$ is  given by \eq{eqmoticasohomogeneo}
which, for convenience, we rewrite here,
\be\label{rhoddotuTT}
\rho\ddot{u}_i=\lambda\pa_i\pa_ju_j
+\mu\pa_j(\pa_i u_j+\pa_j u_i)\, .
\ee
If the body has a finite extent, we must supplement this with the boundary condition 
(\ref{bcDyson}), which  we also rewrite here in the present notation,
\be
\label{bcrhoddotuTT}
n_j \[ \lambda\delta_{ij}\pa_ku_k +\mu (\pa_i u_j+\pa_j u_i+ h_{ij})\]=0\, .
\ee
The equation of motion for $u_{\rm lab}^i$ is given by \eq{rhoddotu}  that, in the present notation, reads
\be\label{rhoddotulabxlab}
\rho\ddot{u}_{\rm lab}^i(t,\vx_{\rm lab})=
\lambda \frac{\pa^2 u_{\rm lab}^j(t,\vx_{\rm lab})}{\pa x_{\rm lab}^i\pa x_{\rm lab}^j}
+\mu  \frac{\pa}{\pa x_{\rm lab}^j} 
\( \frac{\pa u_{\rm lab}^j(t,\vx_{\rm lab})}{\pa x_{\rm lab}^i} + \frac{\pa u_{\rm lab}^i(t,\vx_{\rm lab})}{\pa x_{\rm lab}^j} \)
+\frac{1}{2}\rho \ddot{h}_{ij}x_{\rm lab}^j\, .
\ee
In this equation we can simply rename $\vx_{\rm lab}$ as $\vx$, and rewrite it as
\be\label{rhoddotulab}
\rho\ddot{u}^{\rm lab}_i(t,\vx)=
\lambda\pa_i\pa_ju^{\rm lab}_j+\mu\pa_j(\pa_i u^{\rm lab}_j+\pa_j u^{\rm lab}_i) +\frac{1}{2}\rho \ddot{h}_{ij}x_j\, .
\ee
For a finite body, in the proper detector frame the boundary condition is given, in terms of the normal modes,  by \eq{2bodybounduNinhom}. Then, from \eq{vusumNuN}, the corresponding boundary condition on
$u^{\rm lab}_i(t,\vx_{\rm lab})$, in the homogeneous case,  is 
\be\label{bcrhoddotulab}
n_j\[ \lambda  \delta_{ij} \pa_j u^{\rm lab}_j +\mu  \( \pa_i u^{\rm lab}_j+\pa_ju^{\rm lab}_i\) \]=0\, ,
\ee
where, again, we renamed $\vx_{\rm lab}$ as $\vx$.
We now require  consistency between \eqst{rhoddotuTT}{bcrhoddotuTT} on the one hand, 
and \eqst{rhoddotulab}{bcrhoddotulab} on the other.
This is obtained setting 
\be\label{ulabuTT1}
u^{\rm lab}_i(t,\vx)=u_i(t,\vx)+\frac{1}{2}h_{ij}(t) x_j\, .
\ee
In this way, both the equation of motion and boundary conditions in the TT frame become equivalent to those in the proper detector frame.\footnote{Note that, for the homogeneous case, \eq{ulabuTT1} was also found in a one-dimensional example in ref.~\cite{Hudelist:2022ixo}, see their eq.~(33), and further discussed in \cite{Spanner:2023owt}, where the interpretation as a transformation between the proper detector frame and the TT frame was  made explicitly.}
Note also that, using \eq{xTTxlab}, we have
\be
u^{\rm lab}_i(t,\vx_{\rm lab})=u^{\rm lab}_i(t,\vx)+\frac{1}{2}h_{kj}(t) x_j\pa_ku^{\rm lab}_i(t,\vx)+{\cal O}(h^2)\, ,
\ee
so, to ${\cal O}(h)$, \eq{ulabuTT1} can also be rewritten as
\be\label{ulabuTT2}
u^{\rm lab}_i(t,\vx_{\rm lab})=u_i(t,\vx)+\frac{1}{2}h_{kj}(t) x_j 
\( \delta_{ki}+\pa_ku_i\)
\, .
\ee
However, for slowly varying fluctuations, we have $|\pa_ku_i|\ll 1$, so we can  write
\be\label{ulabuTT3}
u^{\rm lab}_i(t,\vx_{\rm lab})\simeq u_i(t,\vx)+\frac{1}{2}h_{ij}(t) x_j\, .
\ee

\vspace{2mm}\noindent
{\bf Relation between $\vu$ and $\vu^{\rm lab}$: inhomogeneous case.}
We now consider the inhomogeneous case. For simplicity, and to make easier the comparison with the literature, we set $\psi=0$ and we use \eq{eqmotinhomincomplete} rather than \eq{equmotioninhomsimplified4}. So, in the notation of this section, in the TT frame we have the equation of motion (\ref{eqmotinhomhijinside}) and
the boundary condition (\ref{bcDysoninhom}), that we rewrite here
\be\label{rhoddotuTTinhom}
\rho(\vx)\ddot{u}_i=\pa_i\[\lambda(\vx)\pa_ju_j\]
+\pa_j\[ \mu(\vx) (\pa_i u_j+\pa_j u_i+h_{ij})\]
\, ,
\ee
\be
\label{bcrhoddotuTT2}
n_j \[ \lambda(\vx)\delta_{ij}\pa_ku_k +\mu(\vx) (\pa_i u_j+\pa_j u_i+ h_{ij})\]=0\, .
\ee
In the proper detector frame  the equation of motion is given by \eq{rhoddotu}, which in the present notation reads
\be\label{rhoddotu2}
\rho(\vx) \ddot{u}_i^{\rm lab}=\pa_i\[\lambda(\vx)\pa_ju^{\rm lab}_j\]
+\partial_j\[ \mu(\vx)  (\partial_i u^{\rm lab}_j+\partial^{\rm lab}_j u_i)\] 
+\frac{1}{2}\rho(\vx) \ddot{h}_{ij}x_j\, .
\ee
and the boundary condition, in terms of the normal modes,  is  given by \eq{2bodybounduNinhom}, which in terms of $u^{\rm lab}_i(t,\vx) $ gives 
\be\label{bcrhoddotulabinhom}
n_j\[ \lambda(\vx)  \delta_{ij} \pa_j u^{\rm lab}_j +\mu (\vx) \( \pa_i u^{\rm lab}_j+\pa_ju^{\rm lab}_i\) \]=0\, .
\ee
To make \eqs{rhoddotuTTinhom}{bcrhoddotuTT2} consistent with
\eqs{rhoddotu2}{bcrhoddotulabinhom}, it is again sufficient to require that
\be\label{ddotulabipajm}
u^{\rm lab}_i(t,\vx)= u_i(t,\vx)+\frac{1}{2}h_{ij}(t)x_j\, .
\ee
Indeed, when inserting this into \eq{rhoddotu2}, on the left-hand side the term $(1/2)h_{ij}x_j$ in
\eq{ddotulabipajm} generates a  term $(1/2)\ddot{h}_{ij}x_j$, which  is needed to cancel that on the right-hand side of \eq{rhoddotu2}. At the same time, from \eq{ddotulabipajm} follows that
\be\label{paiulabpaiuhij}
\pa_i u^{\rm lab}_j+\pa_ju^{\rm lab}_i= \pa_i u_j+\pa_j u_i+ h_{ij}\, ,
\ee
which produces the correct  $\mu$-dependent term in \eq{rhoddotuTTinhom} (while, using $h_{ii}=0$, we have  
$\pa_i u^{\rm lab}_i=\pa_i u_i$ and therefore the $\lambda$-dependent terms also match). Similarly, inserting
\eq{ddotulabipajm} into \eq{bcrhoddotulabinhom} and using \eq{paiulabpaiuhij}, we get
\eq{bcrhoddotuTT2}, so also the boundary conditions match. \Eq{ddotulabipajm} therefore gives the correct transformation
for the variables describing the displacement from equilibrium of volume element of an elastic medium, between the TT frame and the proper detector frame, even for an inhomogeneous elastic medium. The result is extremely simple, and it just has the same form as the transformation of the coordinates between the two frames, \eq{xTTxlab}.
Once again, we recall that these results only hold at linear order in $\hmn$, and in the limit in which the wavelength of the GW is much larger than the length-scale over which we consider these relations, so that the spatial dependence of $\hmn$ can be neglected.\footnote{When this work was being completed, appeared on the arxiv the paper~\cite{Yan:2024jio}. In this paper it is correctly recognized that, to compare the field-theoretical computation of Dyson to the normal mode computation, one must take into account the transformation between TT frame and proper detector frame. We compare our approaches in App.~\ref{app:Yan}.}

\section{Summary and conclusions}\label{sect:Conclusions}

We finally summarize the main findings of this paper.
The coupling of elastic media to GWs is a classic problem, that has been studied at least since the 1960s, both for its relevance for resonant-mass detectors, and for its conceptual interest in General Relativity. As we have reviewed, the simplest approach consists in working in the proper detector frame, where the effect of a GW can be described as a Newtonian force in flat space, which drives the normal modes of the elastic body. This approach is appropriate when the GW has a frequency of the order of the fundamental mode of the system, but has intrinsic practical and conceptual limitations at high frequencies; conceptual, because   the Newtonian description is only valid when the wavelength of the GW is  much larger  that the size of the elastic body; and  practical, because at higher frequencies we would need to know the frequencies and damping times of the full set of highly excited normal modes. As we discussed, a first principle approach is rather obtained  working in the TT gauge (which has no intrinsic limit of validity, apart from the linearization in $\hmn$) and using a field-theoretical description of the dynamics of the elastic body.

In this paper we have developed a relativistic formulation of elasticity theory, using the methods of low-energy effective field theory, and we have used it to study the coupling of an elastic medium  to GWs. As we have  stressed, even when the elastic vibrations of a medium are non-relativistic (which  is normally the case, although with some exceptions, such as  in neutron stars), a relativistic elasticity theory is anyhow required to  couple of an elastic medium to GWs in full generality (i.e., away from the low-frequency limit where the proper detector frame description is adequate). This is a consequence of the fact   that the metric perturbation $\hmn$ is a relativistic, and in fact a general-relativistic quantity, which couples to a covariantly conserved energy--momentum tensor $\Tmn$, and the latter can be consistently determined only in the context of a relativistic elasticity theory. In this paper we have developed such a relativistic theory, building on the elegant EFT formalism developed for classical and quantum fluids in refs.~\cite{Dubovsky:2005xd,Endlich:2010hf,Endlich:2012pz,Nicolis:2013lma},
and we have used it to determine the corresponding coupling to GWs. An important aspect of the analysis has been the generalization to non-homogeneous elastic media since, as was already well-known in the literature, and as we have confirmed with our formalism, in the TT gauge GWs couple to elastic media only through inhomogeneities in the bulk, or through boundary discontinuities.

Our prototype action, for the isotropic but non-homogeneous case, is given by \eq{Swabcdisotropic} (although we have also studied the anisotropic case). As we have stressed, this is just an example of a relativistic action, and there is an infinite family of choices (within the constraints provided by the EFT symmetries). However, the study of the consequences of this action already allows us to understand the general structure of the results.
A main difference, compared to the ``naive" non-relativistic approach pioneered by Dyson~\cite{Dyson:1969zgf}, is that the basic action is now Lorentz-invariant, and the non-relativistic action emerges as a consequence of spontaneous symmetry breaking, i.e. of the fact that the basic field variable $\xi^a(x)$,  introduced in Section~\ref{sect:defxi}, which is a Lorentz scalar, is expanded around a symmetry-breaking background solution. In the homogeneous case this background solution is very simple, and given by \eq{xiaequilibrium}; however, in the inhomogeneous case, the background solution is more complicated, and involves at least a new function $\psi(\vx)$,  see \eq{xixpapsi}. Within the specific action that we have used, this function is determined in terms of the density $\rho(\vx)$ and the space-dependent Lam\'e  functions $\lambda(\vx)$ and $\mu(\vx)$. In particular, for the theory that we have considered, the relation is given in \eq{solpsi}, together with \eqst{rhoinhom}{muinhom}. However, if we do not want to tie our analysis to a specific Lorentz-invariant action, the function $\psi(\vx)$ must basically be taken as a new phenomenological function describing the elasticity theory, at the same level as 
$\rho(\vx)$, $\lambda(\vx)$ and $\mu(\vx)$. As a consequence, a first general result of our analysis is that the coupling of an inhomogeneous elastic medium to GWs does not depend only on $\mu(\vx)$, as first found in  \cite{Dyson:1969zgf}, but on both  $\mu(\vx)$ and $\psi(\vx)$. The exact form of this coupling, already for the case of small and slowly-varying inhomogeneities over a homogeneous background, is quite complicated, and is given in \eq{equmotioninhom}, that reduces to Dyson's result for  $\psi=0$, but otherwise has many extra terms. With some approximation, we can however use
\eqs{defnu1nu2}{equmotioninhomsimplified4}, to deal with a simpler expression.

Finally, we have stressed that, to compare the field theoretical results to those obtained in the proper detector frame, we must be careful to take into account the relation between the coordinates in the two frames. We found that the correct relation, even in the non-homogeneous case, is given by \eq{ddotulabipajm}.

In conclusion, the relativistic EFT formalism, applied to  elastic bodies, provides significant conceptual insights, allowing us to  see how the non-relativistic theory emerges through spontaneous symmetry breaking, and how the variables describing the perturbations around equilibrium position can be understood in the language of Goldstone bosons; at the same time, it provides a first-principle approach to the computation of the coupling to GWs, putting it on  a firmer footing, and providing modifications to the classic results given in \cite{Dyson:1969zgf}.

\vspace{5mm}
\noindent
{\bf Acknowledgements}. The work of M.M.  is supported by the
Swiss National Science Foundation (SNSF) grants 200020$\_$191957  and CRSII5$\_$213497, and by the SwissMap National Center for Competence in Research.  E.B. is supported by the SNSF
grant CRSII5$\_$213497.
MM thanks for the hospitality the IFAE  in Barcelona, where part of this work was done.
We thank Jan Harms for stimulating our interest in the problem, and Stefano Foffa, Francesco Iacovelli and Niccol\`o Muttoni for  discussions.

\appendix

\section{Geometric interpretation of $n(x)$}\label{app:n}

In this Appendix, following \cite{Magli:1992,Beig:2002pk,Wernig-Pichler:2006xdj,Hudelist:2022ixo}, we discuss a different expression for the quantity $n(x)$ introduced in \eqs{defnpq}{detB}, which emphasizes its geometrical nature and also provides an expression which is more convenient for explicit computations. In this appendix we work in a space-time ${\cal M}$ endowed with a generic metric $\gmn(x)$.

We begin by observing that, by definition, the map (\ref{mapXxi}) is constant on the world-line of a given molecule, since  $X^a$ identifies abstractly the molecule in question  so, as time $t$ progresses and the position $\vx(t)$ of a given molecule evolves, $\xi^a(t,\vx(t) )$ still has the same value $X^a$ that it had at an initial time, say $t=0$. 
In other words, the inverse image of points in ${\cal B}$ form a time-like congruence in ${\cal M}$.
This means that there exists a time-like vector field $v^{\mu}(x)$ such that
\be\label{vmupamxi}
v^{\mu}\pam \xi^a=0\, .
\ee
This vector field, which is called the velocity field of the elastic body, is uniquely specified by further fixing its normalization
\be\label{gmnvmvn}
\gmn v^{\mu}v^{\nu}=-1\, ,
\ee
and choosing its orientation as future-directed, $v^0\geq 0$. For instance, in the simple case of the static equilibrium configuration  given in \eq{xiaequilibrium}, we have  $\pam\xi^a=\delta_{\mu}^a$ and $v^{\mu}=(1,0,0,0)$. In relativistic elasticity theory, the function $\xi^a(x)$  is called a ``configuration", and $\pam\xi^a$ is  called the ``configuration gradient". Observe also that $v^{\mu}(x)$, defined by \eq{vmupamxi} together with the above normalization conditions, is actually a functional of the configuration $\xi^a(x)$.


In the material space ${\cal B}$ we use $\delta_{ab}$ as a metric, and we introduce the  three-form volume element,
\be
\omega = d\xi^1\wedge d\xi^2 \wedge d\xi^3=\frac{1}{3!}\eps_{abc}\, d\xi^a\wedge d\xi^b \wedge d\xi^c\, .
\ee
Since $\xi^a(x)$ defines a map from ${\cal M}$ to ${\cal B}$, we can associate to $\omega$ a three-form $\Omega$ in ${\cal M}$ (the ``pullback" of $\omega$) given by
\be
\Omega = \omega_{\mu\nu\rho}\,  dx^{\mu}\wedge dx^{\nu} \wedge dx^{\rho}\, ,
\ee
where
\be\label{omunurho}
\omega_{\mu\nu\rho}=\frac{1}{3!}\eps_{abc}\pam\xi^a\pan\xi^b\parho\xi^c=
\pam\xi^1\pan\xi^2\parho\xi^3
\, .
\ee
We next introduce the Levi-Civita tensor in ${\cal M}$. We denote by 
$\bar{\epsilon}^{\mu\nu\rho\sigma}$ the Levi-Civita symbol, defined by the conditions that $\bar{\epsilon}^{\mu\nu\rho\sigma}=1$ if $\mu\nu\rho\sigma$ is an even permutation of $0123$, $-1$ if it is an odd permutation, and zero if $\mu\nu\rho\sigma$ are not all different. In a generic curved space, 
$\bar{\epsilon}^{\mu\nu\rho\sigma}$ is not a tensor but a tensor density; the actual   Levi-Civita tensor
$\epsilon^{\mu\nu\rho\sigma}$ 
is given by  (see e.g. eq.~(8.10) of \cite{MTW} or Section~4.4 of \cite{Weinberg:1972kfs})
\be\label{epsilonsqrtgup}
\epsilon^{\mu\nu\rho\sigma}=\frac{1}{\sqrt{-g}}\, \bar{\epsilon}^{\mu\nu\rho\sigma}\, ,
\ee
where $g=|\det (\gmn)|$. Lowering all its indices with $\gmn$, one gets 
\be\label{epsilonsqrtgdown}
\epsilon_{\mu\nu\rho\sigma}\equiv g_{\mu\alpha}g_{\nu\beta}g_{\rho\gamma}g_{\sigma\delta}
\epsilon^{\alpha\beta\gamma\delta}=
-\sqrt{-g}\, \bar{\epsilon}_{\mu\nu\rho\sigma}\, ,
\ee
where $\bar{\epsilon}_{\mu\nu\rho\sigma}$ is {\em defined} so that, numerically, 
$\bar{\epsilon}_{\mu\nu\rho\sigma}=\bar{\epsilon}^{\mu\nu\rho\sigma}$, i.e. 
$\bar{\epsilon}_{0123}=+1$.\footnote{Note that $\bar{\epsilon}_{\mu\nu\rho\sigma}$ is not obtained lowering from  $\bar{\epsilon}^{\mu\nu\rho\sigma}$ lowering the indices with $\emn$, otherwise we would get $\bar{\epsilon}_{0123}=-\bar{\epsilon}^{0123}=-1$. In order not to fall into the mistake of raising and lowering the indices of the Levi-Civita symbol with $\emn$, ref.~\cite{MTW} denotes both
$\bar{\epsilon}^{\mu\nu\rho\sigma}$ and $\bar{\epsilon}_{\mu\nu\rho\sigma}$ as $[\mu\nu\rho\sigma]$. Here we will rather use the notation $\bar{\epsilon}^{\mu\nu\rho\sigma}$ and $\bar{\epsilon}_{\mu\nu\rho\sigma}$, keeping however in mind the above caveat.}
We now define the quantity
\bees
J^{\mu}&\equiv& \epsilon^{\mu\nu\rho\sigma} \omega_{\nu\rho\sigma}\nn\\
&=&\frac{1}{\sqrt{-g}}\, \bar{\epsilon}^{\mu\nu\rho\sigma} \pan\xi^1\parho\xi^2\pas\xi^3\, .
\label{Jmuepsilon}
\ees
By construction this quantity is a four-vector, since $\epsilon^{\mu\nu\rho\sigma}$ and $ \omega_{\nu\rho\sigma}$ are tensors. Its covariant divergence is therefore given by
\bees
\n_{\mu} J^{\mu}&=&\frac{1}{\sqrt{-g}}\, \pam\( \sqrt{-g}\, J^{\mu} \)\nn\\
&=&\frac{1}{\sqrt{-g}}\, \bar{\epsilon}^{\mu\nu\rho\sigma}  \pam\( \pan\xi^1\parho\xi^2\pas\xi^3\)\, ,
\ees
and this vanishes because, e.g. $\pam\pan\xi^1$ is symmetric in $\mu,\nu$ and therefore gives zero when contracted with $\bar{\epsilon}^{\mu\nu\rho\sigma}$, and similarly for the other terms generated by the action of $\pam$ on  $\pan\xi^1\parho\xi^2\pas\xi^3$.  Therefore
\be
\n_{\mu} J^{\mu}=0\, ,
\ee
so $J^{\mu}$ is a conserved current (note that it is conserved for purely geometrical reasons, without the need of using the equations of motion).\footnote{In a more geometric language, the vanishing of the divergence of $J^{\mu}$  is due to the fact that the exterior derivative of the one-form associated to $J_{\mu}$ is the pullback of the exterior derivative of $\omega$, and this vanishes since $\omega$ is a three-form in a three-dimensional space~\cite{Magli:1992}.} We also observe that
\be
J^{\mu}\pam\xi^a=0\, ,
\ee
since, in $ \pam\xi^a\pan\xi^1\parho\xi^2\pas\xi^3$, the index $a$ is necessarily equal to one of the other indices 1, 2 or 3;  if, e.g.,  $a=1$, then $ \pam\xi^a\pan\xi^1\parho\xi^2\pas\xi^3$ is symmetric in $(\mu,\nu)$, for $a=2$,
it is symmetric in $(\mu,\rho)$, and for $a=3$ it is symmetric in $(\mu,\sigma)$; in all cases, the contraction with $ \bar{\epsilon}^{\mu\nu\rho\sigma}$ gives zero. Comparison with \eq{vmupamxi} then shows that $J^{\mu}$ and $v^{\mu}$ are parallel, and therefore exists a function $n(x)$ such that
\be\label{Jmunvmu}
J^{\mu}(x)= n(x) v^{\mu}(x)\, .
\ee
Since $J^{\mu}(x)$ and $v^{\mu}(x)$ are both four-vector fields, $n(x)$ is a Lorentz scalar.
We have called this function $n(x)$ since, indeed, is the same as the function $n(x)$ defined in  
\eqs{detB}{defnpq}. This can be shown observing, from \eq{gmnvmvn}, that
\be\label{n2gmn}
n^2=-\gmn J^{\mu} J^{\nu}= -J^{\mu} J_{\mu}\, .
\ee
From \eqst{epsilonsqrtgup}{Jmuepsilon}, and writing $J_{\mu}= \epsilon_{\mu\nu\rho\sigma} \omega^{\nu\rho\sigma}$,
we then have
\bees
n^2&=&-J^{\mu}J_{\mu}\nn\\
&=&-\epsilon^{\mu\nu\rho\sigma}\omega_{\nu\rho\sigma}
\epsilon_{\mu\alpha\beta\gamma}\omega^{\alpha\beta\gamma} \nn\\
&=&+\bar{\epsilon}^{\mu\nu\rho\sigma}\,  \bar{\epsilon}_{\mu\alpha\beta\gamma}\, \omega_{\nu\rho\sigma}\omega^{\alpha\beta\gamma}
\ees
We now use the identity
\be
\bar{\epsilon}^{\mu\nu\rho\sigma}\,  \bar{\epsilon}_{\mu\alpha\beta\gamma}
= \left|
\begin{array}{ccc}
\delta^{\nu}_{\alpha} & \delta^{\nu}_{\beta} & \delta^{\nu}_{\gamma}\\
\delta^{\rho}_{\alpha} & \delta^{\rho}_{\beta} & \delta^{\rho}_{\gamma}\\
\delta^{\sigma}_{\alpha} & \delta^{\sigma}_{\beta} & \delta^{\sigma}_{\gamma}\\
\end{array}
\right| \nn\\
=
\delta^{\nu}_{\alpha}\delta^{\rho}_{\beta}\delta^{\sigma}_{\gamma}
-\delta^{\nu}_{\alpha} \delta^{\rho}_{\gamma} \delta^{\sigma}_{\beta}
- \delta^{\nu}_{\beta}\delta^{\rho}_{\alpha}\delta^{\sigma}_{\gamma}
+\delta^{\nu}_{\beta} \delta^{\rho}_{\gamma}\delta^{\sigma}_{\alpha}
+\delta^{\nu}_{\gamma}\delta^{\rho}_{\alpha}\delta^{\sigma}_{\beta}
-\delta^{\nu}_{\gamma}\delta^{\rho}_{\beta} \delta^{\sigma}_{\alpha} \, .
\ee
Then, using the explicit expression of $ \omega_{\mu\nu\rho}$ given in \eq{omunurho}, we get
\be
n^2=  \( \delta^{\nu}_{\alpha}\delta^{\rho}_{\beta}\delta^{\sigma}_{\gamma}
-\delta^{\nu}_{\alpha} \delta^{\rho}_{\gamma} \delta^{\sigma}_{\beta}
- \delta^{\nu}_{\beta}\delta^{\rho}_{\alpha}\delta^{\sigma}_{\gamma}
+\delta^{\nu}_{\beta} \delta^{\rho}_{\gamma}\delta^{\sigma}_{\alpha}
+\delta^{\nu}_{\gamma}\delta^{\rho}_{\alpha}\delta^{\sigma}_{\beta}
-\delta^{\nu}_{\gamma}\delta^{\rho}_{\beta} \delta^{\sigma}_{\alpha}  \)
\pan\xi^1\parho\xi^2\pas\xi^3
\pa^{\alpha}\xi^1\pa^{\beta}\xi^2\pa^{\gamma}\xi^3
.
\ee
Performing the contractions we see that this expression is the same as
\be
\det(\pam\xi^a\paM\xi^b)=
 \left|
\begin{array}{ccc}
\pam\xi^1\paM\xi^1 \hspace*{1mm}& \pam\xi^1\paM\xi^2 \hspace*{1mm}& \pam\xi^1\paM\xi^3 \\
\pam\xi^2\paM\xi^1 \hspace*{1mm}& \pam\xi^2\paM\xi^2 \hspace*{1mm}& \pam\xi^2\paM\xi^3 \\
\pam\xi^3\paM\xi^1 \hspace*{1mm}& \pam\xi^3\paM\xi^2 \hspace*{1mm}& \pam\xi^3\paM\xi^3
\end{array}
\right| \, .\nn\\
\ee
Therefore $n$, defined from \eq{Jmunvmu}, satisfies 
\be
n^2 =\det(B)\, ,
\ee 
and (together with the condition that $J^{\mu}$ and $v^{\mu}$ are both future-directed, so that $n>0$) is therefore the same as $n$ defined in \eq{defnpq}.

The expression (\ref{n2gmn}) is also convenient to perform the linearization of $n(x)$, whose result we gave in \eq{nvsu}. To this purpose, we now restrict to $\gmn=\emn$. We begin by expanding $J^{\mu}$ up to quadratic order in $u^i$, inserting \eq{xiaandui} into \eq{Jmuepsilon}. We write
\be
J^{\mu}=J^{\mu}_{(0)}+ J^{\mu}_{(1)}+J^{\mu}_{(2)}\, ,
\ee
where the lower index denotes the order in the expansion in  $\pa u$. The first two terms are easily computed,
\be\label{Jmu0and1}
J^{\mu}_{(0)}=\delta^{\mu}_0\, ,\qquad J^{\mu}_{(1)}=(-\n{\bf\cdot}\vu, \dot{\vu})\, .
\ee
Before computing $J^{\mu}_{(2)}$ we observe that, since $J^{\mu}_{(0)}=\delta^{\mu}_0$, up to quadratic order
\be\label{JmuJMu}
n^2 = -\emn J^{\mu} J^{\nu}=1+2 J^{0}_{(1)}+2 J^{0}_{(2)}-\emn J^{\mu}_{(1)} J^{\nu}_{(1)}\, .
\ee
Therefore, to quadratic order we only need the $\mu=0$ component of $J^{\mu}_{(2)}$, which is given by
\be\label{Jmu2}
J^{0}_{(2)} = \frac{1}{2}\[ (\n{\bf\cdot}\vu)^2-\pa_iu_j\pa_ju_i\]\, .
\ee
Inserting \eqs{Jmu0and1}{Jmu2} into \eq{JmuJMu} and taking the square root we then obtain \eq{nvsu}.

\section{Equations of motion for inhomogeneous elastic bodies}\label{app:contoeqm}

In this appendix we perform explicitly the computation leading to the equations of motion (\ref{eqmfinal}).  Inserting the Lagrangian (\ref{Lalphabetagamma}) into \eq{eqmotionpaj} we get
\bees
0&=& -\frac{\pa\alpha}{\pa\xi^a} (\det B)^{1/2} -\frac{\pa\beta}{\pa\xi^a}   ({\rm Tr}\,  B)^2 
  -\frac{\pa\gamma}{\pa\xi^a} \,  {\rm Tr} (B^2)\label{eqmotinhom1}\\
 && +\pa_j \[ \alpha(\xi) \frac{1}{2}(\det B)^{-1/2}\frac{\d\det B}{\delta(\pa_j\xi^a)}
  +2 \beta(\xi)  {\rm Tr}\,  B  \frac{\d {\rm Tr}\,  B}{\delta(\pa_j\xi^a)}  +
  \gamma(\xi) \frac{\d {\rm Tr} (B^2)}{\delta(\pa_j\xi^a)} 
  \]\, .\nn
  \ees
The variations are computed observing that, in the static situations that we are considering,
\eq{defBab} reduces to
\be\label{Babpahxi}
B^{ab} =\pa_k\xi^a\pa_k\xi^b\, ,
\ee
(with the sum over the spatial index $k$ understood). Then
\be
\frac{\d B^{cd}}{\delta(\pa_j\xi^a)}  =\delta^{ac}\pa_j\xi^d+\delta^{ad}\pa_j\xi^c\, .
\ee
The variation of the terms in the second line in \eq{eqmotinhom1} are computed using
\be
\frac{\d\det B}{\delta(\pa_j\xi^a)}= \frac{\d\det B}{\delta B^{cd}}\frac{\d B^{cd}}{\delta(\pa_j\xi^a)}\, ,
\ee
and similarly for ${\rm Tr}\,  B $ and ${\rm Tr} (B^2)$, together with
\bees
\frac{\d\det B}{\delta B^{cd}}&=& (\det B) (B^{-1})_{cd}\, ,\label{vardetB}\\
 \frac{\d\, {\rm Tr}\,  B}{\delta B^{cd}}&=& \delta_{cd}\, ,\\
 \frac{\d{\rm Tr} (B^2)}{\delta B^{cd}}&=&2 B_{cd}\, .
 \ees
 This gives
\bees
&&\hspace*{-5mm}\pa_j \[ \alpha(\xi) (\det B)^{1/2} (B^{-1})_{ab}\pa_j\xi^b+4 \beta(\xi)  ({\rm Tr}\,  B)  \pa_j\xi^a  +
4 \gamma(\xi) B_{ab}\pa_j\xi^b\] \nn\\
&&=   \frac{\pa\alpha}{\pa\xi^a} (\det B)^{1/2} +\frac{\pa\beta}{\pa\xi^a}   ({\rm Tr}\,  B)^2 
  +\frac{\pa\gamma}{\pa\xi^a} \,  {\rm Tr} (B^2)\label{eqmotinhom2}\, .
\ees
A useful simplification takes place developing the derivative in the first term as
\be
\pa_j \[ \alpha(\xi) (\det B)^{1/2} (B^{-1})_{ab}\pa_j\xi^b \]=
(\pa_j\alpha) (\det B)^{1/2} (B^{-1})_{ab}\pa_j\xi^b+\alpha \pa_j\[ (\det B)^{1/2} (B^{-1})_{ab}\pa_j\xi^b \]\, .
\ee
Since the dependence of $\alpha$ on $\vx$ is through $\xi^a(\vx)$, we have
\be
\pa_j\alpha =\frac{\pa\alpha}{\pa \xi^c}\pa_j\xi^c\, .
\ee
Then,  
\bees
(\pa_j\alpha) (\det B)^{1/2} (B^{-1})_{ab}\pa_j\xi^b
&=&\frac{\pa\alpha}{\pa \xi^c} (\det B)^{1/2} (B^{-1})_{ab}\pa_j\xi^b\pa_j\xi^c\nn\\
&=&\frac{\pa\alpha}{\pa \xi^c} (\det B)^{1/2} (B^{-1})_{ab} B_{bc}=\frac{\pa\alpha}{\pa \xi^c} (\det B)^{1/2}\delta_{ac}\nn\\
&=&\frac{\pa\alpha}{\pa \xi^a} (\det B)^{1/2}
\ees
where, in the second line, we made use of  \eq{Babpahxi}.
This term cancels exactly the first term on the right-hand side of \eq{eqmotinhom2}, so the dependence on the spatial derivative of $\alpha$ disappears. Expanding similarly the other two terms on the left-hand side of \eq{eqmotinhom2}, we get different structures that do not cancel agains the terms in the right-hand side, and we finally obtain \eq{eqmfinal} of the main text, that we repeat here,
\bees
&&\alpha(\xi) \pa_j\[ (\det B)^{1/2} (B^{-1})_{ab}\pa_j\xi^b \] 
+4\beta(\xi)\pa_j\(  {\rm Tr}\,  B\, \pa_j\xi^a\) 
+4\gamma(\xi)\pa_j\(  B_{ab}\pa_j\xi^b\) \nn\\
&&=\frac{\pa\beta}{\pa\xi^b}   {\rm Tr}\,  B \[ \delta_ {ab}{\rm Tr}\,  B-4B_{ab}\]
+\frac{\pa\gamma}{\pa\xi^b}    \[\delta_ {ab}{\rm Tr} (B^2) - 4 B_{ac}B_{cb}\]\, .
\label{eqmfinalApp}
\ees
To linearize it, we write $\xi^a(\vx)$ as in \eq{xiofxlin},
$\xi^a(x)=\delta_i^a \xi_i(\vx)$, where
$\xi_i(\vx)= x_i + \eps_i(\vx)$. Then 
\be
\pa_j\xi_i=\delta_{ij}+\pa_j\eps_i\, .
\ee
We also write $B_{ij}=\delta_i^a \delta_j^b B_{ab}$ for easy of notation. Then,
to linear order in $\pa_j\eps_i$, we have
\bees
B_{ij}&=&\d_{ij}+(\pa_i\eps_j+\pa_j\eps_i)\, ,\\
(B^{-1})_{ij}&=&\d_{ij}-(\pa_i\eps_j+\pa_j\eps_i)\, ,\\
\det B&=& 1+2\pa_i\eps_i\, ,\\
{\rm Tr}\, B&=& 3+2\pa_i\eps_i\, ,\\
{\rm Tr}\, B^2&=& 3+4\pa_i\eps_i\, .
\ees
Inserting these expressions  into \eq{eqmfinalApp}, we get \eq{eqepslin0} of the main text.

\section{Solutions in the proper detector frame as a sum over normal modes}\label{app:NMsolutions}

In this appendix we provide a more explicit discussion of the solution of
\eq{eqforxiNdiss}, both in frequency space and in time domain, in a way that also allows us to study the limit of large frequencies (still, within the limit of validity of the detector-frame description).
\Eq{eqforxiNdiss} is solved writing
\be\label{defFourier}
\xi_N(t)=\int_{-\infty}^{\infty}\, \frac{d\omega}{2\pi}\, \tilde{\xi}_N(\omega) e^{-i\omega t}\, ,
\ee
so that \eq{eqforxiNdiss} becomes
\be
(-\omega^2-i\omega\gamma_N+\omega_N^2) \tilde{\xi}_N(\omega) =
-\omega^2 \kappa_{N,ij} \tilde{h}^{\rm TT}_{ij}(\omega)\, .
\ee
The response  of the $N$-th normal mode  in Fourier space is therefore given by
\be\label{tildexiN1}
\tilde{\xi}_N(\omega)=\frac{\omega^2 }{\omega^2+i\omega\gamma_N-\omega_N^2}
 \kappa_{N,ij} \tilde{h}^{\rm TT}_{ij}(\omega)\, .
\ee
It is convenient to introduce the quality factor of the $N$-th normal mode, $Q_N\equiv \omega_N/\gamma_N$, and  express the result in terms of   $f=\omega/2\pi$, $f_N=\omega_N/(2\pi)$. Then, writing $\tilde{\xi}_N(\omega)$ as $\tilde{\xi}_N(f)$ and $\tilde{h}^{\rm TT}_{ij}(\omega)$ as $\tilde{h}^{\rm TT}_{ij}(f)$, \eq{tildexiN1} becomes
\be\label{xiNfQN}
\tilde{\xi}_N(f)=\frac{f^2 }{f^2-f_N^2+i\frac{f f_N}{Q_N} }\, 
 \kappa_{N,ij} \tilde{h}^{\rm TT}_{ij}(f)\, ,
\ee
and, from \eq{vusumNuN},\footnote{To compare with eq.~(1) of \cite{LGWA:2020mma}, we observe that, there, the term $ff_N/Q_N$ has been written as $f_N^2/Q_N$, which is only valid close to the resonance. The different sign in front of this term must be due to a different sign convention on the Fourier transform, with respect to our definition (\ref{defFourier}). We have also included a generic polarization structure in $\tilde{h}^{\rm TT}_{ij}(f)$, rather than writing it as a scalar quantity $\tilde{h}(f)$, which results in a corresponding tensorial structure in $ \kappa_{N,ij}$.}
\be\label{vusumN}
\tilde{\vu}(f,\vx)=f^2 \tilde{h}^{\rm TT}_{ij}(f) \sum_N
\frac{\kappa_{N,ij}\, \vu_N(\vx) }{f^2-f_N^2+i\frac{f f_N}{Q_N} }\, ,
\ee
corresponding to a transfer function
\be
T_{ij}(f)=f^2\, \sum_N
\frac{\kappa_{N,ij}\, \vu_N(\vx) }{f^2-f_N^2+i\frac{f f_N}{Q_N} }\, .
\ee
Before proceeding further, let us stress again the limit of validity of this computation. If the external perturbation were not a GW, but an actual Newtonian force, as in the most typical situations considered in the theory of elasticity, the above expression would be valid in principle at all frequencies, and we could use it to study the limit $f\ra\infty$. This limit is not trivial mathematically because $f$ appears inside the sum, where also $f_N$ appears, and for arbitrarily large $N$ we have $f_N\ra\infty$, so, even for large $f$, one cannot set a priori $f\gg f_N$ inside the sum. However, we can observe that the coupling $\kappa_{N,ij}$ involves a spatial integral over the mode function
$\vu^*_{N} (\vx)$, see \eq{kappaNij}, which, for highly excited modes, is fast oscillating. Therefore, we expect that highly excited modes do not really contribute, and 
\eq{vusumN} can be approximated with a sum over a finite number of modes, that we generically indicate as a sum up to $N_{\max}$,
\be\label{vusumNmax}
\tilde{\vu}(f,\vx)=f^2 \tilde{h}^{\rm TT}_{ij}(f) \sum_{N=0}^{N_{\rm max}}
\frac{\kappa_{N,ij}\, \vu_N(\vx) }{f^2-f_N^2+i\frac{f f_N}{Q_N} }\, ,
\ee
(where we denote generically by $N=0$ the fundamental mode) with $N_{\rm max}$ chosen so that the inclusion of higher modes 
do not affect the result, within a chosen accuracy.  In this case, the high-frequency limit $f\ra\infty$ is easily taken, and gives~\cite{LGWA:2020mma}
\be\label{highf-limit}
\tilde{\vu}(f,\vx)\simeq \tilde{h}^{\rm TT}_{ij}(f)
\sum_{N=0}^{N_{\rm max}} \kappa_{N,ij}\, \vu_N(\vx) \, ,
\ee
so the transfer function becomes flat in frequency.
More precisely, we expect this to hold for $f\gg f_{N_{\rm max}}$, where $ f_{N_{\rm max}}$ is the frequency of the highest normal modes that need to be included in the sum, within the accuracy requested.
However, we must recall that our external perturbation is not a Newtonian force but a GW, and the interaction of  a GW with a test mass admits a description as a Newtonian force only  if we work in the proper detector frame, and as long as its wavelength  satisfies $\lbar\gg L$, i.e. $f\ll c/(2\pi L)$, where $L$ is the typical linear size of the elastic body. Therefore \eq{highf-limit} is only valid for frequencies $f$ such that
\be
f_{N_{\rm max}}\ll f\ll \frac{c}{2\pi L}\, ,
\ee
as long as this range is non-empty, i.e. as long as $f_{N_{\rm max}}\ll c/(2\pi L)$.

It is also interesting to study the solution in time domain.
The solution for $\xi_N(t)$ could be obtained Fourier-transforming back \eq{xiNfQN}, but we find 
more instructive to directly solve  \eq{eqforxiN} or \eq{eqforxiNdiss} in the time domain. Let us start from \eq{eqforxiN}.
We introduce the Green's functions $G_N(t)$ of the harmonic oscillator with frequency $\omega_N$, defined by
\be\label{ddotGN}
\ddot{G}_N+\omega_N^2 G_N(t)=\delta(t)\, ,
\ee
so that 
\be
\xi(t)=\kappa_{N,ij} \int_{-\infty}^{\infty} dt\, G_N(t-t')\ddot{h}_{ij}^{\rm TT}(t')\, ,
\ee
and we select the retarded Green's function. To find the retarded Green's function we Fourier transform, writing
\be
G_N(t)=\int_{-\infty}^{\infty}\, \frac{d\omega}{2\pi}\, \tilde{G}_N(\omega) e^{-i\omega t}\, .
\ee
Then \eq{ddotGN} gives
$(-\omega^2+\omega_N^2) \tilde{G}_N(\omega)=1$,
so 
\be
\tilde{G}_N(\omega)=\frac{1}{\omega_N^2-\omega^2}\, .
\ee
The retarded Green's function is obtained moving both poles at $\omega=\pm\omega_N$ in the lower complex plane and Fourier-transforming back to get $G(t)$, which gives
\be
G_N(t)=\frac{1}{\omega_N}\theta(t) \sin(\omega_Nt)\, ,
\ee
where $\theta(t)$ is the Heaviside theta function, $\theta(t)=1$ for $t>0$ and $\theta(t)=0$ for $t<0$.
Then, the solution of \eq{eqforxiN} (with the boundary conditions that the modes are not excited before the arrival of the perturbation, i.e. $\xi_N(t)\ra 0$ for $t\ra -\infty$) is 
\be
\xi_N(t)=\frac{\kappa_{N,ij}}{\omega_N}
\int_{-\infty}^t dt'\, \sin[\omega_N(t-t')]\,  \ddot{h}_{ij}^{\rm TT}(t') \, .
\ee
Observe that the high-frequency modes are less and less coupled to the external perturbation, because of the factor $1/\omega_N$ in this expression, as well as because of the factor $\sin[\omega_N(t-t')]$, which oscillates faster and faster, and because of the fact, already observed above, that   $\kappa_{N,ij}$ vanishes for large $N$.

Including now dissipation, \eq{eqforxiNdiss} can be solved  writing 
\be
\xi_N(t)=e^{-\gamma_N t/2}\chi_N(t)\, .
\ee
Then, $\chi_N(t)$ satisfies the equation
\be\label{eqforxiNdisschi}
\ddot{\chi}_N+\tilde{\omega}_N^2\chi_N=\kappa_{N,ij}\, e^{\gamma_N t/2}\ddot{h}_{ij}^{\rm TT}(t) \, ,
\ee
where
\be\label{defomegatilde}
\tilde{\omega}_N^2=\omega_N^2-\frac{\gamma_N^2}{4}\, .
\ee
\Eq{eqforxiNdisschi} can be solved as before, using the Green's function $G_N(t-t')$. Then,   re-expressing the result in term of $\xi_N(t)$, we  get
\bees
\xi_N(t)&=&\frac{\kappa_{N,ij}}{\tilde{\omega}_N}
 \int_{-\infty}^t dt'\, e^{-\gamma_N (t-t')/2} \sin[\tilde{\omega}_N(t-t')]\, \ddot{h}_{ij}^{\rm TT}(t') \\
&=&\frac{\kappa_{N,ij}}{\tilde{\omega}_N}   \int_{-\infty}^t dt'\, 
{\rm Im}\, \[ e^{i \(\tilde{\omega}_N+i\frac{\gamma_N}{2}\) (t-t')} \] \, \ddot{h}_{ij}^{\rm TT}(t')
 \, ,
\ees
where the rewriting in the last line emphasizes that  $\tilde{\omega}_N$ and $\gamma_N$ combine into a complex frequency $\tilde{\omega}_N+i\gamma_N/2$.
The  solution for $\vu(t,\vx)$ is then obtained inserting this expression into \eq{vusumNuN},
\be\label{solvutimedomdiss}
\vu(t,\vx)=
\int_{-\infty}^t dt'\,  \ddot{h}_{ij}^{\rm TT}(t')
 \sum_N \[ \frac{\kappa_{N,ij}\, \vu_N(\vx) }{\tilde{\omega}_N}
 e^{-\gamma_N (t-t')/2} \sin[\tilde{\omega}_N(t-t')]  \]
 \, .
\ee
Let us stress that this result has been obtained by performing a Fourier transform, which involves an integral over arbitrary large frequencies. Once again, our treatment of GWs as a force is only valid for $f\ll c/(2\pi L)$. Therefore, \eq{solvutimedomdiss} is only valid if the contribution to the time-domain result from the high frequency modes is negligible, i.e. if it is dominated by the
modes with frequencies such that 
$f_N\ll c/(2\pi L)$. The argument discussed below \eq{vusumN}, however, indicates that, within a given required accuracy,  the sum can  be truncated to a finite $N_{\rm max}$, so 
\eq{solvutimedomdiss}  is a good approximation to the exact result if $f_{N_{\rm max}}\ll c/(2\pi L)$.

It is interesting to observe  that, when we include dissipation, $\omega_N$ is replaced by $\tilde{\omega}_N$, defined in \eq{defomegatilde}. For the long-lived modes this is a minor change, since
$\gamma_N\ll \omega_N$. However, in general this is no longer true for the highly excited modes, for which $\omega_N$ is large but also $\gamma_N$ is large, i.e. the decay time $\tau_N=2/\gamma_N$ is small, so 
$\omega_N$ and $\gamma_N$ could in principle become comparable. Then, for the highly excited modes,
\eq{defomegatilde} could even give a negative value for $\tilde{\omega}^2_N$, i.e. a purely imaginary expression for $\tilde{\omega}_N$, which basically signals a breakdown of the formalism based on the normal mode expansion, when supplemented by the simple prescription (\ref{eqforxiNdiss}) to account for dissipation.

\section{Comparison with Dyson's result}\label{sect:Dyson}

Dyson, in ref.~\cite{Dyson:1969zgf}, starts from the Lagrangian density (\ref{LDyson})  and states, without derivation, that it gives rise to a conserved energy--momentum tensor, whose components are\footnote{In this appendix we use units $c=1$, as we did in  Section~\ref{sect:EFT_Rel}.}
\bees
T^{\rm Dyson}_{00}&=&
\frac{1}{2}\rho \dot{u}_i^2 
+ \frac{1}{2} c_{ijkl}\pa_i u_j\pa_k u_l \, , \label{T00Dyson}\\
T^{\rm Dyson}_{0i}&=&T^{\rm Dyson}_{i0}=
\rho \dot{u}_i\, ,\label{T0iDyson}\\
T^{\rm Dyson}_{ij}&=&-c_{ijkl}\pa_ku_l\, . \label{TijDyson}
\ees
He then couples this energy--momentum tensor to GWs, adding a term $-(1/2)\hMN\Tmn$ to the Lagrangian density. This results has become quite standard in the literature, and we find  useful to look at it closely.  

First of all, it is puzzling that a symmetric Lorentz--covariant energy--momentum tensor could emerge from the non-relativistic Lagrangian used by Dyson, see the discussion of point (3) on page~\pageref{point3}.   
Comparing with the result that we obtained with our fully covariant formalism, we see that, in the homogeneous case,  Dyson's result is almost correct, but there are differences. Comparing  \eqs{T00Dyson}{T00final} we see that $T_{00}$ is similar; in our result we find an  extra term (the term proportional to $\tilde{w}_{jk}$) that is of the same order as the $c_{ijkl}$ term and therefore, in a generic anisotropic setting, cannot be neglected;  however, being proportional to $\tilde{w}_{jk}$, this term  vanishes in the isotropic limit. Quite importantly,
our result also includes a total derivate term $\pa_i K_i$; the latter does not contribute to the total energy $\int d^3x\, T^{00}$;  however, when considering the coupling of $\Tmn$  to a Newtonian gravitational field with a non-vanishing component $h_{00}$,  this terms  produces a non-vanishing contribution, which in this case would in fact be the leading contribution, since it is linear in $u$. Furthermore, as discussed below \eq{T0ihom}, this term is essential to ensure the conservation equation $\pa_0 T^{00}+\pa_i T^{0i}=0$ at linear order in $u$. Without this term, the $\nu=0$ component of the conservation equation $\pam T^{\nu\mu}=0$ is not satisfied.\footnote{Interestingly, Dyson must have realized that, with his energy--momentum tensor, there was a problem with the $\nu=0$ component of the conservation equation,  since  in his eq.~(2.11) he only writes $\pam T^{j\mu}=0$. However, 
to obtain a consistent coupling to GWs, the energy--momentum tensor must be covariantly conserved, $\pam T^{\nu\mu}=0$, so that, after integration by parts, a term in the action proportional to $\hmn\TMN$ is invariant under linearized diffeomorphisms, $\hmn\ra\hmn-(\pam\theta_{\nu}+\pan\theta_{\mu})$.}

Comparing \eqs{T0iDyson}{T0ifinal} we see that for the  component $T_{0i}$ we agree except for the sign, that we believe is incorrect in \cite{Dyson:1969zgf}. Indeed, as we already saw below  \eq{T0ihom},
our sign, together with the sign of $T^{00}$ (which is the same as in Dyson), is necessary to satisfy the 
$\nu=0$ component of conservation equation $\pam T^{\nu\mu}=0$, i.e.
$\pa_0 T^{00}+\pa_i T^{0i}=0$ [where $T^{00}$ includes the $\pa_iK_i$ term, \eq{T00final}]; furthermore, as we will see below, it is also required to satisfy the 
$\nu=i$ component of conservation equation, i.e. $\pa_0 T^{i0}+\pa_j T^{ij}=0$.

For $T_{ij}$, comparing   \eqs{TijDyson}{Tijfinal} we see that we agree on the term $-c_{ijkl}\pa_ku_l $, but we also have extra terms. Again, these terms must be included in the general anisotropic case, but vanish in the isotropic limit. We therefore agree on the isotropic limit given by \eq{Tijlambdamu}.

Eventually, this energy--momentum tensor is coupled to GWs and, in the field--theoretical approach, one  chooses the TT gauge. Then, from this point of view, in the end only $T_{ij}$ is relevant for the coupling to GWs, so the results obtained in the literature  in the isotropic and homogeneous case using Dyson's expression (\ref{Tijlambdamu}) are correct.\footnote{When comparing with the results of Dyson, or when using them, one must be aware that he defines
$\hmn$ from $\gMN=\eMN+\hMN+{\cal O}(h^2)$, see his eq.~(2.14). In contrast, we use
$\gmn=\emn+\hmn+{\cal O}(h^2)$, so, for us, $\gMN=\eMN-\hMN+{\cal O}(h^2)$. Therefore his definition of $\hmn$ differs from ours by a sign. For this reasons, 
he correctly writes the coupling to GWs as $(-1/2)\hMN\Tmn$, see his eq.~(2.23), while we write
$(+1/2)\hMN\Tmn$.\label{foot:signhmn}
}
However, as we repeatedly stressed, when studying GWs we need to consider also the inhomogeneous situation, since GWs couple to an elastic medium either through discontinuities at the boundary, or to bulk inhomogeneities. In ref.~\cite{Dyson:1969zgf} this is performed simply promoting $\rho$ and $c_{ijkl}$ (or, in the isotropic case, $\mu$ and $\lambda$) to function of $\vx$. Our formalism shows that the full result is significantly more complicated, see \eqst{T00inhomo}{Tijinhom}.

Given that our results differ already in the homogeneous case, it is interesting to try to understand how Dyson might have got his result. In ref.~\cite{Dyson:1969zgf}, Dyson states that his result  comes from Noether's theorem applied to the non-relativistic Lagrangian that he considers, but does not provide any explicit detail.\footnote{Apart for the rather cryptic statement: 
{\em ``If we applied the usual
field-theoretical recipe (Wentzel 1949) to the Lagrangian ${\cal L}$, considering $z^j(x)$ as a
classical field, we would obtain the wrong energy-momentum tensor. The usual procedure
is based on translation invariance applied to the coordinates $x_j$, and is invalid in a situation
where the coordinates $x_j$ are anchored to a particular object."}, see below his eq.~(2.13). This presumably refers to the fact that we must use the translations in field space rather than in space-time,  which we will indeed do next.}
Anyhow, let us   compute all  conserved currents associated to the Lagrangian density (\ref{LDyson}), restricting for simplicity to the homogeneous case, and compare with \eqst{T00Dyson}{TijDyson}. We have already done this exercise in Section~\ref{sect:symmetriesNR}, where we found that the global shift symmetry (\ref{transu}) gives rise to three conserved currents
$j_i^{\mu}$, with $i=1,2,3$, and, from \eqss{ji0}{jij}{Sijexpl},
\be\label{j0jjDyson}
j^0_{i}=-\rho \dot{u}_i\, ,\qquad
j^j_i=c_{ijkl}\pa_ku_l\, .
\ee
The conservation equation $\pam j^{\mu}_i=0$  then reads
\be\label{patpaj0}
\pa_0\( -\rho \dot{u}_i\)+\pa_j\( c_{ijkl}\pa_ku_l \)=0\, ,
\ee
that we rewrite as 
\be\label{patpaj}
\pa_0\( \rho \dot{u}_i\)+\pa_j\( -c_{ijkl}\pa_ku_l \)=0\, .
\ee
Dyson uses the signature $\emn=(-,+,+,+)$, as we do, and therefore
$ (T^{\rm Dyson})^{0i}=- (T^{\rm Dyson})_{0i}$. Then, \eq{patpaj}
would equivalent to
$\pa_0 (T^{\rm Dyson})^{0i}+ \pa_j (T^{\rm Dyson})^{ji}=0$ if we had $(T^{\rm Dyson})^{0i}=+\rho \dot{u}_i$, and therefore $(T^{\rm Dyson})_{0i}=-\rho \dot{u}_i$, contrary to
the sign in \eq{T0iDyson}. Therefore, after correcting this sign [in agreement with our result (\ref{T0ifinal})], we indeed have
\be\label{panTinuDyson}
\pam (T^{\rm Dyson})^{\mu i}=0\, .
\ee
However, the quantity that is conserved here is a tensor $(T^{\rm Dyson})^{\mu i}$ with a spatial index $i$ and a Lorentz index $\mu$, rather than a full  tensor with two Lorentz indices $\mu,\nu$. This is a consequence of the fact that this conservation equation is related to the symmetry under shifts in field space (\ref{transu}), which has three parameters $a^i$ labeled by a spatial index (rather than four labeled by a Lorentz index).

In contrast, the quantity
denoted $T^{\rm Dyson}_{00}$ in \eq{T00Dyson} is the same as the quantity denoted $\theta_{00}$ in \eq{T00timetransl}. At the level of the  non-relativistic Lagrangian (\ref{LDyson}),
the conservation of $\theta^{\mu\nu}$ is related to a different symmetry, space-time translations (rather than translations in field space), and  the conservation law involving $\theta^{00}$ is  \eq{pa0T00paiTi0}; the `partner'
of $\theta^{00}$, in this conservation equation, is the quantity $\theta^{i0}$ given in \eq{Tio}, which has nothing to do with any component of 
$(T^{\rm Dyson})^{\mu\nu}$. Indeed, $\theta^{00}$ is quadratic in $u$, and so is also its partner $\theta^{i0}$ in the conservation equation. In contrast, $(T^{\rm Dyson})^{i0}$ is linear in $u$. 

In conclusion, the full energy--momentum tensor given in \eqst{T00Dyson}{TijDyson} does not follow from Noether's theorem applied to the Lagrangian (\ref{LDyson}). It appears that Dyson had determined
$T^{\mu i}$, or equivalently  ${T^0}_i$ and ${T^j}_i$, identifying them with  the three currents $j_i^{\mu}$ associated to
the invariance under translation in field space, and has patched them up with a $T^{00}$ assumed to be the same as the $\theta^{00}$ associated to time translation invariance, given by \eq{T00timetransl} (whose partner, in the conservation equation, is actually $\theta^{i0}$ given in \eq{Tio}, see \eq{pa0T00paiTi0}, and not $T^{i0}$). 
Finally, he must have determined $T^{i0}$ by requiring 
that $T^{i0}=T^{0i}$. This, however, is another arbitrary  step: as discussed  in point (3) on page~\pageref{point3}, this apparently innocuous equality actually
states  that the energy flux is the same as the momentum density, and this only holds in a relativistic theory, where  energy and momentum form a four-vector, so it could not follow from his non-relativistic Lagrangian.  As we saw in \eqs{Tio}{Toi}, for instance, the energy--momentum tensor associated to space--time translations of his non-relativistic Lagrangian is not symmetric,  and in this case $\theta_{0i}$ has nothing to do with $\theta_{i0}$, to the extent that one is proportional to $\rho$ and the other to $c_{ijkl}$.

In conclusion,  eventually the result (\ref{T00Dyson}), at least for homogeneous backgrounds,  is almost correct, apart from a sign in $T_{0i}$, a missing factor $\pa_iK_i$ in $T_{00}$ (relevant for the conservation equation and for the coupling to Newtonian fields described by $h_{00}$), and some corrections to $T_{00}$ and $T_{ij}$ that vanish in the isotropic limit. However, its derivation was heuristic at best. Furthermore, in the non-homogenous case the result is not simply obtained by promoting $c_{ijkl}$ to a function of $\vx$. Indeed, already in the isotropic case, many other terms appear, see \eqss{T00inhomo}{T0iinhompapapsi}{Tijinhom} or, with some approximation, the simpler forms given in 
\eqs{equmotioninhomsimplified3}{equmotioninhomsimplified4}.

\section{Comparison with Yan et al.}\label{app:Yan}

Shortly before completion of this paper, appeared on the arxiv  ref.~\cite{Yan:2024jio}. In that work the authors correctly recognize that, to compare the field-theoretical computation of Dyson to the normal mode computation, one must take into account the transformation between the TT frame and the proper detector frame. 
To determine the relation between the two frames they  start from the equations 
\bees
\rho \ddot{u}_i&=&\pa_j\bar{\sigma}_{ij}-h_{ij}\pa_j\mu\, ,\label{Yan1}\\
\rho \ddot{u}^{\rm lab}_i&=&\pa_j\bar{\sigma}^{\rm lab}_{ij}
+\frac{1}{2}\rho \ddot{h}_{ij}x_j\, , \label{Yan2}
\ees
see their eqs. (6') and (10), rewritten  in the  notation of our Section~\ref{sect:Comparison}.
At the notation level, we have further added an overbar on $\sigma_{ij}$ to stress that this is the tensor computed in the free theory, i.e. (restricting to the isotropic case),
\bees
\bar{\sigma}_{ij}(\vx)&=& \lambda(\vx)\delta_{ij}\pa_ku_k +\mu(\vx) (\pa_i u_j+\pa_j u_i)\, ,
\label{sigmaij1}\\
\bar{\sigma}^{\rm lab}_{ij}(\vx)&=& 
\lambda(\vx)\delta_{ij}\pa_ku^{\rm lab}_k +\mu(\vx) (\pa_i u^{\rm lab}_j+\pa_j u^{\rm lab}_i)\, ,
\label{sigmaij2}
\ees
(see eq.~(5) of \cite{Yan:2024jio})
while, when adding the coupling to GWs, we have reserved the notation $\sigma_{ij}$ to denote 
the tensor obtained from \eq{defsigmaij} with the full Lagrangian, including the coupling to $\hmn$, which, in the TT gauge, gives the result for $\sigma_{ij}$ given in \eq{sigmaijhijTT}.

They now impose
the relation
\be\label{Yanuu}
\ddot{u}^{\rm lab}_i(t,\vx)=\ddot{u}_i(t,\vx)+\frac{1}{2} \ddot{h}_{ij}(t)  x_j +{\cal O}(h^2)\, ,
\ee 
 (their eq.~(11), rewritten in our notation), justifying it somehow as ``the rule of addition of acceleration" [note that they never work out explicitly the relation between TT and proper detector frame coordinates, our \eq{xTTxlab}, nor the integrated form of \eq{Yanuu}, our \eq{ddotulabipajm}]. To get consistency between \eqss{Yan1}{Yan2}{Yanuu}
they are then forced to impose  
\be\label{Yancond}
\pa_j\bar{\sigma}_{ij}^{\rm lab}=\pa_j\bar{\sigma}_{ij}-h_{ij}(t)\pa_j\mu\, ,
\ee
see their eq.~(12), that they present as an extra condition that must be required. However, this spurious extra condition only emerges because, unfortunately, the sign in front of $h_{ij}$ in \eq{Yan1} is wrong. This is because  \eq{Yan1} is taken from ref.~\cite{Dyson:1969zgf} which,  as we already discussed in footnote~\ref{foot:signhmn},  defines the metric perturbation from $\gMN=\eMN+\hMN$, while
ref.~\cite{Yan:2024jio} uses the same definition of $\hmn$ as ours, $\gmn=\emn+\hmn$, and therefore (to linear order), $\gMN=\eMN-\hMN$. The correct system of equations is in fact
\bees
\rho \ddot{u}_i&=&\pa_j\bar{\sigma}_{ij}+h_{ij}\pa_j\mu\, ,\label{Yan1correct}\\
\rho \ddot{u}^{\rm lab}_i&=&\pa_j\bar{\sigma}^{\rm lab}_{ij}
+\frac{1}{2}\rho \ddot{h}_{ij}x_j\, . \label{Yan2correct}
\ees
Using the relation (\ref{Yanuu}), the ``consistency condition" now reads
\be\label{Yancondcorr}
\pa_j\bar{\sigma}_{ij}^{\rm lab}=\pa_j\bar{\sigma}_{ij}+h_{ij}(t)\pa_j\mu\, .
\ee
However, if we plug \eq{ddotulabipajm} into the left-hand side of \eq{Yancondcorr},
with $\bar{\sigma}_{ij}^{\rm lab}$ given by \eq{sigmaij2}
(and we recall that $h_{ij}$ is in the TT gauge)  we get automatically the expression in the right-hand side, 
so \eq{Yancondcorr} is automatically satisfied [which was the content of our discussion in \eqst{rhoddotuTTinhom}{paiulabpaiuhij}]. Therefore, there is no extra condition to be imposed. On the other hand, if one used \eqs{Yan1}{Yan2}, including the wrong sign in \eq{Yan1}, and repeats the same steps, one would find a condition on $h_{ij}$ that depends on $\pa_j\mu$,  that is never satisfied by $h_{ij}$, which is an external GWs and knows nothing about $\mu$. 

We finally observe that ref.~\cite{Yan:2024jio} did not include the boundary conditions in the analysis. Our discussion in Section~\ref{sect:Comparison} shows that the same transformation of variables, given in \eq{ddotulabipajm}, also makes consistent the boundary conditions.


\bibliographystyle{utphys}
\bibliography{elasticGW.bib} 

\end{document}